 \pdfoutput=1

\RequirePackage{fix-cm}
\documentclass[twocolumn,epjc3]{svjour3}
\smartqed
\RequirePackage{graphicx}
\RequirePackage{amsmath,amssymb}
\RequirePackage{array,makecell,multirow}
\RequirePackage[colorlinks,citecolor=blue,urlcolor=blue,linkcolor=blue]{hyperref}

\journalname{Eur. Phys. J. C}
\makeatletter\def\makeheadbox{}\makeatother

\newcommand{\hcell}[1]{\Gape[3pt]{\makecell{#1}}}
\makeatletter
\AtBeginDocument{%
  \def\hline{\noalign{\ifnum0=`}\fi\hrule \@height \arrayrulewidth
    \futurelet\reserved@a\@xhline}%
  \def\@xhline{\ifx\reserved@a\hline
    \vskip\doublerulesep\vskip-\arrayrulewidth\fi
    \ifnum0=`{\fi}}%
}
\makeatother

\begin{document}

\title{Echoes of the Maldacena--Milekhin--Popov traversable wormhole}

\author{Rajdeep Mondal\thanksref{e1,addr1,addr2}
        \and
        Abhishake Sadhukhan\thanksref{e2,addr2}
}

\thankstext{e1}{e-mail: rajdeep20263412@students.iisertirupati.ac.in}
\thankstext{e2}{e-mail: abhishake.physics@presiuniv.ac.in}

\institute{Department of Physics, Indian Institute of Science Education and Research Tirupati, Tirupati 517619, India \label{addr1}
           \and
           Department of Physics, Presidency University, Kolkata 700073, India \label{addr2}
}

\date{}

\maketitle

\begin{abstract}
The traversable Maldacena--Milekhin--Popov (MMP) wormhole is held open by quantum effects of charged fermions. Its mouths lie below extremality, which defines its binding energy. We ask whether a massless scalar or a test U(1) field produces echoes. Each mouth acts as the barrier of an extremal charged black hole of radius $r_e$, and the throat as a cavity. A quantum of frequency $\omega$ carries $256\pi\omega r_e/g^2$ times the binding energy, with $g$ the gauge coupling, independent of charge. Echoes form near the barrier top, where one quantum carries more than $50/g^2$ times binding energy. Such a quantum is expected to turn the wormhole into two black holes, and echo trains cannot form. At low frequency a wave crosses one barrier with probability $K_l(\omega r_e)^{4l+2}S_l$, with $l$ the angular momentum. To our knowledge the Coulomb factor $S_l$ is new. Below the binding energy the first signal through the throat has crossed the barrier twice, with probability below $4\times10^{-11}$. For mouths of radius $2.8\,\mathrm{cm}$ a signal crosses the throat in $4.7\times10^{17}\,\mathrm{yr}$, and the cosmic microwave background destroys the wormhole within $10^{-16}\,\mathrm{s}$. For mouths of radius $2.8\times10^{-8}\,\mathrm{cm}$ the crossing takes $47\,\mathrm{yr}$. With a throat held by a dark sector it survives the CMB for $10^{19}\,\mathrm{yr}$. X-ray background photons then limit its life to about $9\times10^{2}\,\mathrm{yr}$, an upper bound from photons alone. In linear theory echo energies fall as $m^{-1-3/(4l+2)}$ with echo number $m$, and one power faster for a pulse sent in from outside.
\end{abstract}

\section{Introduction}
\label{sec:intro}

Can a traversable wormhole held open by quantum effects produce echoes? We answer this question for the wormhole of Maldacena, Milekhin and Popov (MMP) \cite{Maldacena:2018gjk}. The answer turns on energy rather than on scattering. The MMP wormhole is bound by a small energy, the amount by which its two mouths lie below extremality. A single quantum in the frequency band where echoes form carries more than $50/g^2$ times this energy, and $238/g^2$ to $304/g^2$ times for the dipole mode of the scalar, with $g$ the gauge coupling, independent of charge (Sec.~\ref{sec:energy}). Such a quantum is expected to turn the wormhole into two black holes. The fixed background then no longer holds, and the echo trains of linear theory cannot form. Quanta below the binding energy are allowed, but one mouth transmits them with probability below $6.3\times10^{-6}$ (Sec.~\ref{sec:energy}). The same energy budget limits the life of the wormhole in thermal radiation.

We compare two parameter sets. Set~1 has the charge $q=10^{33}$ and centimetre-sized mouths. A signal needs $4.7\times10^{17}$~yr to cross its throat, and the wormhole survives about $10^{-16}$~s in the cosmic microwave background (Sec.~\ref{sec:lifetime}). Set~2 has $q=10^{25}$ and mouths of atomic size, and it fits inside a universe with our cosmological constant. Section~\ref{sec:matching} defines both sets. A signal crosses the throat of set~2 in $47$~yr. If the throat is held by a dark sector, set~2 survives the cosmic microwave background for $4\times10^{26}$~s, longer than the age of the Universe. Photons of the cosmic X-ray background shorten this to about $9\times10^{2}$~yr. This is an upper bound from photons alone. Our lifetimes count massless quanta only, and slow massive particles can only shorten them (Sec.~\ref{sec:lifetime}).

These conclusions need the echo band, the decay of the echoes and the low-frequency transmission of the actual MMP barrier. Most of this paper computes them. The low-frequency law holds for the extremal black hole itself, and the decay laws for any cavity without absorption between two such barriers. They do not depend on the energy argument.

In classical general relativity a traversable wormhole needs matter that violates the null energy condition (NEC), $T_{\mu\nu}k^\mu k^\nu\geq0$ for null $k^\mu$ \cite{Rubakov:2014jja}. Static and dynamic wormholes both violate it near the throat \cite{Morris:1988cz,Hochberg:1998ii}, and quantum fields can violate it locally. The first traversable wormhole consistent with causality couples the two boundaries of an eternal AdS black hole \cite{Gao:2016bin}. MMP constructed a four-dimensional one in Einstein--Maxwell theory with charged massless fermions \cite{Maldacena:2018gjk}. The one-loop Casimir energy of the fermions, which move in the magnetic field of two near-extremal black holes, holds the $AdS_2\times S^2$ throat open. Nearly-$AdS_2$ gravity matches the SYK model at low energy \cite{Sachdev:1992fk,Kitaev:2015k,Maldacena:2016hyu}. This links MMP to eternal traversable wormholes \cite{Maldacena:2018lmt} and to two coupled SYK models \cite{Chen:2019qqe}. Humanly traversable wormholes exist in a Randall--Sundrum II model \cite{Maldacena:2020sxe}, and Einstein--Maxwell--Dirac wormholes are studied in \cite{Blazquez-Salcedo:2020czn,Konoplya:2021hsm,Churilova:2021tgn}. Other models include the Morris--Thorne metric \cite{Morris:1988cz}, braneworlds \cite{Dai:2020rnc} and Casimir wormholes in modified gravity \cite{Battista:2024gud}. The MMP geometry is fixed by microscopic physics, and a calculation on it has no model reflectivity to choose.

A black hole absorbs incoming waves and rings down in quasinormal modes (QNMs) \cite{Berti:2009kk}. A horizonless object has a potential barrier instead \cite{Cardoso:2016oxy}. Its prompt ringdown can still be that of the black hole \cite{Cardoso:2016rao}. Part of the wave is trapped behind the barrier and leaks out as echoes, separated by the light-travel time across the cavity \cite{Cardoso:2016oxy,Cardoso:2016rao,Bueno:2017hyj,Konoplya:2016hmd}. Their shape depends on the cavity and on the barrier \cite{Mark:2017dnq,Zhang:2025ygb}. Tentative evidence for echoes in LIGO/Virgo data \cite{Abedi:2016hgu} has been disputed \cite{Westerweck:2017hus}, and later searches found no significant signal \cite{Uchikata:2023zcu}.

Low-frequency scattering on the MMP background has recently been studied analytically in \cite{Freivogel:2026ujn}. There the two mouths act as a resonant cavity, and the accumulated late-time transmission approaches half the black-hole cross-section. The analysis also covers charged scalars and charged massless fermions, which cross the throat almost freely through the lowest Landau level. It is perturbative in $\omega r_e$, with $r_e$ the radius of a mouth, does not reach the barrier top, and names quasinormal modes and echoes as an open direction.

We solve the scattering numerically, which reaches the barrier top where echoes form. We treat a massless scalar and a test U(1) field. The latter is a second U(1) field and not the MMP photon (Sec.~\ref{sec:em_pert}), and we call it the gauge field. Each mouth is an extremal Reissner--Nordstr\"om exterior up to corrections of relative order $|\hat\epsilon|$, the fractional distance from extremality (Sec.~\ref{sec:worm}), which is $2\times10^{-70}$ for set~1 and $2\times10^{-54}$ for set~2. The single-barrier problem is then the absorption problem of the extremal black hole, whose low-frequency limit is known \cite{Maldacena:1997ih,Gubser:1997cm,Crispino:2000jx,Crispino:2010fd,Oliveira:2011zz,Benone:2014qaa}. The Couch--Torrence inversion symmetry makes the barrier even in the tortoise coordinate \cite{Couch:1984,Bizon:2012we,Lucietti:2012xr}. With it and the closed-form static solutions, the low-frequency transmission follows in closed form. We recover its known prefactor and find its first correction, a Coulomb factor from the logarithm in the tortoise coordinate on both sides of the barrier. This factor fixes the late-time decay of the echoes.

In the tortoise coordinate the barriers are a distance $2D\simeq1.6\times10^{35}r_e$ apart for set~1 and $1.6\times10^{27}r_e$ for set~2, too far for a time-domain evolution. We compute the single-barrier amplitudes by Numerov integration with exact boundary conditions and sum the reflections in closed form. In linear theory a pulse in the throat leaves as an echo train at $T_m=(2m+1)D$, just below the light-ring frequency, with energies that fall as a power of $m$. A pulse sent in from outside gives a train that falls one power faster.

Section~\ref{sec:energy} tests these results against the energy budget. MMP noted that waves can cross the wormhole but small energetic objects cannot \cite{Maldacena:2018gjk}. The mouths lie below extremality by the binding energy of \cite{Maldacena:2020sxe}.

Sections~\ref{sec:worm} to~\ref{sec:global_potentials} set up the geometry, the wave equations and the global potentials. Section~\ref{sec:numerov} treats one mouth, Sec.~\ref{sec:echoes} the echoes and the cavity modes, and Sec.~\ref{sec:energy} the energy budget and the lifetime. The appendices give the master equations, the contour method, the numerics, the low-frequency matching and the tortoise length of the throat.

In Secs.~\ref{sec:perturbations} to~\ref{sec:echoes} and in the appendices we set $G_N=c=1$. Lengths and times are then measured in units of the mouth radius $r_e$, and frequencies in units of $1/r_e$. We quote frequencies as the dimensionless product $\omega r_e$. Section~\ref{sec:energy} restores $\hbar$. Section~\ref{sec:units} collects the conventions and the notation. Numerical values quoted without a table, a figure or an equation are printed by the notebook \cite{Mondal:2026code}, which also draws every data figure.

\section{Wormhole geometry}
\label{sec:worm}
The complete spacetime geometry of the MMP traversable wormhole arises as a solution of the Einstein--Maxwell equations coupled to massless charged fermions \cite{Maldacena:2018gjk}. The asymptotically flat near-extremal Reissner--Nordstr\"om geometry describes each mouth. The $AdS_{2}\times S^{2}$ throat develops near the extremal horizon. The throat is supported by the Casimir energy of massless fermions propagating in the background of a magnetic flux. The integer $q$ denotes the magnetic charge, while $g$ is the $U(1)$ gauge coupling constant. These parameters determine both the extremal radius $r_{e}$ and the characteristic length of the throat. MMP find the throat length scale \cite{Maldacena:2018gjk}
\begin{equation}
    L_{\text{wh}} = \frac{16 \pi^{3/2} q^2 \, \ell_p}{g^3},
    \label{eq:Lwh_def}
\end{equation}
where $\ell_p$ is the Planck length. This scale sets the size of the $AdS_{2}$ region.

\subsection{The near-extremal wormhole mouth}

The wormhole mouth is described by the Reissner--Nordstr\"om metric of a magnetically charged black hole \cite{Maldacena:2018gjk} of mass $M$, with $G_N$ Newton's constant,
\begin{align}
ds^2 = &-\left(1-\frac{2MG_{N}}{r} +\frac{r_{e}^2}{r^2}\right) dt^2 \nonumber\\
&+ \left(1-\frac{2MG_{N}}{r}+\frac{r_{e}^2}{r^2}\right)^{-1} dr^2+r^2d\Omega^2 .
\label{eq:RN_metric}
\end{align}
The horizons of the metric \eqref{eq:RN_metric} are located at $r_{\pm}= MG_{N} \pm \sqrt{M^2 G_{N}^2 -r_{e}^2}$, where $r_+$ and $r_-$ are the outer and inner horizons, respectively. The extremal condition, where the two horizons coincide ($r_+ = r_- = r_e$), corresponds to $MG_N = r_e$.

The MMP mouth is close to, but not exactly at, extremality. We measure the offset by a small parameter $\epsilon$, with dimensions of length,
\begin{equation}
G_{N}M = r_e + \epsilon .
\label{eq:epsilon_def}
\end{equation}
When $\epsilon > 0$, the object is a black hole with two distinct horizons. When $\epsilon < 0$, there is no horizon, and the geometry can be interpreted as the mouth of a wormhole. The dimensionless measure of near-extremality is $\hat\epsilon=\epsilon/r_e$. With \eqref{eq:epsilon_def} the mouth metric \eqref{eq:RN_metric} becomes
\begin{align}
ds^2 = &-\left[\left(1-\frac{r_{e}}{r}\right)^2 - \frac{2\epsilon}{r}\right]dt^2 \nonumber\\
&+ \left[\left(1-\frac{r_{e}}{r}\right)^2 - \frac{2\epsilon}{r}\right]^{-1} dr^2 + r^2 d\Omega^2 .
\label{eq:mouth_metric}
\end{align}

\subsection{The $AdS_2 \times S^2$ wormhole throat}

In the near-horizon limit ($r \to r_e$), the geometry approaches that of $AdS_2 \times S^2$. Following MMP \cite{Maldacena:2018gjk}, the idealised throat metric can be written in $(\tilde\tau,\rho)$ coordinates as
\begin{equation}
    ds^2 = r_e^2 \left[-(\rho^2+1)\,d\tilde\tau^2 + \frac{d\rho^2}{\rho^2+1} + d\Omega^2\right].
    \label{eq:throat_metric}
\end{equation}

To account for the backreaction from the quantum fermions that support the wormhole throat, MMP introduce small perturbations $\gamma(\rho)$ and $\phi(\rho)$ of the metric \eqref{eq:throat_metric} \cite{Maldacena:2018gjk}. The perturbed throat geometry then takes the form
\begin{align}
    ds^2 = r_e^2 \Big[&-(\rho^2+1+\gamma)\,d\tilde\tau^2 + \frac{d\rho^2}{\rho^2+1+\gamma} \nonumber\\
    &+ (1+\phi)\,d\Omega^2\Big].
    \label{eq:perturbed_throat_metric}
\end{align}

Following \cite{Maldacena:2018gjk}, the throat coordinates $(\tilde\tau, \rho)$ are related to the global $(t,r)$ coordinates of the mouth region by
\begin{equation}
    \rho = \frac{L_{\text{wh}}(r-r_{e})}{r_{e}^2}, \qquad \tilde\tau = \frac{t}{L_{\text{wh}}}.
    \label{eq:coord_rel}
\end{equation}
In the coordinates $t$ and $r$ of the mouth, the perturbed throat metric \eqref{eq:perturbed_throat_metric} becomes
\begin{align}
    ds^2 = &-\frac{r_{e}^2}{L_{\text{wh}}^2}\,(1+\rho^2+\gamma(\rho))\,dt^2
           + \frac{L_{\text{wh}}^2}{r_{e}^2}\,\frac{dr^2}{1+\rho^2+\gamma(\rho)} \nonumber\\
           &+ (1+\phi(\rho))\,r_{e}^2\, d\Omega^2,
    \label{eq:throat_metric_tr}
\end{align}
with $\rho(r)$ defined in Eq.~\eqref{eq:coord_rel}. The functions
\begin{equation}\label{eq:gam}
\gamma(\rho)=-\alpha\left[\rho^2+\rho(3+\rho^2)\arctan \rho-\ln(1+\rho^2)\right]
\end{equation}
and $\phi(\rho)=\alpha(1+\rho \arctan \rho)$ arise from the quantum stress-energy tensor supporting the throat \cite{Maldacena:2018gjk}. At large $\rho$ Eq.~\eqref{eq:gam} becomes
\begin{equation}\label{eq:largeg}
   \gamma(\rho) \;\sim\; \alpha\left[-\tfrac{\pi}{2}\rho^3 - \tfrac{3\pi}{2}\rho + 2\ln\rho + \tfrac{8}{3} + O(\rho^{-2})\right].
\end{equation}
Here
\begin{equation}
   \alpha = \frac{q\,G_N}{4\pi r_e^2}
   \label{eq:alpha_def}
\end{equation}
is the constant of MMP \cite{Maldacena:2018gjk}. The $O(\alpha)$ throat with the trace part of the fermion stress tensor kept is derived in \cite{Sadhukhan:2026czl}. Its $\phi$ and $\gamma$ differ from the ones above by the constant $\alpha/3$ and the term $-\alpha\rho^2/3$. These pieces change the $AdS_2$ and $S^2$ radii by the same factor, $r_e^2\to r_e^2(1+\alpha/3)$, and MMP absorb them into the definition of $r_e$ \cite{Maldacena:2018gjk}. In our calculation they enter only the tortoise length of the throat, at relative order $\alpha$, which is $2.5\times10^{-35}$ for set~1 and $2.5\times10^{-27}$ for set~2 (Table~\ref{tab:bg_params}). We use the forms of MMP throughout.

\subsection{Global structure}
\label{sec:global_structure}

The two mouths of the MMP solution lie in the same asymptotically flat space, at a distance $d$ from each other. In this section and in Sec.~\ref{sec:lambda} we set $c=1$ except where $c$ is written. A signal through the throat takes the time $\pi L_{\rm wh}$ seen from outside. It must not arrive before a signal sent through the ambient space \cite{Maldacena:2018gjk,Maldacena:2020sxe},
\begin{equation}
    \pi L_{\rm wh}>d .
    \label{eq:no_shortcut}
\end{equation}
The mouths carry opposite magnetic charges and attract each other \cite{Maldacena:2018gjk}. Each mouth is close to an extremal black hole of mass $r_e/G_N$, and its magnetic attraction equals its gravitational one. Two mouths at rest therefore approach with the relative acceleration
\begin{equation}
    \ddot d=-\frac{4r_e}{d^2},
    \label{eq:ddot_flat}
\end{equation}
twice the Newtonian value. On a circular orbit of frequency $\Omega$ this acceleration is $-\Omega^2d$, which gives the Kepler formula of MMP \cite{Maldacena:2018gjk},
\begin{equation}
    \Omega^2=\frac{4r_e}{d^3}.
    \label{eq:kepler}
\end{equation}
The orbital frequency must stay well below the energy gap of the throat, which is of order $1/L_{\rm wh}$ seen from outside \cite{Maldacena:2018gjk,Maldacena:2020sxe},
\begin{equation}
    \Omega L_{\rm wh}\ll1 .
    \label{eq:slow_orbit}
\end{equation}
With Eq.~\eqref{eq:kepler} this is a lower bound on the separation,
\begin{equation}
    d\gg d_{\min}=\left(4r_eL_{\rm wh}^2\right)^{1/3}.
    \label{eq:dmin}
\end{equation}
MMP also assume $d\ll L_{\rm wh}$ \cite{Maldacena:2018gjk}. In flat space the allowed separations are therefore
\begin{equation}
    d_{\min}\ll d\ll L_{\rm wh}.
    \label{eq:window_flat}
\end{equation} In the scattering problem below, each barrier opens onto its own exterior region. This is adequate for propagation through the throat. The orbital motion is slower than every time-scale of the barrier, and radiation that leaves through the far mouth simply enters the common exterior at a distance $d$ from the near one.

\subsection{Cosmological constant}
\label{sec:lambda}

MMP construct the wormhole in asymptotically flat space, and we use their solution as it stands. Our Universe approaches de Sitter space with the Hubble rate
\begin{equation}
    H_\Lambda=H_0\sqrt{\Omega_\Lambda},\qquad \frac{c}{H_\Lambda}=1.7\times10^{28}~{\rm cm},
    \label{eq:HLambda}
\end{equation}
for the Hubble constant $H_0=67.4$~km\,s$^{-1}$\,Mpc$^{-1}$ and the dark-energy fraction $\Omega_\Lambda=0.685$ \cite{Planck:2018vyg}. Around a compact object the static patch of de Sitter space is flat up to relative corrections $(H_\Lambda r)^2$ for $r\ll1/H_\Lambda$. The flat-space construction is then the leading order in $H_\Lambda L_{\rm wh}$. MMP give the analogous conditions in anti-de Sitter space, $d<R$ and $\Omega R<1$ with $R$ the curvature radius \cite{Maldacena:2018gjk}. In de Sitter space two free bodies recede from each other with the relative acceleration $H_\Lambda^2d$. Equation~\eqref{eq:ddot_flat} becomes
\begin{equation}
    \ddot d=-\frac{4r_e}{d^2}+H_\Lambda^2d .
    \label{eq:ddot_lambda}
\end{equation}
The attraction wins, and the mouths stay bound, only for
\begin{equation}
    d\ll d_\Lambda=\left(\frac{4r_e}{H_\Lambda^2}\right)^{1/3}.
    \label{eq:dLambda}
\end{equation}
With Eq.~\eqref{eq:dmin} the allowed separations are
\begin{equation}
    d_{\min}\ll d\ll d_\Lambda,\qquad \frac{d_{\min}^3}{d_\Lambda^3}=\left(H_\Lambda L_{\rm wh}\right)^2 .
    \label{eq:window_lambda}
\end{equation}
This window is open only for
\begin{equation}
    L_{\rm wh}\ll\frac{1}{H_\Lambda}.
    \label{eq:Lwh_lambda}
\end{equation}
The same condition keeps the crossing time $\pi L_{\rm wh}$ below the Hubble time. It also keeps the de Sitter temperature \cite{Gibbons:1977mu} below the energy gap of the throat,
\begin{equation}
    T_{\rm dS}=\frac{\hbar H_\Lambda}{2\pi k_B}\ll\frac{\hbar c}{k_BL_{\rm wh}}.
    \label{eq:TdS}
\end{equation}
For $g=1$ Eq.~\eqref{eq:Lwh_lambda} means $q\ll3\times10^{29}$, or $r_e\ll10^{-3}$~cm. Set~1 of Sec.~\ref{sec:matching} does not satisfy it, since it gives $L_{\rm wh}=8.6\times10^{6}c/H_\Lambda$. It describes the wormhole in asymptotically flat space, the setting of \cite{Maldacena:2018gjk}. Set~2 satisfies it, with
\begin{align}
    L_{\rm wh}&=8.6\times10^{-10}\,c/H_\Lambda,\nonumber\\
    d_{\min}&=2.8\times10^{10}~{\rm cm},\qquad d_\Lambda=3.1\times10^{16}~{\rm cm}.
    \label{eq:set2_window}
\end{align}
Its mouths can be placed at any separation in this window, a ratio of $10^{6}$. The energy gap of its throat, $\hbar c/(k_BL_{\rm wh})=1.6\times10^{-20}$~K, is far above the de Sitter temperature $T_{\rm dS}=2.2\times10^{-30}$~K. For set~1 the time-scales of the cavity in Sec.~\ref{sec:cavity} exceed the Hubble time and are results of linear theory in asymptotically flat space. For set~2 the crossing time is $47$~yr.

The conditions of this section concern the geometry alone. They do not include the radiation that fills the Universe. Photons of the X-ray background alone destroy set~2 after about $t_X=8.7\times10^{2}$~yr, Eq.~\eqref{eq:tX} of Sec.~\ref{sec:lifetime}. This time is an upper bound on the life of set~2, because Sec.~\ref{sec:lifetime} counts massless quanta only. It is $5.0\times10^{-8}$ of the Hubble time $1/H_\Lambda$, and it covers $18$ crossings of the throat. By Eq.~\eqref{eq:ddot_lambda} the cosmological acceleration changes the separation of the mouths in this time by
\begin{equation}
    \frac{\delta d}{d}=\frac{(H_\Lambda t_X)^2}{2}=1.2\times10^{-15},
\end{equation}
and the bound \eqref{eq:dLambda} plays no role. By Eqs.~\eqref{eq:kepler} and~\eqref{eq:dmin} the orbital period is
\begin{equation}
    \frac{2\pi}{\Omega}=2\pi L_{\rm wh}\left(\frac{d}{d_{\min}}\right)^{3/2}.
    \label{eq:period}
\end{equation}
It is $95$~yr at $d_{\min}$ and exceeds $t_X$ for $d>4.4\,d_{\min}$. Over most of the window the mouths complete less than one orbit before the wormhole collapses. The sky is also at $2.7$~K, far above the energy gap of the throat. Section~\ref{sec:lifetime} states the assumption under which this does not heat the throat.

\section{Field perturbations and wave dynamics}
\label{sec:perturbations}

We reduce the equations of a massless scalar (spin 0) and a test vector field (spin 1) to one-dimensional Schr\"odinger-like equations. Both the mouth and the throat have the static, spherically symmetric form
\begin{equation}
    ds^2=-f\,dt^2+\frac{dr^2}{f}+g_{\theta\theta}\,d\Omega^2 .
    \label{eq:static_form}
\end{equation}
In the mouth Eq.~\eqref{eq:mouth_metric} gives
\begin{equation}
    f=\left(1-\frac{r_e}{r}\right)^2-\frac{2\epsilon}{r},\qquad g_{\theta\theta}=r^2 .
\end{equation}
In the throat Eq.~\eqref{eq:throat_metric_tr} gives
\begin{equation}
    f_{\rm th}=\frac{r_e^2}{L_{\rm wh}^2}\left(1+\rho^2+\gamma\right),\qquad g_{\theta\theta}=h_{\rm th}=r_e^2(1+\phi).
\end{equation}
We use the tortoise coordinate $dr_*=dr/f$. A prime on $f$, $f_{\rm th}$, $h_{\rm th}$ and $g_{\theta\theta}$ denotes $d/dr$.

\subsection{Scalar field}
\label{sec:scalar_pert}
The massless scalar $\Phi$ obeys $\Box\Phi=0$. It has no background value, and its back-reaction on the metric is of second order. We keep the metric fixed. With $\Phi=R(r)Y_{lm}e^{-i\omega t}$ and the master function $\chi=g_{\theta\theta}^{1/2}R$ the equation becomes $d^2\chi/dr_*^2+[\omega^2-V_s]\chi=0$ (\ref{app:scalar}). In the mouth
\begin{equation}
    V_s^{\text{mouth}}(r) = f(r)\left[\frac{f'(r)}{r} + \frac{l(l+1)}{r^2}\right],
    \label{eq:mouth_potential_scalar}
\end{equation}
and in the throat
\begin{align}
    V_s^{\text{throat}}(r) = f_{\text{th}} \Bigg[& \frac{l(l+1)}{h_{\text{th}}} - \frac{f_{\text{th}}\,(h'_{\text{th}})^2}{4 h_{\text{th}}^2} \nonumber\\
    &+ \frac{f_{\text{th}}\,h''_{\text{th}}}{2 h_{\text{th}}} + \frac{f'_{\text{th}}\,h'_{\text{th}}}{2 h_{\text{th}}} \Bigg].
    \label{eq:throat_potential_scalar}
\end{align}

\subsection{Test vector field}
\label{sec:em_pert}
To investigate the spin dependence of trapping and echoes we add a second, minimally coupled U(1) field $\mathcal{A}_\mu$ with field strength $\mathcal{F}_{\mu\nu}$. It has no background value, and no matter in the MMP solution carries its charge. Its perturbations decouple from the metric at linear order, exactly as for the scalar field. We call it the gauge field below. The MMP photon behaves differently. Its perturbation couples to a metric perturbation of the opposite parity, because the background magnetic field is odd under parity. The lowest-Landau-level fermions respond to it as well. When the perturbation has an electric field along the background magnetic field lines, the Schwinger current of these fermions screens it. This gives it the mass found by MMP \cite{Maldacena:2018gjk}. This response does not switch off in the mouth \cite{Maldacena:2018gjk,Sadhukhan:2026mcu}. The coupled system is reduced to master equations in \cite{Sadhukhan:2026czl,Sadhukhan:2026mcu}, and its quasinormal modes can be obtained from them.

We expand $\mathcal{A}_\mu$ in the axial and polar vector harmonics of \cite{Zerilli:1974ai}. For a mode of frequency $\omega$ and multipole $l$ both parities obey $d^2\Psi_g/dr_*^2+[\omega^2-V_g]\Psi_g=0$ with $V_g=f\,l(l+1)/g_{\theta\theta}$ (\ref{app:gp}). It lacks the term $ff'/r$ of $V_s$,
\begin{align}
    V_g^{\text{mouth}}(r) &= f(r) \, \frac{l(l+1)}{r^2}, \label{eq:gauge_mpot} \\
    V_g^{\text{throat}}(\rho) &= f_{\text{th}}(\rho) \, \frac{l(l+1)}{r_e^2 (1+\phi(\rho))}. \label{eq:gauge_thpot}
\end{align}

\subsection{Geometric and potential matching at the junction}
\label{sec:matching}

The mouth metric \eqref{eq:mouth_metric} and the throat metric \eqref{eq:throat_metric_tr} approximate the same geometry in complementary regions. We join them at a radius
\begin{equation}
    \tilde r \equiv r_e(1+\delta), \qquad \delta \ll 1,
    \label{eq:junction_def}
\end{equation}
inside the region where both descriptions hold. We write $\tilde\rho=\rho(\tilde r)$. We take MMP's closed-form results for $r_e$ and $\epsilon$ and $L_{\rm wh}$ of Eq.~\eqref{eq:Lwh_def} \cite{Maldacena:2018gjk}. The constant $\alpha$ then follows from Eq.~\eqref{eq:alpha_def} with $G_N=\ell_p^2$. All background quantities follow without root-finding,
\begin{align}
    r_e &= \frac{\sqrt{\pi}\,q\,\ell_p}{g}, \qquad
    L_{\text{wh}} = \frac{16\pi^{3/2}q^2\ell_p}{g^3}, \nonumber\\
    \alpha &= \frac{q\,\ell_p^2}{4\pi r_e^2}, \qquad
    \epsilon = -\frac{g^3\ell_p}{512\pi^{3/2}q}.
    \label{eq:closed_form_bg}
\end{align}
We use two parameter sets. Set~1 has $q=10^{33}$, $\ell_p=1.6\times10^{-33}$~cm, $g=1$ and $\delta=10^{-6}$. Set~2 has $q=10^{25}$ with the same $\ell_p$, $g$ and $\delta$. With it the wormhole also fits inside a universe with our cosmological constant (Sec.~\ref{sec:lambda}). Table~\ref{tab:bg_params} lists the resulting background quantities in physical units for both sets. The throat is long compared to the mouth, $L_{\rm wh}/r_e=16\pi q/g^2$. The backreaction is weak, $\alpha=g^2/(4\pi^2q)$.

The construction leaves a wide choice of parameters. For $g=1$ and $\delta=10^{-6}$ the conditions listed below and the cosmological-constant condition hold for $2\times10^{4}\ll q\ll3\times10^{29}$. The lower end comes from $\tilde\rho\gg1$. A larger $\delta$ lowers it, and the quantum-fluctuation bound $L_{\rm wh}\ll q^3\ell_p$ alone needs only $q\gg89$. The upper end is the cosmological-constant condition of Sec.~\ref{sec:lambda}. Set~1 lies above this range and fails only that condition. In units of $r_e$ every single-barrier result and every echo shape depends only on $\omega r_e$ and $g$. The charge enters only through $D/r_e\simeq8\pi^2q/g^2$ and through $r_e=\sqrt\pi\,q\ell_p/g$ in cm. Lengths then scale as $q$ and the time-scales of the throat as $q^2$, and the results for any $q$ in this window follow from either set by scaling. We use two sets to show both ends of this freedom. Set~1 is the set of the first version of this work, in asymptotically flat space. Set~2 lies inside the cosmological window and has a crossing time of $47$~yr.

\begin{table}
\centering
\renewcommand{\arraystretch}{1.3}
\setlength{\arrayrulewidth}{0.6pt}
\caption{Closed-form background parameters for $\ell_p=1.6\times10^{-33}$~cm, $g=1$, $\delta=10^{-6}$. Set~1 has $q=10^{33}$ and set~2 has $q=10^{25}$. Lengths are in cm. The other entries are dimensionless. $D$ is the tortoise half-length of Eq.~\eqref{eq:D_def}.}
\label{tab:bg_params}
\begin{tabular}{|l|l|l|}
\hline
Quantity & Set 1 & Set 2 \\
\hline
$q$ & $10^{33}$ & $10^{25}$ \\
$r_e$ & $2.8359$ & $2.8359\times10^{-8}$ \\
$L_{\rm wh}$ & $1.4255\times10^{35}$ & $1.4255\times10^{19}$ \\
$L_{\rm wh}/r_e$ & $5.0265\times10^{34}$ & $5.0265\times10^{26}$ \\
$\alpha$ & $2.5330\times10^{-35}$ & $2.5330\times10^{-27}$ \\
$\epsilon$ & $-5.61\times10^{-70}$ & $-5.61\times10^{-62}$ \\
$\hat\epsilon$ & $-1.9789\times10^{-70}$ & $-1.9789\times10^{-54}$ \\
$\tilde\rho=\rho(\tilde r)$ & $5.0265\times10^{28}$ & $5.0265\times10^{20}$ \\
$\alpha\tilde\rho$ & $1.2732\times10^{-6}$ & $1.2732\times10^{-6}$ \\
$D/r_e$ & $7.8957\times10^{34}$ & $7.8957\times10^{26}$ \\
\hline
\end{tabular}
\end{table}

Direct substitution shows that the closed forms \eqref{eq:closed_form_bg} obey two exact identities,
\begin{equation}
    L_{\text{wh}} = \frac{4r_e}{\pi\alpha}, \qquad
    \epsilon = -\frac{r_e^3}{2L_{\text{wh}}^2}.
    \label{eq:bg_identities}
\end{equation}
At $r=r_e$ the mouth gives $f=-2\epsilon/r_e$. The throat gives $f_{\rm th}=r_e^2/L_{\rm wh}^2$, since $\gamma(0)=0$. The second identity of Eq.~\eqref{eq:bg_identities} makes the two equal. The mouth and throat metric functions agree exactly at the throat centre $\rho=0$. At the junction $\tilde r$ of Eq.~\eqref{eq:junction_def} they agree only up to the terms each description leaves out. With $u=(r-r_e)/r_e$ and $\epsilon$ dropped, the mouth gives
\begin{equation}
    f=\frac{u^2}{(1+u)^2}=u^2-2u^3+3u^4+O(u^5).
\end{equation}
In the throat $\rho=L_{\rm wh}u/r_e$. The term $\rho^2$ gives $u^2$, and the term $-\tfrac{\pi}{2}\alpha\rho^3$ of Eq.~\eqref{eq:largeg} gives
\begin{equation}
    -\frac{\pi}{2}\,\alpha L_{\rm wh}\,\frac{u^3}{r_e}=-2u^3
\end{equation}
by the first identity of Eq.~\eqref{eq:bg_identities}. The throat then gives $f_{\rm th}=u^2-2u^3$ up to terms suppressed by $r_e/L_{\rm wh}$. The first difference is $3u^4$. At $u=\delta$ it gives the relative mismatch $3u^4/u^2=3\delta^2=3\times10^{-12}$ in $f$. In $df/du$ the difference is $12u^3$ against $df/du\simeq2u$, a relative mismatch $6\delta^2$. In the same way $\phi\simeq\tfrac{\pi}{2}\alpha\rho$ gives
\begin{equation}
    h_{\rm th}'=r_e^2\,\frac{d\phi}{d\rho}\,\frac{d\rho}{dr}=\frac{\pi}{2}\,\alpha L_{\rm wh}=2r_e,
\end{equation}
against $2r_e(1+\delta)$ for $r^2$. The relative mismatch in $g_{\theta\theta}'$ is $\delta=10^{-6}$. A direct numerical evaluation confirms all three values.

The potentials built from them are continuous across $\tilde r$ to between $2\times10^{-12}$ and $4\times10^{-12}$ in relative terms for $1\le l\le10$. The scalar potential at $l=0$ has no centrifugal term. It is then dominated by $ff'g_{\theta\theta}'/2g_{\theta\theta}$, and the mismatch $\delta$ in $g_{\theta\theta}'$ together with the jump of $g_{\theta\theta}''$ from about zero to $2$ gives the relative mismatch $1.5\delta=1.5\times10^{-6}$. All these values are the same for both parameter sets. None of this enters the scattering calculation of Sec.~\ref{sec:numerov}. The barrier is computed from the mouth metric alone, and the throat enters only through its tortoise length.

Both sets of Table~\ref{tab:bg_params} satisfy the flat-space conditions under which the MMP solution is a controlled, horizonless wormhole. The parameter $\epsilon$ is negative, and the mouth has no horizon. The charge is large, $q\gg1$, as the one-loop Casimir treatment requires. The junction lies in the overlap region, with $\delta\ll1$ and $\tilde\rho\gg1$. The throat corrections stay small up to the junction, $\alpha\tilde\rho\ll1$. The throat length obeys the quantum-fluctuation bound $L_{\text{wh}}\ll q^3\ell_p$ \cite{Maldacena:2018gjk}. The ratio $L_{\rm wh}/q^3\ell_p$ is $8.9\times10^{-32}$ for set~1 and $8.9\times10^{-24}$ for set~2. The separation of the mouths must lie in the window of Eq.~\eqref{eq:window_flat}. For set~1 Eq.~\eqref{eq:dmin} gives $d_{\min}=6.1\times10^{23}$~cm, about $200$~kpc, against $L_{\rm wh}=1.4\times10^{35}$~cm. For set~2 the window runs from $d_{\min}$ to $d_\Lambda$ of Eq.~\eqref{eq:set2_window}.

The throat is never a shortcut, as Eq.~\eqref{eq:no_shortcut} requires. Light that goes through it takes $\pi L_{\rm wh}/c$, while light through the ambient space takes $d/c$. For set~1 the throat takes $1.5\times10^{25}$~s, or $4.7\times10^{17}$~yr, and light crosses $d_{\min}$ in $2.0\times10^{13}$~s. Since $d\ll L_{\rm wh}$, the ratio $\pi L_{\rm wh}/d$ is much larger than $\pi$ for every allowed $d$ and reaches $7\times10^{11}$ at $d_{\min}$. For set~2 the throat takes $47$~yr. Light crosses this window in $0.95$~s to $1.1\times10^{6}$~s, or $12$~days. The ratio $\pi L_{\rm wh}/d$ lies between $1.4\times10^{3}$ and $1.6\times10^{9}$. A signal through the throat therefore always arrives after one sent outside. MMP describe their solution as a long wormhole that does not violate causality in the ambient space \cite{Maldacena:2018gjk}. Maldacena and Milekhin show that the trip through such a wormhole always takes longer than the trip outside, and note that it cannot be turned into a time machine \cite{Maldacena:2020sxe}.

\subsection{Units and notation}
\label{sec:units}

Sections~\ref{sec:perturbations} to~\ref{sec:echoes} and the appendices use $G_N=c=1$. Then $\hbar=\ell_p^2$. Every length and time is measured in units of the extremal radius $r_e$. This includes $r$, $r_*$, $x$, $t$, $\tau$, $D$, the grid steps and the pulse widths. Frequencies $\omega$, and the real and imaginary parts $\omega_R$ and $\omega_I$ of a quasinormal frequency, are in units of $1/r_e$, and potentials $V$ in units of $1/r_e^2$. The tables, figures and text quote a frequency as the dimensionless product $\omega r_e$. The same value of $\omega r_e$ is a different physical frequency $\omega$ for the two parameter sets. For set~1 $r_e=2.8359$~cm and $c/r_e=1.06\times10^{10}$~s$^{-1}$. For set~2 $r_e=2.8359\times10^{-8}$~cm and $c/r_e=1.06\times10^{18}$~s$^{-1}$. The metric functions $f$ and $h_{\rm th}/r_e^2$, the transmission and reflection probabilities, the quality factors and all ratios are dimensionless. Field amplitudes are normalised to the unit height of the initial pulse.

Section~\ref{sec:energy} restores $\hbar$ and uses $\hbar=c=1$ with $G_N=\ell_p^2$. It quotes energies in eV, temperatures in K and times in s. Physical units otherwise appear only in Secs.~\ref{sec:lambda} and~\ref{sec:matching} with Table~\ref{tab:bg_params}, in Sec.~\ref{sec:global_potentials}, in the ringdown frequency of Sec.~\ref{sec:qnm_photonsphere}, in the separation of the mouths in Sec.~\ref{sec:exterior} and in the time-scales of Secs.~\ref{sec:cavity} and~\ref{sec:qnm_cavity}.

A few letters have a second use. $\Delta\omega$ is the step of the frequency grid in Secs.~\ref{sec:numerovB} and~\ref{sec:bounces} and the spacing of the cavity resonances in Sec.~\ref{sec:cavity}. $\phi$ is a throat function and, in the appendices, the azimuthal angle. In Sec.~\ref{sec:global_potentials} and \ref{app:D} $\theta$ is defined by $\rho=\tan\theta$. $h$ is the Numerov step and $h_{\rm th}$ the sphere function of the throat. $m$ counts echoes and is also the azimuthal number, and $\tilde\tau$ is the throat time of MMP.

\section{Global potentials and spin comparison in tortoise coordinates}
\label{sec:global_potentials}

To assemble the global effective potential seen by an incoming wave, we must express $V_s(r)$ and $V_g(r)$ as functions of the tortoise coordinate $r_*$, defined by $dr_*=dr/f(r)$ with the convention $r_*(r_e)=0$ at the throat centre. The function $f_{\text{th}}(r)$ equals $r_e^2/L_{\text{wh}}^2$ at $\rho=0$, that is $4.0\times10^{-70}$ for set~1 and $4.0\times10^{-54}$ for set~2 (Table~\ref{tab:bg_params}). It rises only to about $\delta^2=10^{-12}$ at $\tilde r$. The throat therefore contributes the tortoise length $J$ of Eq.~\eqref{eq:D_def}, about $7.9\times10^{34}r_e$ for set~1 and $7.9\times10^{26}r_e$ for set~2, although it spans only $\delta\,r_e=10^{-6}r_e$ in $r$.

In the throat we tabulate $r_*$ by high-precision (50-digit) quadrature of $dr_*=L_{\text{wh}}\,d\rho/(1+\rho^2+\gamma(\rho))$. We split the range into 200 subintervals with log-spaced endpoints up to $\tilde\rho$, which is $5\times10^{28}$ for set~1 and $5\times10^{20}$ for set~2. Over this range $f_{\rm th}$ grows by the factor $1+\tilde\rho^2$, $2.5\times10^{57}$ for set~1. \ref{app:D} evaluates the same integral in closed form.

In the mouth, by contrast, $\epsilon$ is negligible. For $r\ge\tilde r$ the term $2\epsilon/r$ of $f$ is smaller than $(1-r_e/r)^2$ by at least the factor $2|\hat\epsilon|/\delta^2$, which is $4\times10^{-58}$ for set~1 and $4\times10^{-42}$ for set~2. With $u\equiv(r-r_e)/r_e$ the metric function is $f=u^2/(1+u)^2$. Integrating $dr_*=r_e(1+u)^2du/u^2=r_e(1+2/u+1/u^2)\,du$ and measuring from the barrier peak gives
\begin{equation}
    x\equiv\frac{r_*-r_*(2r_e)}{r_e} = u + 2\ln u - \frac{1}{u},
    \label{eq:mouth_tortoise_closed_form}
\end{equation}
which vanishes at the peak $u=1$. Substituting $r=r_e(1+u)$, $f$ and $f'=2u/[r_e(1+u)^3]$ into Eqs.~\eqref{eq:mouth_potential_scalar} and~\eqref{eq:gauge_mpot} gives the two potentials in the same variable,
\begin{equation}
    r_e^2V_g=\frac{l(l+1)\,u^2}{(1+u)^4},\qquad r_e^2V_s=r_e^2V_g+\frac{2u^3}{(1+u)^6}.
    \label{eq:V_of_u}
\end{equation}
We invert \eqref{eq:mouth_tortoise_closed_form} by Newton iteration in $\ln u$. This evaluates $V(x)$ to machine precision at any $x$, with no interpolation. The barrier region sits at $r_*-r_*(\tilde r)\sim r_e/\delta$, negligible compared to the throat crossing itself.

The map $u\to1/u$ sends $x\to-x$, since $x(1/u)=1/u-2\ln u-u=-x(u)$. It leaves both potentials \eqref{eq:V_of_u} unchanged, since $u^2/(1+u)^4$ and $u^3/(1+u)^6$ are invariant under it. This is the Couch--Torrence inversion symmetry of the extremal Reissner--Nordstr\"om geometry \cite{Couch:1984}. It fixes the photon sphere $u=1$ and exchanges the near-horizon region with spatial infinity. The barrier is therefore even in the tortoise coordinate, up to corrections of order $|\hat\epsilon|$. Numerically it holds to $1.2\times10^{-15}$. The same symmetry is used in the study of wave equations on the extremal black hole \cite{Bizon:2012we,Lucietti:2012xr}. At large $u$ Eq.~\eqref{eq:mouth_tortoise_closed_form} gives $u\simeq x-2\ln x$, and $V\simeq l(l+1)/u^2$. The potentials then fall as $l(l+1)/x^2$ with corrections of order $\ln|x|/|x|^3$. By the symmetry the same holds on the throat side. The single-barrier problem of Sec.~\ref{sec:numerov} is the scattering problem of the extremal Reissner--Nordstr\"om black hole. The wormhole differs from the black hole only in what lies beyond the barrier. The extremal black hole also has the Aretakis instability on its horizon \cite{Aretakis:2011ha,Lucietti:2012xr}. It needs a horizon and does not arise here, since the throat replaces the horizon by a region of finite tortoise length.

Both potentials rise to a single peak in the mouth and fall off as $r^{-2}$ at large $r$. For both fields and every $l$ the peak sits at $r_{\rm peak}=2r_e$, the photon-sphere radius $r_{ph}$ of the extremal Reissner--Nordstr\"om black hole \cite{Chandrasekhar:1985kt}. This value follows from the inversion symmetry. The potentials \eqref{eq:V_of_u} obey $V(u)=V(1/u)$. Differentiating this identity with respect to $u$ at $u=1$ gives $dV/du=-dV/du$ there, which forces $dV/du=0$. The single maximum sits at $u=1$, that is at $r=2r_e$. A numerical maximisation confirms it to $2\times10^{-8}r_e$. At $u=1$ the potentials \eqref{eq:V_of_u} give the barrier heights
\begin{align}
    V_{g,\max}&=\frac{l(l+1)}{16},\nonumber\\
    V_{s,\max}&=\frac{l(l+1)}{16}+\frac{1}{32}=\frac{2l(l+1)+1}{32}
    \label{eq:heights}
\end{align}
in units of $1/r_e^2$. At the peak $dV/du=0$ and $f=1/4$, and $d^2V/dx^2=f^2\,d^2V/du^2$. The curvatures at the peak in the tortoise coordinate are $d^2V_g/dx^2=-l(l+1)/256$ and $d^2V_s/dx^2=-[4l(l+1)+3]/1024$ in units of $1/r_e^4$.

By Eq.~\eqref{eq:V_of_u} the scalar potential exceeds the gauge potential by $ff'/r=2u^3/[r_e^2(1+u)^6]$ for every $l$. At the peak the offset is $1/(32\,r_e^2)$, Eq.~\eqref{eq:heights}.

\begin{figure}
    \centering
    \includegraphics[width=\columnwidth]{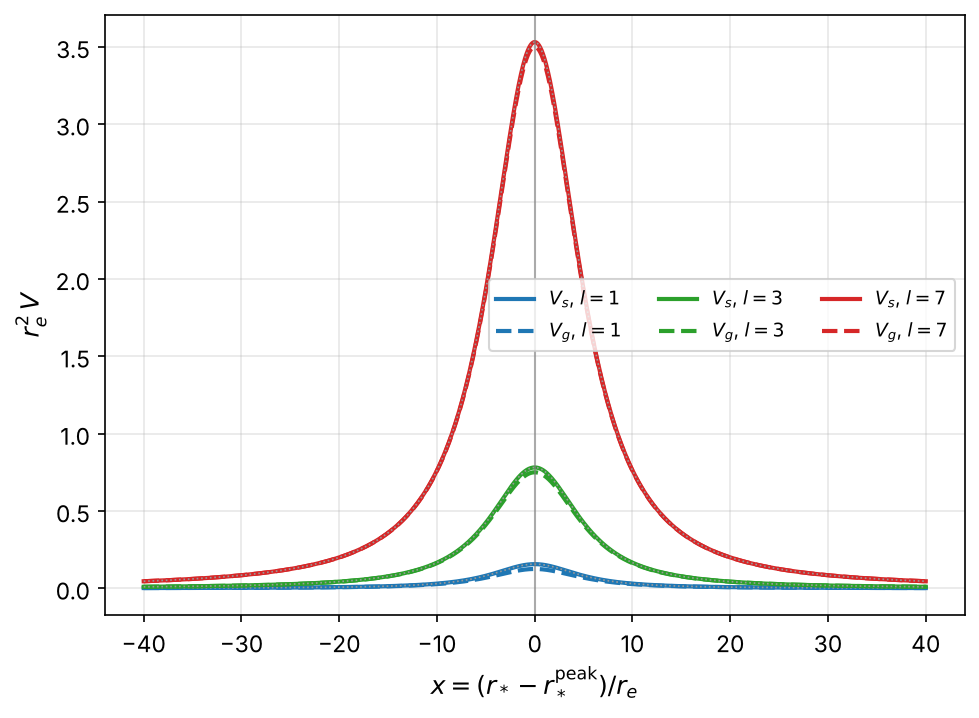}
    \caption{Barrier profile against $x=(r_*-r_*^{\rm peak})/r_e$, scalar (solid) and gauge (dashed), for $l=1,3,7$ (different colours). The potentials are in units of $1/r_e^2$. The grey dotted curves show $V_s(-x)$ and lie on top of $V_s(x)$. In units of $r_e$ it is the same for both parameter sets.}
    \label{fig:Vrtortplot}
\end{figure}

Figure~\ref{fig:Vrtortplot} shows the resulting barrier profile in $x$ for the scalar (solid) and gauge (dashed) potentials at $l=1,3,7$. The profile is strongly skewed in $r$, with a fast rise from $\tilde r$ followed by a long $1/r^2$ tail. In $x$ it is exactly symmetric.

The centrifugal term grows with $l$ while the offset stays fixed. At high multipole the two fields see nearly the same barrier, as Fig.~\ref{fig:Vrtortplot} shows for $l=7$.

\begin{figure}
    \centering
    \includegraphics[width=\columnwidth]{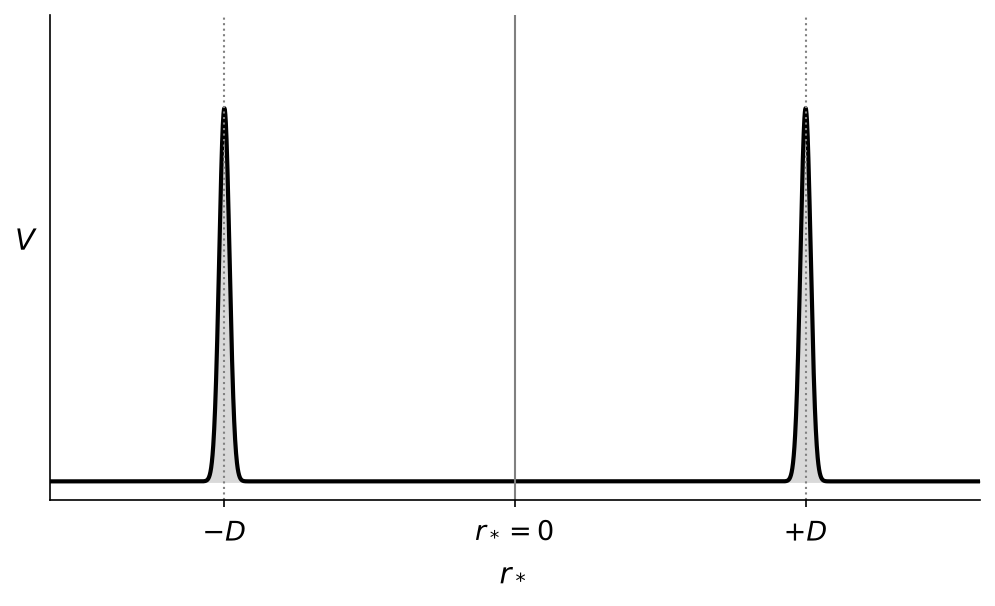}
    \caption{Schematic (not to scale) of the global potential against the tortoise coordinate. Two needle-like barriers at $r_*=\pm D$ are separated by a cavity that is flat on this scale. The picture is symmetric about the throat centre $r_*=0$. The shape is the same for both parameter sets, and only $D$ differs.}
    \label{fig:schematic}
\end{figure}

The throat and mouth contributions to $r_*$, $J$ and $z_{\rm peak}\simeq10^{6}r_e$ of Eq.~\eqref{eq:D_def} below, differ by 29 orders of magnitude for set~1 and 21 for set~2. The full two-mouth potential cannot be plotted to scale. Figure~\ref{fig:schematic} shows it schematically. Two barriers, each about $10\,r_e$ wide at $l=1$ (Sec.~\ref{sec:transmission}), sit at $r_*=\pm D$ and are separated by a cavity of half-width $J$.

The cavity is not exactly flat. Up to corrections of relative order $\alpha\tilde\rho\sim10^{-6}$ the throat has $f_{\rm th}=(r_e^2/L_{\rm wh}^2)(1+\rho^2)$ and $h_{\rm th}=r_e^2$. With $dr=r_e^2d\rho/L_{\rm wh}$ the throat potential is $V_g=l(l+1)(1+\rho^2)/L_{\rm wh}^2$, and the tortoise coordinate is $r_*=L_{\rm wh}\arctan\rho$. Since $1+\rho^2=1/\cos^2(r_*/L_{\rm wh})$, together these give
\begin{equation}
    V_g=\frac{l(l+1)}{L_{\rm wh}^2\cos^2(r_*/L_{\rm wh})},
    \label{eq:throat_PT}
\end{equation}
the P\"oschl--Teller well of global $AdS_2$. Near each end, where $\pi L_{\rm wh}/2-|r_*|\ll L_{\rm wh}$, it reduces to $l(l+1)/(\pi L_{\rm wh}/2-|r_*|)^2$ and continues the centrifugal tail of the mouth. Its floor is $l(l+1)/L_{\rm wh}^2$, that is $4.0\times10^{-70}\,l(l+1)/r_e^2$ for set~1 and $4.0\times10^{-54}\,l(l+1)/r_e^2$ for set~2, and at the junction it has risen to about $\delta^2l(l+1)/r_e^2$. A wave with $\omega L_{\rm wh}\gg l$ crosses this well as a free wave, apart from the phase that the tails already carry. Every frequency at which we compute scattering amplitudes has $\omega r_e\ge5\times10^{-4}$, that is $\omega L_{\rm wh}\ge2.5\times10^{31}$ for set~1 and $2.5\times10^{23}$ for set~2. The thermal integrals of Sec.~\ref{sec:lifetime} are dominated by $\omega\gtrsim k_BT/\hbar$, which for $T\ge10^{-12}$~K means $\omega L_{\rm wh}\ge6\times10^{23}$ for set~1 and $6\times10^{7}$ for set~2. 

The well matters only at $\omega L_{\rm wh}\sim l$, where it fixes the resonances of Sec.~\ref{sec:cavity}. With $\theta=r_*/L_{\rm wh}$ and $\rho=\tan\theta$ the mode equation in the well is
\begin{equation}
    -\frac{d^2\psi}{d\theta^2}+\frac{l(l+1)}{\cos^2\theta}\,\psi=(\omega L_{\rm wh})^2\,\psi .
    \label{eq:well_modes}
\end{equation}
Its solutions regular at $\theta=\pm\pi/2$ are
\begin{equation}
    \psi_n=\cos^{l+1}\theta\;C^{(l+1)}_n(\sin\theta),\qquad \omega L_{\rm wh}=n+l+1,
    \label{eq:ads2_modes}
\end{equation}
with $n=0,1,2,\dots$ and the Gegenbauer polynomials $C^{(l+1)}_n$ \cite[Ch.~22]{Abramowitz:1964}. One checks the frequencies by substitution. These are the normal modes of a field of dimension $l+1$ in global $AdS_2$. Their spacing $1/L_{\rm wh}$ is the energy gap of the throat seen from outside \cite{Maldacena:2018gjk}. We recover them in Sec.~\ref{sec:cavity}.

Both potentials are positive everywhere, in the mouths and in the throat. For the gauge field this is manifest in Eqs.~\eqref{eq:gauge_mpot} and~\eqref{eq:gauge_thpot}. For the scalar the mouth term $f'/r$ of Eq.~\eqref{eq:mouth_potential_scalar} is positive for $r\ge\tilde r$, since $r_ef'(\tilde r)\simeq2(\delta+\hat\epsilon)$ and $\delta\gg|\hat\epsilon|$. In the throat we take $\rho\ge0$, since the throat is symmetric under $\rho\to-\rho$. There $\partial_\rho\phi\ge0$ and $\partial_\rho^2\phi=2\alpha/(1+\rho^2)^2>0$. The two terms of Eq.~\eqref{eq:throat_potential_scalar} with $h''_{\rm th}$ and $f'_{\rm th}h'_{\rm th}$ are positive and of first order in $\alpha$. The negative term $-f_{\rm th}(h'_{\rm th})^2/4h_{\rm th}^2$ is of second order. It is smaller than the term with $f'_{\rm th}h'_{\rm th}$ by the factor $(1+\rho^2)\partial_\rho\phi/4\rho$, which grows to $\pi\alpha\tilde\rho/8=5\times10^{-7}$ at the junction. A wave equation $\partial_t^2\Psi=\partial_{r_*}^2\Psi-V\Psi$ with $V\ge0$ conserves the positive energy
\begin{equation}
    E=\frac12\int\left[(\partial_t\Psi)^2+(\partial_{r_*}\Psi)^2+V\Psi^2\right]dr_* ,
    \label{eq:energy}
\end{equation}
and none of its solutions grows exponentially. The test fields are therefore stable on the fixed background. The stability of the wormhole itself is a separate question. It concerns perturbations of the metric, the MMP gauge field and the fermions (Sec.~\ref{sec:em_pert}).

The two barriers of Fig.~\ref{fig:schematic} sit at $r_*=\pm D$ with
\begin{align}
    D &\equiv r_*(r_{\rm peak}) = J + z_{\rm peak} \simeq \frac{\pi}{2}L_{\text{wh}}, \nonumber\\
    z_{\rm peak} &= r_*(2r_e)-r_*(\tilde r) = 1.000028\times10^{6},
    \label{eq:D_def}
\end{align}
in units of $r_e$. Table~\ref{tab:bg_params} gives $D=7.896\times10^{34}$ for set~1 and $7.896\times10^{26}$ for set~2. Here $J=r_*(\tilde r)$ is the tortoise length of the throat, which we compute by the 50-digit quadrature described at the start of this section. We set $r_e=1$. The peak is at $u=1$ and the junction at $u=\delta$. Eq.~\eqref{eq:mouth_tortoise_closed_form} then gives exactly
\begin{equation}
    z_{\rm peak}=x(1)-x(\delta)=0-\Big(\delta+2\ln\delta-\frac1\delta\Big)=\frac1\delta-2\ln\delta-\delta .
\end{equation}
\ref{app:D} evaluates $J$ to first order in $\alpha$. Adding $z_{\rm peak}$, the terms in $1/\delta$ cancel. Restoring $r_e$,
\begin{equation}
    D=\frac{\pi}{2}L_{\rm wh}+2r_e\ln\frac{L_{\rm wh}}{4r_e}+3r_e,
    \label{eq:D_closed}
\end{equation}
up to terms of order $\delta\,r_e$. The junction drops out, as it should. For set~1 $L_{\rm wh}/r_e$ of Table~\ref{tab:bg_params} gives the correction $2r_e\ln(L_{\rm wh}/4r_e)+3r_e=160.03\,r_e$ to $\pi L_{\rm wh}/2$. For set~2 it is $123.19\,r_e$. For both sets the 50-digit quadrature exceeds it by $3\times10^{-6}r_e$. \ref{app:D} shows that this is the term $3\delta\,r_e$, which comes from $-\delta$ in $z_{\rm peak}$ and from the second-order term of the throat integral.

\section{Scattering by one mouth}
\label{sec:numerov}

A direct time-domain evolution of a wave packet across the cavity is not feasible. The cavity has the width $2D\simeq1.6\times10^{35}r_e$ for set~1 and $1.6\times10^{27}r_e$ for set~2. At the step used below this would need $2D/h\simeq3\times10^{36}$ and $3\times10^{28}$ grid points. This section treats the scattering of a monochromatic wave off a \emph{single} barrier, a one-dimensional problem on the domain $|x|\le X(\omega)$ around $r_*=D$, with $X\ge2000\,r_e$ (Sec.~\ref{sec:numerovB}). Section~\ref{sec:echoes} sums the multiple reflections between the two identical barriers in closed form. Following \cite{Riley:2026avn}, we compute the single-barrier reflection and transmission amplitudes $\beta(\omega)$ and $\kappa(\omega)$ of the actual MMP potential by Numerov integration. We compare them throughout with the two standard leading-order WKB estimates. The same amplitudes give the ringdown of one mouth (Sec.~\ref{sec:qnm_photonsphere}) and the absorption that sets the lifetime in Sec.~\ref{sec:energy}. Everything in this section depends only on $\omega r_e$ and is the same for both parameter sets. Only the conversion of the ringdown to Hz and s differs.

\subsection{The single-barrier scattering problem}
\label{sec:single_barrier}

Let $x$ be the tortoise coordinate centred on the barrier peak, Eq.~\eqref{eq:mouth_tortoise_closed_form}. In the frequency domain the master equations \eqref{eq:master_general_scalar} and \eqref{eq:axial_master} both take the form
\begin{equation}
    \frac{d^2\chi}{dx^2}+\left[\omega^2-V(x)\right]\chi=0,
    \label{eq:master_x}
\end{equation}
with $V=V_s$ or $V_g$. By the symmetry of Sec.~\ref{sec:global_potentials} the potential is even in $x$. It approaches $l(l+1)/x^2$ on both sides, with corrections of order $\ln|x|/|x|^3$ for both fields. The throat lies at $x<-z_{\rm peak}$. There the potential continues the same tail and flattens at $l(l+1)/L_{\rm wh}^2$, Eq.~\eqref{eq:throat_PT}.

A wave that has crossed the throat arrives at the barrier from the left ($x\to-\infty$). The scattering solution satisfies
\begin{equation}
    \chi(x)\to
    \begin{cases}
        e^{+i\omega x}+\beta(\omega)\,e^{-i\omega x}, & x\to-\infty,\\[2pt]
        \kappa(\omega)\,e^{+i\omega x}, & x\to+\infty,
    \end{cases}
    \label{eq:bc}
\end{equation}
with the time dependence $e^{-i\omega t}$. The first line is the wave incident from the throat together with its reflection, and the second is the wave transmitted to the asymptotic region. The reflection and transmission amplitudes obey the exact unitarity relation $|\beta|^2+|\kappa|^2=1$. They are complex, with frequency-dependent phases. Since the barrier is even, a wave incident from the asymptotic region sees the same amplitudes. The modulus $|\kappa(\omega)|^2$ is the transmission probability.

\subsection{Numerov integration}
\label{sec:numerovB}

Equation \eqref{eq:master_x} has no first-derivative term, which makes it suited to the Numerov scheme of Eq.~\eqref{eq:numerov} on a uniform grid $x_j=-X+jh$ (\ref{app:numerics}). We integrate inwards, from $x=+X$ to $x=-X$, starting from the purely outgoing solution at $+X$.

The boundary conditions use the exact solutions of the $l(l+1)/x^2$ tails. These are the Riccati--Hankel functions, which we normalise as
\begin{equation}
    w_+(x)=i^{\,l+1}\,\omega|x|\,h^{(1)}_l(\omega|x|),\qquad w_-(x)=\overline{w_+(x)},
    \label{eq:tail_basis}
\end{equation}
with $h^{(1)}_l$ the spherical Hankel function \cite[Ch.~10]{Abramowitz:1964}. Since $h^{(1)}_l(z)\to(-i)^{l+1}e^{iz}/z$, they tend to $e^{\pm i\omega|x|}$ as $|x|\to\infty$. On the right the outgoing wave is $w_+$. On the left the incident wave $e^{i\omega x}$ is $w_-$ and the reflected wave $e^{-i\omega x}$ is $w_+$. We seed the recursion with $\chi_N=w_+(x_N)$ and $\chi_{N-1}=w_+(x_{N-1})$. At $x=-X$ we decompose the solution as $\chi=A\,w_-+B\,w_+$ using the two adjacent grid values there, by solving the $2\times2$ linear system
\begin{equation}
    \begin{pmatrix}w_-(x_0) & w_+(x_0)\\ w_-(x_1) & w_+(x_1)\end{pmatrix}
    \begin{pmatrix}A\\B\end{pmatrix}
    =
    \begin{pmatrix}\chi_0\\\chi_1\end{pmatrix}.
    \label{eq:AB_system}
\end{equation}
Dividing the whole solution by $A$ gives the normalisation of Eq.~\eqref{eq:bc}, with $\beta=B/A$ and $\kappa=1/A$.

The functions \eqref{eq:tail_basis} are exact for the pure centrifugal tail and need no large-$\omega X$ expansion. Using them at $x=\pm X$ amounts to replacing $V$ by its centrifugal part $l(l+1)/x^2$ beyond $|x|=X$. The difference, of order $\ln|x|/|x|^3$, is the only part of the potential we neglect. We take $X(\omega)=\max(X_{\min},N_\omega/\omega)$ with $X_{\min}=2000\,r_e$ and the dimensionless $N_\omega=400$. Then $\omega X\geq400\gg l$, and the neglected part acts only where the solution is already oscillatory. The step is $h=0.05\,r_e$. The potential on the grid is evaluated exactly through the inversion of Sec.~\ref{sec:global_potentials}. We use the frequencies $\omega_i=i\,\Delta\omega$ with $\Delta\omega=0.002/r_e$ and $i=1,\dots,3000$.

\subsubsection{Validation}

On the exactly solvable P\"oschl--Teller barrier the scheme reproduces $|\kappa|^2$ to $6\times10^{-12}$ and unitarity to $2\times10^{-11}$. For the MMP barriers the unitarity defect stays below $5\times10^{-10}$ over the whole frequency grid. Since it cannot test $|\kappa|^2$ far below $10^{-10}$, we also test the relative accuracy of $|\kappa|^2$ directly. Halving or doubling the step, the domain or $N_\omega$ changes $|\kappa|^2$ by less than $7\times10^{-7}$. An adaptive integration in a different coordinate agrees to $2\times10^{-6}$ down to $|\kappa|^2\sim10^{-17}$, and a 20-digit recomputation agrees to $2.5\times10^{-7}$ down to $|\kappa|^2\sim10^{-30}$. A leapfrog evolution of a Gaussian pulse through the barrier shares no code with the Numerov solver. It reproduces the transmitted and reflected waveforms predicted from $\beta(\omega)$ and $\kappa(\omega)$ to $0.6\%$ of the peak. This confirms the dispersive phases of $\beta$ and $\kappa$, which the real-phase WKB treatment below discards. \ref{app:numerics} gives the details, the convergence table and the waveforms.

\subsection{WKB estimates}
\label{sec:wkb}

We compare the Numerov reflection and transmission amplitudes, $\beta(\omega)$ and $\kappa(\omega)$, with the two standard semi-analytic approximations. The Schutz--Will formula \cite{Schutz:1985km,Iyer:1986np} replaces the barrier by the inverted parabola fitted to its peak,
\begin{equation}
    |\kappa_{\rm par}(\omega)|^2=\frac{1}{1+e^{2\pi\Lambda(\omega)}},\qquad
    \Lambda(\omega)=\frac{V_0-\omega^2}{\sqrt{-2V_0''}},
    \label{eq:wkb_par}
\end{equation}
with $V_0$ the height of the barrier and $V_0''=d^2V/dx^2$ at the peak. Both $V_0$ and $V_0''$ are given in closed form in Sec.~\ref{sec:global_potentials}. This formula gives $|\kappa_{\rm par}|^2=1/2$ exactly at the barrier top for any $V_0,V_0''$. It is the leading order of the WKB treatment of black-hole quasinormal modes and echoes \cite{Konoplya:2019hlu,Cardoso:2016rao}. Higher orders \cite{Iyer:1986np,Konoplya:2019hlu} improve it near the top but still expand about the peak. Our statements about WKB refer to the leading order. The second formula is the full tunnelling-action formula \cite{Riley:2026avn},
\begin{align}
    |\kappa_{\rm act}(\omega)|^2&=\frac{1}{1+e^{2I(\omega)}},\nonumber\\
    I(\omega)&=\int_{x_<}^{x_>}\sqrt{V(x)-\omega^2}\,dx,
    \label{eq:wkb_act}
\end{align}
where $x_<$ and $x_>$ are the classical turning points of the actual potential. It reduces to $e^{-2I}$ deep below the barrier. Above the barrier there are no turning points, and we set it to $1$. It does not describe above-barrier reflection. Eq.~\eqref{eq:wkb_par} is accurate only where the barrier is close to its quadratic expansion. Eq.~\eqref{eq:wkb_act} is accurate only where $I\gg1$. In both cases we adopt the real-phase convention $\beta=-\sqrt{1-|\kappa|^2}$, $\kappa=+\sqrt{|\kappa|^2}$, which preserves unitarity exactly but drops the dispersion phases.

\subsection{Transmission through the barrier}
\label{sec:transmission}

\begin{figure*}
    \centering
    \includegraphics[width=0.32\textwidth]{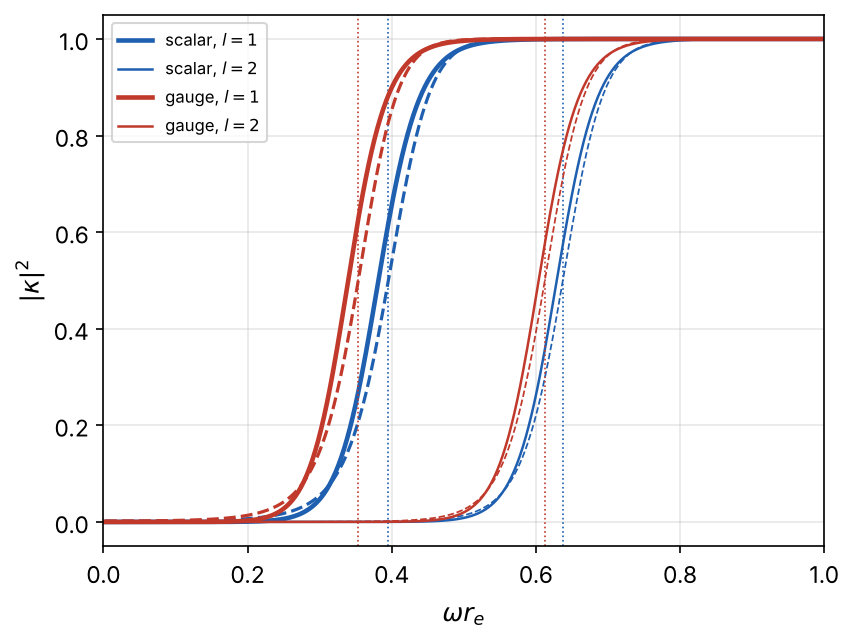}\hfill
    \includegraphics[width=0.32\textwidth]{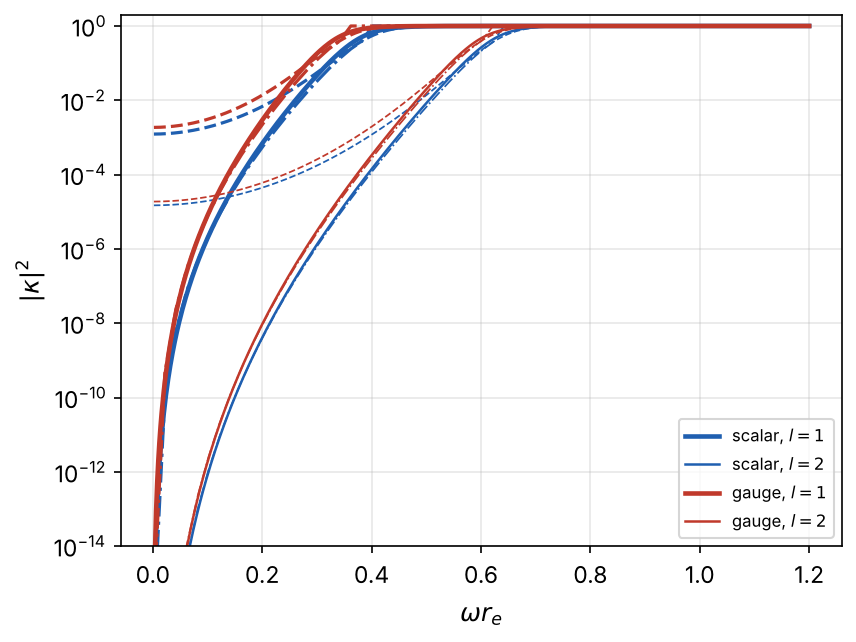}\hfill
    \includegraphics[width=0.32\textwidth]{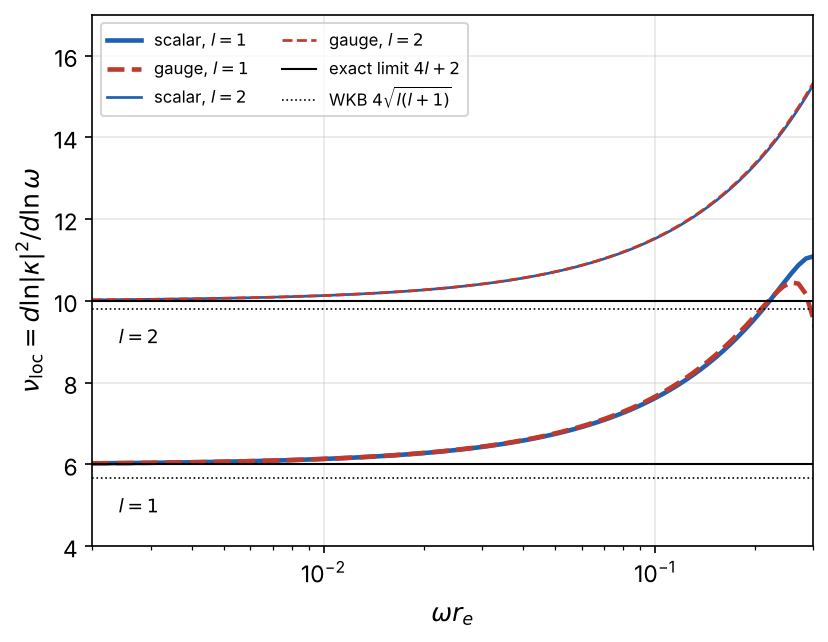}
    \caption{Single-barrier transmission probability $|\kappa(\omega)|^2$ for the scalar (blue) and gauge (red) fields at $l=1$ (thick) and $l=2$ (thin). Frequencies are in units of $1/r_e$. Left, linear scale with the barrier-top frequencies $\sqrt{V_0}$ dotted. Solid lines are Numerov and dashed lines the parabolic (Schutz--Will) WKB formula, Eq.~\eqref{eq:wkb_par}. Middle, logarithmic scale showing the sub-barrier suppression, with the tunnelling-action WKB formula, Eq.~\eqref{eq:wkb_act}, added (dash-dotted). The parabolic formula saturates at a spurious constant value as $\omega\to0$, while the tunnelling-action formula tracks the Numerov result. Right, local exponent $d\ln|\kappa|^2/d\ln\omega$ from Numerov. Solid horizontal lines mark the limit $4l+2$ of Eq.~\eqref{eq:lowfreq} at $\omega\to0$ and dotted lines the WKB estimate $4\sqrt{l(l+1)}$ without the Langer correction. In units of $r_e$ it is the same for both parameter sets.}
    \label{fig:transmission}
\end{figure*}

Figure~\ref{fig:transmission} and Table~\ref{tab:transmission} summarise the single-barrier transmission for both fields at $l=1$ and $l=2$.

\begin{table*}
\centering
\renewcommand{\arraystretch}{1.3}
\setlength{\arrayrulewidth}{0.6pt}
\caption{Transmission probability $|\kappa(\omega)|^2$ through one barrier at $l=1$ (upper block) and $l=2$ (lower block), from Numerov, the parabolic WKB formula \eqref{eq:wkb_par} and the tunnelling-action WKB formula \eqref{eq:wkb_act}. The left block is the scalar field and the right block the gauge field. Entries are dimensionless. The barrier tops are at $\sqrt{V_0}=0.3953$ (scalar) and $0.3536$ (gauge) for $l=1$, and $0.6374$ and $0.6124$ for $l=2$. Frequencies are given as $\omega r_e$. The entries are the same for both parameter sets.}
\label{tab:transmission}
\begin{tabular}{|c|c|c|c|c|c|c|}
\hline
\multirow{2}{*}{$\omega r_e$} & \multicolumn{3}{c|}{scalar field} & \multicolumn{3}{c|}{gauge field} \\
\cline{2-7}
 & Numerov & \hcell{parabolic\\WKB} & \hcell{action\\WKB} & Numerov & \hcell{parabolic\\WKB} & \hcell{action\\WKB} \\
\hline
\multicolumn{7}{|c|}{$l=1$} \\
\hline
0.05 & $1.39\times10^{-8}$ & $1.37\times10^{-3}$ & $1.79\times10^{-8}$ & $5.62\times10^{-8}$ & $2.11\times10^{-3}$ & $7.03\times10^{-8}$ \\
0.10 & $1.95\times10^{-6}$ & $1.89\times10^{-3}$ & $1.97\times10^{-6}$ & $7.97\times10^{-6}$ & $3.08\times10^{-3}$ & $7.78\times10^{-6}$ \\
0.20 & $6.87\times10^{-4}$ & $6.81\times10^{-3}$ & $5.43\times10^{-4}$ & $2.93\times10^{-3}$ & $1.38\times10^{-2}$ & $2.20\times10^{-3}$ \\
0.25 & $6.45\times10^{-3}$ & $1.77\times10^{-2}$ & $4.71\times10^{-3}$ & $2.76\times10^{-2}$ & $4.14\times10^{-2}$ & $1.91\times10^{-2}$ \\
0.30 & $0.047$ & $0.055$ & $0.033$ & $0.181$ & $0.147$ & $0.123$ \\
0.35 & $0.244$ & $0.191$ & $0.173$ & $0.597$ & $0.469$ & $0.468$ \\
0.40 & $0.656$ & $0.540$ & $1$ & $0.899$ & $0.853$ & $1$ \\
0.45 & $0.913$ & $0.879$ & $1$ & $0.980$ & $0.980$ & $1$ \\
0.55 & $0.996$ & $0.998$ & $1$ & $0.999$ & $1.000$ & $1$ \\
\hline
\multicolumn{7}{|c|}{$l=2$} \\
\hline
0.10 & $8.11\times10^{-13}$ & $1.96\times10^{-5}$ & $9.08\times10^{-13}$ & $1.84\times10^{-12}$ & $2.51\times10^{-5}$ & $2.04\times10^{-12}$ \\
0.20 & $4.10\times10^{-9}$ & $4.45\times10^{-5}$ & $3.96\times10^{-9}$ & $9.39\times10^{-9}$ & $5.99\times10^{-5}$ & $9.01\times10^{-9}$ \\
0.30 & $1.31\times10^{-6}$ & $1.75\times10^{-4}$ & $1.16\times10^{-6}$ & $3.05\times10^{-6}$ & $2.56\times10^{-4}$ & $2.67\times10^{-6}$ \\
0.40 & $1.40\times10^{-4}$ & $1.18\times10^{-3}$ & $1.16\times10^{-4}$ & $3.32\times10^{-4}$ & $1.95\times10^{-3}$ & $2.72\times10^{-4}$ \\
0.50 & $8.37\times10^{-3}$ & $1.37\times10^{-2}$ & $6.59\times10^{-3}$ & $1.99\times10^{-2}$ & $2.59\times10^{-2}$ & $1.55\times10^{-2}$ \\
0.60 & $0.262$ & $0.220$ & $0.211$ & $0.465$ & $0.393$ & $0.391$ \\
0.70 & $0.922$ & $0.908$ & $1$ & $0.967$ & $0.966$ & $1$ \\
0.85 & $0.999$ & $1.000$ & $1$ & $1.000$ & $1.000$ & $1$ \\
\hline
\end{tabular}
\end{table*}

\subsubsection{Barrier top}
\label{sec:barrier_top}

At $\omega=\sqrt{V_0}$ the numerical transmission is $|\kappa|^2=0.619$ (scalar) and $0.628$ (gauge) for $l=1$, and $0.574$ and $0.576$ for $l=2$. All exceed the value $1/2$ of a parabolic barrier, as for the Ellis--Bronnikov throat, where \cite{Riley:2026avn} finds $0.68$--$0.70$. The wavelength at the top, $2\pi/\sqrt{V_0}\simeq15.9\,r_e$ (scalar) and $17.8\,r_e$ (gauge) at $l=1$, is comparable to the width of the barrier core where $V\gtrsim V_0/2$, $10.4\,r_e$ and $11.0\,r_e$. The wave samples the full profile with its $1/x^2$ tails rather than the quadratic expansion. The transmission is $1/2$ at $0.965\sqrt{V_0}$ (scalar) and $0.960\sqrt{V_0}$ (gauge) for $l=1$, and at $0.987$ and $0.986$ for $l=2$. At the top it falls towards $1/2$ as $l$ grows, from $0.75$ at $l=0$ to $0.54$ at $l=4$ for the scalar. The parabolic formula \eqref{eq:wkb_par} is in error there by $0.119$ (scalar) and $0.128$ (gauge) at $l=1$, and by $0.074$ and $0.076$ at $l=2$.

\subsubsection{Low-frequency limit}

Below the top the transmission falls steeply, but as a power law rather than exponentially, because the classically forbidden region is bounded by the long-range $1/x^2$ tails. For $\omega r_e\ll1$ the field near the barrier obeys Eq.~\eqref{eq:master_x} with $\omega=0$. For the scalar its solutions are $(1+u)\,u^{l}$ and $(1+u)\,u^{-l-1}$. For the gauge field they are $u^l(lu+l+1)$ and its Couch--Torrence image $u^{-l-1}(l+(l+1)u)$. Let $\Xi(x)$ be the static solution that decays as $x^{-l}$ for $x\to+\infty$. For $x\to-\infty$ it grows as $\mu_l|x|^{l+1}$, with $\mu_l=1$ for the scalar and $\mu_l=l/(l+1)$ for the gauge field. Far from the barrier, to leading order in $\omega r_e$, the solutions are Riccati--Bessel functions of $\omega x$. Matching the outgoing wave at $x>0$ to $\Xi$ and continuing $\Xi$ to $x<0$ gives (\ref{app:lowfreq})
\begin{equation}
    |\kappa|^2\simeq K_l\,(\omega r_e)^{4l+2},\quad K_l=\frac{4}{\big[(2l-1)!!\,(2l+1)!!\,\mu_l\big]^2}.
    \label{eq:lowfreq}
\end{equation}
Here $(-1)!!=1$. The gauge value holds for $l\ge1$, since $\mu_0=0$ and the gauge field has no $l=0$ mode (\ref{app:gp}). For the $l=0$ scalar Eq.~\eqref{eq:lowfreq} is $4(\omega r_e)^2$, the absorption probability of an extremal black hole whose cross-section equals the horizon area. The prefactor $K_l$ is not new. For the scalar it is the zero-temperature limit of the general-$l$ cross section of \cite{Maldacena:1997ih}. For the test Maxwell field it is the result of \cite{Crispino:2000jx} in four dimensions, and its $l=1$ value $16/9$ is given in \cite{Gubser:1997cm}. Our derivation uses the closed-form static solutions and the Couch--Torrence symmetry, and it extends to the Coulomb factor below. The law of Eq.~\eqref{eq:lowfreq} is semiclassical. For a near-extremal black hole, quantum fluctuations of the throat modify the low-frequency cross-section \cite{Emparan:2025sao}. We work on the fixed MMP background and do not include such effects. The tunnelling action $I$ of Eq.~\eqref{eq:wkb_act} shows where the exponent comes from. For $\omega r_e\ll1$ the action is dominated by the two tails, where $V\simeq l(l+1)/x^2$. Each tail runs from $|x|\sim r_e$ to the turning point $|x|=\sqrt{l(l+1)}/\omega$, and away from the turning point $\sqrt{V-\omega^2}\simeq\sqrt{l(l+1)}/|x|$. To logarithmic accuracy
\begin{align}
    I&\simeq2\int_{r_e}^{\sqrt{l(l+1)}/\omega}\frac{\sqrt{l(l+1)}}{x}\,dx\simeq2\sqrt{l(l+1)}\,\ln\frac{1}{\omega r_e},\nonumber\\
    |\kappa|^2&\simeq e^{-2I}\propto(\omega r_e)^{4\sqrt{l(l+1)}} .
\end{align}
The Langer replacement $l(l+1)\to(l+\tfrac12)^2$ \cite{Langer:1937qr} turns the exponent into $4(l+\tfrac12)=4l+2$. Each tail thus suppresses the amplitude by a factor of order $(\omega r_e)^{l+1/2}$.

Our numerical transmission confirms both the exponent and the prefactor (Figs.~\ref{fig:transmission} and~\ref{fig:lowfreq}, Table~\ref{tab:powerlaw}). The local exponent $d\ln|\kappa|^2/d\ln\omega$ is $6.03$ at $\omega r_e=0.002$ for $l=1$ and $10.03$ for $l=2$. Fits over the higher window $0.015\le\omega r_e\le0.06$ give $6.46$ and $10.44$ (Table~\ref{tab:powerlaw}). Exponents fitted at finite frequency are steeper than the true limit. Most of the excess comes from the Coulomb factor $S_l$ of Sec.~\ref{sec:coulomb}, which adds about $4\pi\omega r_e$ to the local exponent. The law $K_l(\omega r_e)^{4l+2}S_l$ gives $6.026$ at $\omega r_e=0.002$, the Numerov value, and $6.39$ over the fit window. The remaining $0.07$ for $l=1$ and $0.05$ for $l=2$ are of order $(\omega r_e)^2$. The ratio of $|\kappa|^2$ to Eq.~\eqref{eq:lowfreq} extrapolates to unity within $3\times10^{-4}$ for $l=0,1,2$ and both fields. Its first correction grows linearly in $\omega$. A fit of the logarithm of the ratio at small $\omega$ gives a slope between $12.6$ and $12.8$ in $\omega r_e$, close to $4\pi\simeq12.57$.

\subsubsection{Coulomb factor}
\label{sec:coulomb}

\begin{figure}
    \centering
    \includegraphics[width=0.92\columnwidth]{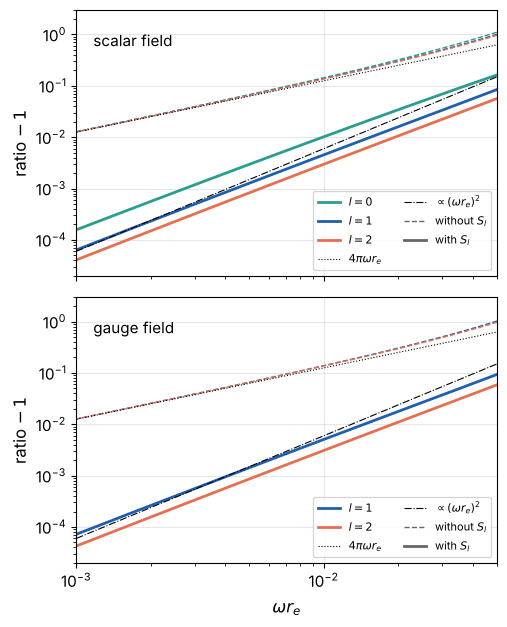}
    \caption{Deviation of the Numerov transmission probability from the low-frequency law, for the scalar field (top) and the gauge field (bottom). Solid curves show $|\kappa|^2/[K_l(\omega r_e)^{4l+2}S_l]-1$, with the Coulomb factor of Eq.~\eqref{eq:sommerfeld}. Dashed curves show $|\kappa|^2/[K_l(\omega r_e)^{4l+2}]-1$, without it. Colours mark $l$. Without $S_l$ the deviation follows $4\pi\omega r_e$ (dotted). With $S_l$ it is smaller by two orders of magnitude at $\omega r_e=10^{-3}$ and falls as $(\omega r_e)^2$ (dash-dotted). In units of $r_e$ it is the same for both parameter sets.}
    \label{fig:lowfreq}
\end{figure}

The first correction to Eq.~\eqref{eq:lowfreq} comes from the logarithm in the tortoise coordinate \eqref{eq:mouth_tortoise_closed_form}. At large $u$ we have $x\simeq u+2\ln u$. The term $2\ln u$ is the tortoise image of the Newtonian part $-2G_NM/r$ of $f$, and it acts on a wave like a long-range attractive $1/r$ potential. We write the master equation \eqref{eq:master_x} in the variable $u$ and set $\hat\chi=f^{1/2}\chi$, which removes the first derivative. This gives Eq.~\eqref{eq:phi_eq} of \ref{app:lowfreq}. At large $u$ the frequency term of that equation is
\begin{equation}
    \omega^2r_e^2\,\frac{(1+u)^4}{u^4}=\omega^2r_e^2\Big(1+\frac4u+\frac{6}{u^2}+\dots\Big),
\end{equation}
and the short-range term $U\simeq2/u^3$ is small compared with $l(l+1)/u^2$. The term $6\omega^2r_e^2/u^2$ shifts $l(l+1)$ by $6\omega^2r_e^2$, an effect of order $\omega^2r_e^2$, and we drop it. With $\xi=\omega r_eu$ we have $d^2/du^2=\omega^2r_e^2\,d^2/d\xi^2$, and
\begin{align}
    0&=\frac{d^2\hat\chi}{du^2}+\Big[\omega^2r_e^2\Big(1+\frac4u\Big)-\frac{l(l+1)}{u^2}\Big]\hat\chi\nonumber\\
    &=\omega^2r_e^2\Big\{\frac{d^2\hat\chi}{d\xi^2}+\Big[1+\frac{4\omega r_e}{\xi}-\frac{l(l+1)}{\xi^2}\Big]\hat\chi\Big\}.
\end{align}
In the standard form
\begin{equation}
    \frac{d^2\hat\chi}{d\xi^2}+\left[1-\frac{2\eta}{\xi}-\frac{l(l+1)}{\xi^2}\right]\hat\chi=0,\qquad \eta=-2\omega r_e,
    \label{eq:coulomb}
\end{equation}
which holds up to terms of order $\omega^2r_e^2$. This is the radial Schr\"odinger equation of a charged particle with angular momentum $l$ in a Coulomb potential, written in the scaled radius $\xi$. Its solutions are the Coulomb wave functions \cite[Ch.~14]{Abramowitz:1964}. The strength $\eta$ is the Sommerfeld parameter. It is negative here because the gravitational term attracts. We call Eq.~\eqref{eq:coulomb} the Coulomb problem of the mouth. Near $\xi=0$ the regular solution behaves as $C_l(\eta)\,\xi^{l+1}$, with the normalisation \cite[Ch.~14]{Abramowitz:1964}
\begin{equation}
    C_l(\eta)=\frac{2^l\,e^{-\pi\eta/2}\,|\Gamma(l+1+i\eta)|}{(2l+1)!} .
    \label{eq:Cl}
\end{equation}
For $\eta=0$ this is $1/(2l+1)!!$, the value for a free wave. An attractive potential pulls the wave in, and $C_l(\eta)>C_l(0)$ for $\eta<0$. This is the Sommerfeld enhancement of nuclear and atomic physics. By the Couch--Torrence symmetry the throat side obeys Eq.~\eqref{eq:coulomb} in the variable $\omega r_e/u$. Each side multiplies $\kappa$ by $C_l(\eta)/C_l(0)$ and $|\kappa|^2$ by the square of this ratio (\ref{app:lowfreq}). With two sides,
\begin{align}
    |\kappa|^2&\simeq K_l\,(\omega r_e)^{4l+2}\,S_l(\omega r_e),\nonumber\\
    S_l&=\left[\frac{C_l(\eta)}{C_l(0)}\right]^4=\left[\frac{e^{2\pi\omega r_e}\,|\Gamma(l+1+2i\omega r_e)|^2}{(l!)^2}\right]^2 .
    \label{eq:sommerfeld}
\end{align}
With $\varpi=2\omega r_e=|\eta|$, the recurrence $\Gamma(z+1)=z\Gamma(z)$ and $|\Gamma(1+i\varpi)|^2=\pi\varpi/\sinh\pi\varpi$ \cite[Ch.~6]{Abramowitz:1964} give
\begin{align}
    |\Gamma(l+1+i\varpi)|^2&=\frac{\pi\varpi}{\sinh\pi\varpi}\prod_{j=1}^{l}(j^2+\varpi^2),\nonumber\\
    S_l&=\Big[\frac{2\pi\varpi}{1-e^{-2\pi\varpi}}\prod_{j=1}^{l}\Big(1+\frac{\varpi^2}{j^2}\Big)\Big]^2 .
\end{align}
For $l=0$ the product is empty and
\begin{equation}
    S_0=\left[\frac{4\pi\omega r_e}{1-e^{-4\pi\omega r_e}}\right]^2 .
    \label{eq:S0}
\end{equation}
For small $\varpi$ we have $2\pi\varpi/(1-e^{-2\pi\varpi})=1+\pi\varpi+O(\varpi^2)$ and the product is $1+O(\varpi^2)$. Hence, for every $l$,
\begin{equation}
    S_l=1+2\pi\varpi+O(\varpi^2)=1+4\pi\omega r_e+O(\omega^2r_e^2).
    \label{eq:Sl_small}
\end{equation} The factor $S_l$ removes the linear growth of the ratio (Fig.~\ref{fig:lowfreq}). With it the fitted slope drops below $0.21$ for $l=0,1,2$ and both fields, and the deviation of the $l=1$ scalar at $\omega r_e=0.01$ falls from $14\%$ to $0.5\%$. At $\omega r_e=0.05$ it falls from $101\%$ to $8.5\%$. The residual slope is not a linear term. When $\omega r_e$ doubles from $10^{-3}$ to $2\times10^{-3}$ and again to $4\times10^{-3}$, the remainder grows by the factors $3.58$ and $3.53$ for the $l=0$ scalar, and by $3.64$ and $3.60$ for the $l=1$ gauge field. A linear term would double. A term $(\omega r_e)^2\ln(1/\omega r_e)$ grows by $3.60$ and $3.55$. The remainder is of order $(\omega r_e)^2$ with a logarithm, which the leading matching does not control.

Coulomb waves in the far region of the extremal black hole are used in \cite{Benone:2014qaa}, which finds one Gamow factor at $l=0$, the square root of $S_0$ in Eq.~\eqref{eq:S0}. The near-horizon side is treated there without it. The inversion symmetry puts an identical Coulomb problem on that side, and the factor appears squared. The $l=0$ ratio of $|\kappa|^2$ to $4(\omega r_e)^2$ is $1.053$ at $\omega r_e=4\times10^{-3}$, $1.0260$ at $2\times10^{-3}$, $1.013$ at $10^{-3}$ and $1.006$ at $5\times10^{-4}$. At $\omega r_e=2\times10^{-3}$ the squared factor gives $1.0254$, and a single factor would give $1.0126$. The numerics support the squared factor.

The tunnelling-action formula, Eq.~\eqref{eq:wkb_act}, tracks the Numerov result in this regime. Below the top it stays within $32\%$ at $l=1$ and $22\%$ at $l=2$ in Table~\ref{tab:transmission}, down to $|\kappa|^2\sim10^{-12}$. The parabolic formula, Eq.~\eqref{eq:wkb_par}, saturates at a finite value at zero frequency, $1.2\times10^{-3}$ for the scalar at $l=1$. At $\omega r_e=0.05$ it overestimates the transmission by five orders of magnitude, $1.37\times10^{-3}$ against $1.39\times10^{-8}$ (Table~\ref{tab:transmission}). The real barrier reflects perfectly as $\omega\to0$ for every $l$.

\begin{table*}
\centering
\renewcommand{\arraystretch}{1.35}
\setlength{\arrayrulewidth}{0.6pt}
\caption{Low-frequency transmission $|\kappa|^2\simeq K_l(\omega r_e)^\nu$. Local exponent at $\omega r_e=0.002$ and exponent fitted over $0.015\le\omega r_e\le0.06$, from Numerov and from the law $K_l(\omega r_e)^{4l+2}S_l$ of Eq.~\eqref{eq:sommerfeld} with the same estimators. They are compared with the limit $4l+2$ of Eq.~\eqref{eq:lowfreq} at $\omega\to0$ and the WKB estimate $4\sqrt{l(l+1)}$ of Sec.~\ref{sec:transmission}. At finite $\omega$ the law with $S_l$ has the local exponent $4l+2+d\ln S_l/d\ln\omega\simeq4l+2+4\pi\omega r_e$ by Eq.~\eqref{eq:Sl_small}. The excess of the Numerov exponents over $4l+2$ comes mostly from $S_l$. The last two columns give $K_l$ from Eq.~\eqref{eq:lowfreq} and from extrapolating the Numerov ratio to $\omega=0$, where $S_l=1$. The entries are the same for both parameter sets.}
\label{tab:powerlaw}
\begin{tabular}{|c|c|c|c|c|c|c|c|c|c|}
\hline
\multirow{2}{*}{field} & \multirow{2}{*}{$l$} & \multicolumn{2}{c|}{$\nu_{\rm loc}$ at $\omega r_e=0.002$} & \multicolumn{2}{c|}{$\nu_{\rm fit}$ over $0.015$--$0.06$} & \multirow{2}{*}{\hcell{$4l+2$\\($\omega\to0$)}} & \multirow{2}{*}{\hcell{$4\sqrt{l(l+1)}$\\(WKB)}} & \multirow{2}{*}{\hcell{$K_l$\\Eq.~\eqref{eq:lowfreq}}} & \multirow{2}{*}{\hcell{$K_l$\\(Numerov)}} \\
\cline{3-6}
 & & Numerov & with $S_l$ & Numerov & with $S_l$ & & & & \\
\hline
scalar & 1 & 6.026  & 6.026  & 6.456  & 6.388  & 6  & 5.66 & $4/9$     & 0.4444 \\
gauge  & 1 & 6.026  & 6.026  & 6.463  & 6.388  & 6  & 5.66 & $16/9$    & 1.778 \\
scalar & 2 & 10.026 & 10.026 & 10.439 & 10.392 & 10 & 9.80 & $4/2025$  & $1.975\times10^{-3}$ \\
gauge  & 2 & 10.026 & 10.026 & 10.441 & 10.392 & 10 & 9.80 & $4/900$   & $4.444\times10^{-3}$ \\
\hline
\end{tabular}
\end{table*}

\subsubsection{Scalar versus gauge}

Since $V_s-V_g=ff'/r>0$ everywhere (Sec.~\ref{sec:global_potentials}), the scalar barrier is higher, by $1/(32r_e^2)$ at the peak, Eq.~\eqref{eq:heights}, and the scalar transmission is smaller than the gauge one at the same $\omega$ and $l$. In the low-frequency limit \eqref{eq:lowfreq} the ratio is exact. The gauge field transmits $(l+1)^2/l^2$ times more than the scalar, four times at $l=1$. The difference is largest near the top in absolute terms and disappears at high $l$, where both potentials are dominated by the common centrifugal term.

\subsection{Photon-sphere quasinormal modes of a single mouth}
\label{sec:qnm_photonsphere}

A distant observer facing one mouth first sees the barrier of that mouth ringing on its own. Up to corrections of order $\hat\epsilon$ the mouth is an extremal Reissner--Nordstr\"om exterior with $G_NM=r_e$ (Sec.~\ref{sec:matching}), and these photon-sphere modes are those of the extremal black hole. They are known. Onozawa et al.\ computed them with a continued-fraction method \cite{Onozawa:1995vu}, a modified continued fraction applies to extremal black holes in general \cite{Richartz:2015saa}, and complex scaling of the tortoise coordinate has recently been applied at extremality \cite{Ogawa:2026veu}. The spectrum depends only on the square of the charge, not on whether it is electric or magnetic \cite{Chandrasekhar:1985kt,DeFelice:2023rra}.

We compute the modes as a check of the single-barrier problem and to fix the observable time-scale. We integrate along a complex contour \cite{Andersson:1992prs,Glampedakis:2003dn}, and the evenness of the barrier turns the outgoing condition on the throat side into $d\psi/dx=0$ at the peak $x=0$ (\ref{app:contour}). Table~\ref{tab:qnm_ps} lists the fundamental modes. We write a mode as $\omega=\omega_R+i\omega_I$, with $\omega_I<0$. At $l=0,1,2$ the scalar values agree with \cite{Onozawa:1995vu} to within $3\times10^{-6}$, the precision to which those are quoted. Our gauge field is a test field, and its modes are not the electromagnetic-type modes of the coupled Einstein--Maxwell problem in \cite{Onozawa:1995vu}. At $l=10$ the damping rates, $0.08841/r_e$ (scalar) and $0.08831/r_e$ (gauge), lie within $0.1\%$ of half the eikonal Lyapunov exponent of the photon orbit, $1/(8\sqrt2\,r_e)=0.08839/r_e$ \cite{Cardoso:2008bp}.

At $l=1$ the quality factor $Q=\omega_R/2|\omega_I|$ is $2.11$ (scalar) and $1.99$ (gauge), and the mouth rings down within about one cycle. This damping is $3\times10^{34}$ times faster than that of the cavity mode at the barrier top for set~1 (Sec.~\ref{sec:qnm_cavity}), and $3\times10^{26}$ times faster for set~2. The $l=1$ scalar mode has the frequency $\omega_Rc/(2\pi r_e)=0.64$~GHz and the decay time $r_e/(c|\omega_I|)=1.1$~ns for set~1, and $6.4\times10^{16}$~Hz and $1.1\times10^{-17}$~s for set~2.

\begin{table*}
\centering
\renewcommand{\arraystretch}{1.3}
\setlength{\arrayrulewidth}{0.6pt}
\caption{Photon-sphere (single-mouth) quasinormal modes, $n=0$, at $l=0,1,2,10$, given as $\omega r_e$. Contour values come from the complex-contour integration of \ref{app:contour}. For the scalar modes at $l=0,1,2$ and the gauge mode at $l=1$ they change by at most $3\times10^{-11}$ in $\omega r_e$ when the step, the start of the ray or its angle is changed. Fit values are the mean and standard deviation of the time-domain ringdown fit described there. WKB values are Eq.~\eqref{eq:qnm_wkb}. $Q=\omega_R/2|\omega_I|$ uses the contour values and is dimensionless. The gauge field has no $l=0$ mode. The notebook \cite{Mondal:2026code} lists every $l\le10$. In units of $1/r_e$ the modes are the same for both parameter sets.}
\label{tab:qnm_ps}
\begin{tabular}{|c|c|c|c|c|c|c|c|c|}
\hline
\multirow{2}{*}{field} & \multirow{2}{*}{$l$} & \multicolumn{3}{c|}{$\omega_R\,r_e$} & \multicolumn{3}{c|}{$|\omega_I|\,r_e$} & \multirow{2}{*}{$Q$} \\
\cline{3-8}
 & & contour & fit & WKB & contour & fit & WKB & \\
\hline
\multirow{4}{*}{scalar} & 0 & $0.133459$ & $0.1338\pm0.0023$ & $0.2008$ & $0.095844$ & $0.0999\pm0.0008$ & $0.0953$ & $0.70$ \\
 & 1 & $0.377642$ & $0.3789\pm0.0011$ & $0.4055$ & $0.089384$ & $0.0891\pm0.0003$ & $0.0904$ & $2.11$ \\
 & 2 & $0.626573$ & $0.6288\pm0.0015$ & $0.6436$ & $0.088748$ & $0.0882\pm0.0004$ & $0.0892$ & $3.53$ \\
 & 10 & $2.625372$ & $2.6338\pm0.0058$ & $2.6295$ & $0.088409$ & $0.0842\pm0.0027$ & $0.0884$ & $14.85$ \\
\hline
\multirow{3}{*}{gauge} & 1 & $0.335302$ & $0.3371\pm0.0012$ & $0.3638$ & $0.084391$ & $0.0843\pm0.0005$ & $0.0859$ & $1.99$ \\
 & 2 & $0.601440$ & $0.6039\pm0.0019$ & $0.6186$ & $0.086973$ & $0.0865\pm0.0004$ & $0.0875$ & $3.46$ \\
 & 10 & $2.619418$ & $2.6278\pm0.0058$ & $2.6235$ & $0.088308$ & $0.0841\pm0.0027$ & $0.0883$ & $14.83$ \\
\hline
\end{tabular}
\end{table*}

\section{Echoes in linear theory}
\label{sec:echoes}

This section computes the echo trains and the cavity modes in linear theory on the fixed background. Their spectra locate the echo band, and their late-time decay comes from the low-frequency end of the spectrum. Section~\ref{sec:energy} tests these results against the energy budget of the wormhole. It shows that a quantum in the echo band carries at least $57/g^2$ times the binding energy (Sec.~\ref{sec:onequantum}). Only quanta below the threshold frequency $\omega_*$ of Eq.~\eqref{eq:omega_star} carry less, and few of their echoes escape (Sec.~\ref{sec:below}). Above the threshold the results of this section describe what linear theory gives for a double-barrier cavity of this shape, and Sec.~\ref{sec:consequences} explains why the MMP wormhole cannot realise them. The echo waveforms, spectra and energies, the decay laws, the finesse and the number of crossings $N_{\rm echo}$ are the same for both parameter sets. The charge enters only through $D$, which sets the arrival times, the spacing of the resonances, $\omega_I$ and $Q$. We give these for both sets.

\subsection{From one barrier to the echo train}
\label{sec:bounces}

By the $\rho\to-\rho$ symmetry of the throat the two barriers are identical. A wave crosses the cavity between them freely (Sec.~\ref{sec:global_potentials}). We launch a time-symmetric Gaussian pulse at the throat centre,
\begin{align}
    \Psi(r_*,0)&=\exp\!\left(-\frac{r_*^2}{2b^2}\right),\qquad \partial_t\Psi(r_*,0)=0,\nonumber\\
    \tilde\Psi_0(\omega)&=b\sqrt{2\pi}\,e^{-b^2\omega^2/2},
    \label{eq:initial_data}
\end{align}
with $b=2r_e$. It splits into equal right- and left-moving halves,
\begin{equation}
    \Psi=\tfrac12\Psi_0(r_*-t)+\tfrac12\Psi_0(r_*+t).
\end{equation}
The conserved energy of the test field is Eq.~\eqref{eq:energy}. In the cavity $V\simeq0$, and the pulse carries
\begin{align}
    E_{\rm pulse}&=\frac12\int\Big(\frac{d\Psi_0}{dr_*}\Big)^2dr_*=\frac{1}{2b^4}\int r_*^2\,e^{-r_*^2/b^2}dr_*\nonumber\\
    &=\frac{\sqrt\pi}{4b}.
    \label{eq:Epulse}
\end{align}
We track every reflection and transmission at $r_*=\pm D$. The signal that leaves through the $+D$ barrier first contains the original right-mover, at $t=D$ with spectral amplitude $\tfrac12\kappa$. Next comes the left-mover after one reflection at $-D$, at $t=3D$ with amplitude $\tfrac12\beta\kappa$. The right-mover after reflections at $+D$ and $-D$ arrives at $t=5D$ with amplitude $\tfrac12\kappa\beta^2$, and so on. The two interleaved series combine into
\begin{equation}
    T_m=(2m+1)\,D,\qquad \tilde\Psi_m(\omega)=\tfrac12\,\kappa(\omega)\,\beta(\omega)^m\,\tilde\Psi_0(\omega),
    \label{eq:echo_series}
\end{equation}
with $m=0,1,2,\dots$. Echoes leave the wormhole at every odd multiple of $D$, separated by $\Delta T=2D\simeq\pi L_{\text{wh}}$, each filtered by one more power of the single-barrier reflection amplitude than the previous one. This spacing reflects the symmetric initial data. Half of the pulse moves towards each barrier, and the two halves reach $+D$ alternately. A pulse that enters through one mouth returns to it after a full round trip, $4D=2\pi L_{\rm wh}$. With $\tau=t-T_m$ the time on the clock of echo $m$, the echo carries the energy
\begin{equation}
    E_m=\int(\partial_\tau\Psi_m)^2\,d\tau=\frac1\pi\int_0^\infty\omega^2|\tilde\Psi_m(\omega)|^2\,d\omega .
    \label{eq:Em_def}
\end{equation}
The first form is the time integral of the energy flux $(\partial_t\Psi)^2$ of an outgoing wave. The second follows from Parseval's theorem and $\tilde\Psi_m(-\omega)=\overline{\tilde\Psi_m(\omega)}$. Since
\begin{equation}
    \sum_m|\beta|^{2m}=\frac{1}{1-|\beta|^2}=\frac{1}{|\kappa|^2},
\end{equation}
we have $\sum_m|\tilde\Psi_m|^2=\tfrac14|\tilde\Psi_0|^2$ at every $\omega$. Then
\begin{equation}
    \sum_mE_m=\frac{1}{4\pi}\int_0^\infty\omega^2|\tilde\Psi_0|^2d\omega=\frac{\sqrt\pi}{8b}=\frac{E_{\rm pulse}}{2}.
    \label{eq:energy_sum}
\end{equation}
Half of the pulse energy eventually escapes through each of the two barriers.

With $b=2r_e$ the pulse bandwidth $1/b=0.5/r_e$ is comparable to the barrier-top frequencies $\sqrt{V_0}\simeq0.35/r_e$--$0.40/r_e$ of the $l=1$ modes (Table~\ref{tab:transmission}). This initial data is a bookkeeping device that isolates the cavity. Its proper width at the throat centre is below the Planck length (Sec.~\ref{sec:energy}), and it cannot be prepared physically. A pulse sent in from outside is treated in Sec.~\ref{sec:exterior}, with the same amplitudes $\beta$ and $\kappa$.

The waveform of echo $m$ on its own clock, $\tau=t-T_m$, is the Fourier synthesis
\begin{equation}
    \Psi_m(\tau)=\int\frac{d\omega}{2\pi}\,\tilde\Psi_m(\omega)\,e^{-i\omega\tau}.
\end{equation}
We evaluate it by FFT on $2^{16}$ frequencies with $\Delta\omega=0.002/r_e$, using $\beta(-\omega)=\bar \beta(\omega)$ and $\kappa(-\omega)=\bar \kappa(\omega)$. Each waveform only requires resolving the pulse's own $\mathcal{O}(1)$ bandwidth. The value of $D$ enters only through the exact time shifts \eqref{eq:echo_series}. Summing the geometric series instead gives the Fabry--P\'erot transfer function of the cavity, Eq.~\eqref{eq:Tcav} of Sec.~\ref{sec:cavity}.

\subsection{The echo train}
\label{sec:echo_results}

\begin{figure*}
    \centering
    \includegraphics[width=0.48\textwidth]{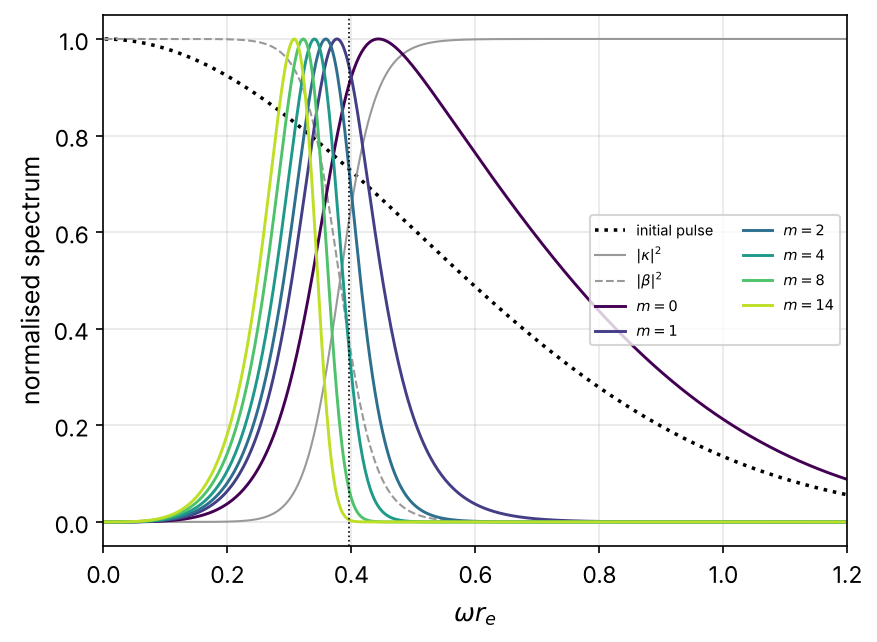}\hfill
    \includegraphics[width=0.48\textwidth]{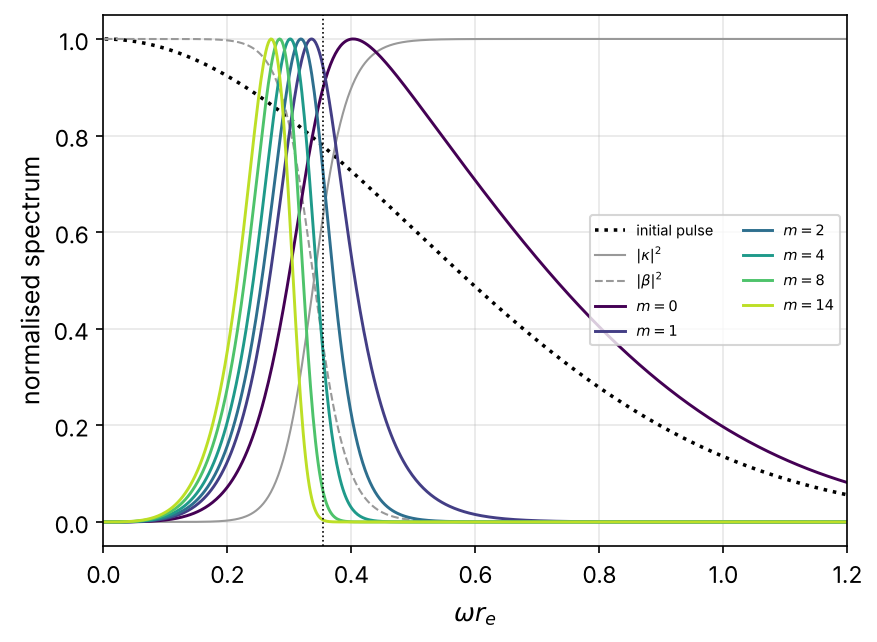}
    \caption{Normalised spectra $|\tilde\Psi_m(\omega)|$ of the echoes $m=0,1,2,4,8,14$ (coloured) at $l=1$, for the scalar (left) and gauge (right) fields, against $\omega r_e$. The spectrum of the initial pulse is dotted, the single-barrier $|\kappa|^2$ and $|\beta|^2$ are grey and the barrier-top frequency is the vertical dotted line. The direct pulse is high-pass filtered by $\kappa(\omega)$. Every subsequent reflection multiplies the spectrum by $\beta(\omega)$, which narrows it around, and slowly pushes it below, the barrier top. In units of $r_e$ it is the same for both parameter sets.}
    \label{fig:echo_spectra}
\end{figure*}

Table~\ref{tab:echoes} and Fig.~\ref{fig:echo_spectra} present the $l=1$ transmitted echo train for both fields. The train is evaluated at the barrier peak $r_* = D$. Since the transmitted wave is purely outgoing beyond the peak, it is identical, up to a rigid time shift, to what would be seen at any point further out.

\begin{table}
\centering
\renewcommand{\arraystretch}{1.3}
\setlength{\arrayrulewidth}{0.6pt}
\caption{Peak amplitude $A_m=\max_\tau|\Psi_m|$ and energy fraction $E_m/E_{\rm pulse}$ of the echoes for $l=1$, $b=2r_e$. $A_m$ is relative to the unit height of the initial pulse, and $E_m/E_{\rm pulse}$ is dimensionless. The arrival times are $T_m=(2m+1)D$, with $D$ of Table~\ref{tab:bg_params}. The values do not depend on the parameter set. The last row gives the cumulative energy fraction that has left the cavity through the $+D$ barrier after fifteen echoes. The maximum possible is $1/2$.}
\label{tab:echoes}
\begin{tabular}{|c|c|c|c|c|}
\hline
$m$ & \hcell{$A_m$\\scalar} & \hcell{$A_m$\\gauge} & \hcell{$E_m/E_{\rm pulse}$\\scalar} & \hcell{$E_m/E_{\rm pulse}$\\gauge} \\
\hline
0 & 0.1970 & 0.2168 & $0.378$ & $0.407$ \\
1 & 0.0474 & 0.0480 & $2.05\times10^{-2}$ & $1.74\times10^{-2}$ \\
2 & 0.0307 & 0.0300 & $9.36\times10^{-3}$ & $7.74\times10^{-3}$ \\
3 & 0.0231 & 0.0227 & $5.91\times10^{-3}$ & $4.82\times10^{-3}$ \\
4 & 0.0187 & 0.0185 & $4.27\times10^{-3}$ & $3.45\times10^{-3}$ \\
6 & 0.0137 & 0.0136 & $2.69\times10^{-3}$ & $2.15\times10^{-3}$ \\
8 & 0.0110 & 0.0108 & $1.94\times10^{-3}$ & $1.53\times10^{-3}$ \\
10 & 0.0092 & 0.0090 & $1.50\times10^{-3}$ & $1.18\times10^{-3}$ \\
14 & 0.0070 & 0.0069 & $1.02\times10^{-3}$ & $7.94\times10^{-4}$ \\
\hline
\multicolumn{3}{|l|}{sum over $m<15$} & 0.436 & 0.455 \\
\hline
\end{tabular}
\end{table}

\subsubsection{Spectral content}

The direct pulse ($m=0$) is high-pass filtered by $\kappa(\omega)$. Its spectral centroid, $\langle\omega\rangle r_e=0.554$ (scalar) and $0.522$ (gauge), lies at $1.4\sqrt{V_0}$ and $1.5\sqrt{V_0}$. Every subsequent echo is multiplied by one more factor of $\beta(\omega)$, which is close to unity below the top and falls to zero above it. The product $\kappa\,\beta^m$ is a band-pass filter peaked just below $\sqrt{V_0}$ that narrows with $m$. For the scalar field the echo centroids are $0.955\sqrt{V_0}$ for $m=1$, $0.835\sqrt{V_0}$ for $m=4$ and $0.749\sqrt{V_0}$ for $m=14$. The rms widths shrink from $0.050/r_e$ to $0.032/r_e$ (Fig.~\ref{fig:echo_spectra}). The echoes are thus quasi-monochromatic wave packets at the light-ring (barrier-top) frequency of the mouth. For $l=1$ their frequency is $\omega r_e\simeq0.30$--$0.38$ for the scalar field and $0.26$--$0.34$ for the gauge field, with periods of about $17$--$24\,r_e$. They slowly redden as the higher-frequency part of the trapped energy leaks out first. Echo trains of compact objects with a reflecting surface show the same filtering \cite{Cardoso:2016rao,Mark:2017dnq}. Here it follows from the actual MMP barrier rather than from a model reflectivity.

\subsubsection{Decay law}
\label{sec:decay_law}

\begin{figure}
    \centering
    \includegraphics[width=\columnwidth]{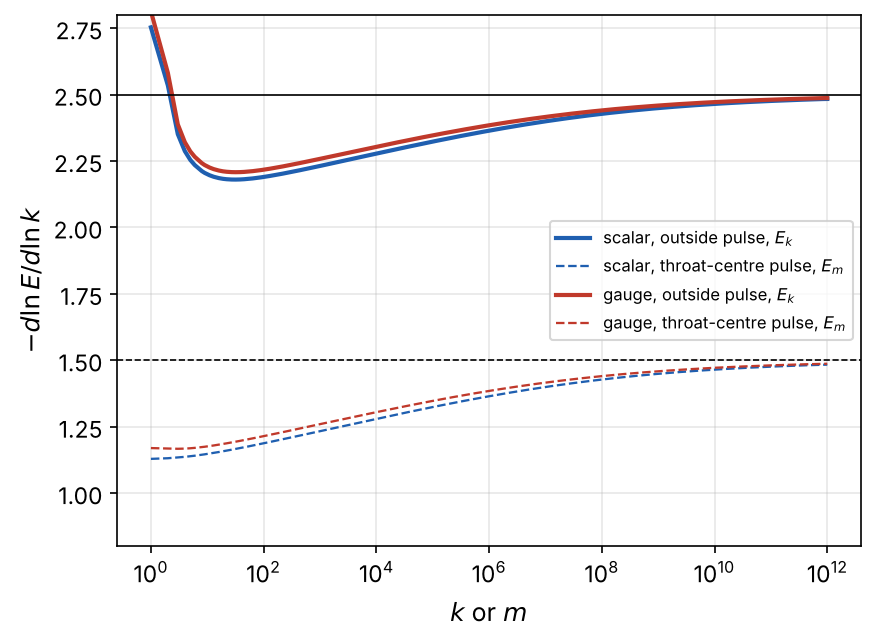}
    \caption{Local decay exponents of the echo energies at $l=1$, $b=2r_e$. Solid lines, $E_k$ for a pulse sent in from outside (Sec.~\ref{sec:exterior}). Dashed lines, $E_m$ for the throat-centre pulse (Sec.~\ref{sec:echo_results}). The horizontal lines are the asymptotic values $5/2$ from Eq.~\eqref{eq:latetime_ext} and $3/2$ from Eq.~\eqref{eq:latetime}, for $\nu=6$. The exponents are the same for both parameter sets.}
    \label{fig:decay}
\end{figure}

The echo amplitudes do not decay geometrically. Since $|\beta(\omega)|\to1$ as $\omega\to0$, the low-frequency part of the trapped energy is released only algebraically. By Eqs.~\eqref{eq:echo_series} and~\eqref{eq:Em_def}, the energy of echo $m$ is
\begin{equation}
    E_m=\frac{1}{4\pi}\int_0^\infty\omega^2|\tilde\Psi_0|^2\,|\kappa|^2\,(1-|\kappa|^2)^m\,d\omega .
\end{equation}
For $m\gg1$ the factor $(1-|\kappa|^2)^m$ is exponentially small wherever $|\kappa|^2$ is not itself small. The integral is then dominated by $\omega\to0$, where $|\kappa|^2\simeq K_l\omega^\nu$ with $\nu=4l+2$ by Eq.~\eqref{eq:lowfreq}, and $|\tilde\Psi_0|^2\simeq2\pi b^2$. In this region $S_l\simeq1$. With $(1-K_l\omega^\nu)^m\simeq e^{-mK_l\omega^\nu}$,
\begin{equation}
    E_m\simeq\frac{b^2K_l}{2}\int_0^\infty\omega^{2+\nu}e^{-mK_l\omega^\nu}\,d\omega .
\end{equation}
The substitution $z=mK_l\omega^\nu$, with $\omega=(z/mK_l)^{1/\nu}$, gives
\begin{align}
    \int_0^\infty\omega^{2+\nu}e^{-mK_l\omega^\nu}d\omega&=\frac{(mK_l)^{-1-3/\nu}}{\nu}\int_0^\infty z^{3/\nu}e^{-z}dz\nonumber\\
    &=\frac{(mK_l)^{-1-3/\nu}}{\nu}\,\Gamma\Big(1+\frac3\nu\Big).
\end{align}
With $K_l(mK_l)^{-1-3/\nu}=K_l^{-3/\nu}m^{-1-3/\nu}$,
\begin{equation}
    E_m\simeq\frac{b^2}{2\nu}\,\Gamma\!\left(1+\frac{3}{\nu}\right)K_l^{-3/\nu}\,m^{-1-3/\nu}.
    \label{eq:latetime}
\end{equation}
For $l=1$ the exponent is $3/2$. This limit is reached only at large $m$. Eq.~\eqref{eq:latetime} needs $mK_l\omega^\nu\sim1$ at frequencies where the power law holds, $\omega r_e\lesssim10^{-2}$, which for $l=1$ means $m\gtrsim10^{12}$. Figure~\ref{fig:decay} shows the local exponent $-d\ln E_m/d\ln m$ for $1\le m\le10^{12}$ (dashed lines). We compute $E_m$ directly from Eq.~\eqref{eq:Em_def}, with
\begin{equation}
    |\tilde\Psi_m|^2=\tfrac14\,|\tilde\Psi_0|^2\,|\kappa|^2\left(1-|\kappa|^2\right)^m .
    \label{eq:Psim_sq}
\end{equation}
This needs no waveforms and works for any $m$. We use the Numerov $|\kappa|^2$ on a logarithmic grid $0.002\le\omega r_e\le6$ and the analytic law $K_l(\omega r_e)^{4l+2}S_l$ of Eq.~\eqref{eq:sommerfeld} below $\omega r_e=0.002$. For the scalar field the local exponent is $1.15$ at $m=10$, $1.36$ at $m=10^6$ and $1.48$ at $m=10^{12}$, where $E_m$ is $0.91$ times Eq.~\eqref{eq:latetime}. Fits over $m=4$--$14$ give $E_m\propto m^{-1.14}$ (scalar) and $m^{-1.17}$ (gauge). The peak amplitudes follow $m^{-0.78}$ (Table~\ref{tab:echoes}).

Power-law echo trains have been found numerically before. Fits give peak amplitudes that fall as $n^{-1.31}$ to $n^{-1.34}$ with echo number $n$ for a perfectly reflecting wall in place of the horizon of a spinning black hole \cite{Wang:2018gin}. For the throat models of \cite{Abedi:2026gxm} they fall as $n^{-p}$ with $p\simeq1.05$--$1.40$. Both exponents come from fits. Eq.~\eqref{eq:latetime} gives the asymptotic exponent in closed form. Its derivation uses only $|\kappa|^2\simeq K_l\omega^\nu$ at low frequency, a pulse spectrum that stays finite at $\omega=0$ and a cavity without absorption. It holds for any such cavity, with $\nu$ set by its barrier. A cavity that absorbs part of each wave adds a factor below one at every bounce, and the decay then turns exponential \cite{Maggio:2019zyv,Abedi:2026gxm}.

After fifteen echoes, at $t=29D$, $43.6\%$ (scalar) and $45.5\%$ (gauge) of the pulse energy has left through the $+D$ barrier, out of a maximum of $50\%$, Eq.~\eqref{eq:energy_sum} (Table~\ref{tab:echoes}). The remainder is still trapped in the cavity at low frequency. The asymptotic value of one half is the same factor found analytically in \cite{Freivogel:2026ujn}. It arises there for the same reason. Radiation trapped in the throat eventually leaks out through either mouth with equal probability.

\subsubsection{Spin, WKB, multipole and pulse width}

The scalar and gauge trains differ mainly in the direct pulse ($A_0$ by $10\%$) and the first echo, since the offset $ff'/r$ acts near the top. A train built from the parabolic-WKB amplitudes overestimates the peak amplitudes by $28\%$ for the direct pulse and by a factor $3.1$ at $m=14$. Its energies decay as $m^{-1.04}$ instead of $m^{-1.14}$, because the parabolic barrier leaks at low frequency. The arrival times are the same for every $l$. For fixed $b$ the direct amplitude $A_0$ of the scalar drops from $0.34$ at $l=0$ to $0.0135$ at $l=4$, since the pulse has less power above the higher barrier top. Relative to it the echoes grow. The ratio $A_1/A_0$ rises from $0.18$ at $l=0$ to $0.40$ at $l=4$ for the scalar, and from $0.22$ at $l=1$ to $0.39$ at $l=4$ for the gauge field. The first version of this work is cited in \cite{Riley:2026avn} for stronger echoes at higher $l$, which holds only relative to the direct pulse. A narrow pulse ($b=r_e$) gives $A_1/A_0=0.10$, and a broad one ($b=6r_e$) gives $0.72$ with $A_0=0.026$.

\subsection{A pulse incident from outside one mouth}
\label{sec:exterior}

An observer near one mouth who sends in a pulse, or who watches a source fall onto the mouth, sees a different train. Let the pulse with spectrum $\tilde\Psi_0$ reach the $+D$ barrier from outside. The observer first receives the prompt reflection $\beta\tilde\Psi_0$. This is the response of an extremal Reissner--Nordstr\"om black hole, ringdown included. The part $\kappa\tilde\Psi_0$ that enters the throat reflects at $-D$ and returns after the round trip $4D$ with spectrum $\kappa^2\beta\tilde\Psi_0$. Echo $k\geq1$ arrives at $4kD$ with spectrum $\kappa^2\beta^{2k-1}\tilde\Psi_0$. The far mouth emits $\kappa^2\beta^{2k}\tilde\Psi_0$ at $2D+4kD$, $k\geq0$. Together the two series contain every power $|\beta|^{2j}$ with $j\ge0$ once. Unitarity then reads
\begin{equation}
    |\beta|^2+|\kappa|^4\sum_{j\ge0}|\beta|^{2j}=|\beta|^2+|\kappa|^2=1 .
\end{equation}
At each frequency the far mouth receives in total the fraction
\begin{equation}
    |\kappa|^4\sum_{k\ge0}|\beta|^{4k}=\frac{|\kappa|^4}{1-|\beta|^4}=\frac{|\kappa|^2}{1+|\beta|^2}.
    \label{eq:far_fraction}
\end{equation} For $|\kappa|^2\ll1$ this is about $|\kappa|^2/2$. The near-side echoes carry in total $|\kappa|^4\sum_{k\ge1}|\beta|^{4k-2}=|\kappa|^2|\beta|^2/(1+|\beta|^2)$, which is also about $|\kappa|^2/2$. The far mouth is at least $d_{\min}$ away, about $200$~kpc for set~1 and $2.8\times10^{10}$~cm for set~2 (Sec.~\ref{sec:matching}). The far-side signal emerges far from the observer at the near mouth.

\begin{table*}
\centering
\renewcommand{\arraystretch}{1.3}
\setlength{\arrayrulewidth}{0.6pt}
\caption{Echoes of a pulse sent at one mouth from outside, $l=1$, $b=2r_e$. For $k\geq1$, $A_k$ and $E_k$ are the peak amplitude and energy of echo $k$ on the near side, and $E^{\rm far}_k$ is the energy that leaves the far mouth at $2D+4kD$. The row $k=0$, set off by a line, gives the prompt reflection ($A_0$, $E_0$) and the direct crossing to the far mouth ($E^{\rm far}_0$). Energies are fractions of the incident pulse energy and amplitudes are relative to the unit height of the incident pulse. The last row sums the near-side echo energies $E_k$ over $1\le k\le15$. The entries are the same for both parameter sets, and only the arrival times scale with $D$.}
\label{tab:exterior}
\begin{tabular}{|c|c|c|c|c|c|c|}
\hline
\multirow{2}{*}{$k$} & \multicolumn{3}{c|}{scalar field} & \multicolumn{3}{c|}{gauge field} \\
\cline{2-7}
 & $A_k$ & $E_k$ & $E^{\rm far}_k$ & $A_k$ & $E_k$ & $E^{\rm far}_k$ \\
\hline
0 & 0.4554 & $0.2443$ & $0.7148$ & 0.4307 & $0.1863$ & $0.7789$ \\
\hline
1 & 0.0588 & $0.0222$ & $6.89\times10^{-3}$ & 0.0595 & $0.0193$ & $5.83\times10^{-3}$ \\
2 & 0.0214 & $3.29\times10^{-3}$ & $1.91\times10^{-3}$ & 0.0214 & $2.75\times10^{-3}$ & $1.58\times10^{-3}$ \\
3 & 0.0125 & $1.24\times10^{-3}$ & $8.67\times10^{-4}$ & 0.0124 & $1.02\times10^{-3}$ & $7.09\times10^{-4}$ \\
4 & 0.0085 & $6.38\times10^{-4}$ & $4.88\times10^{-4}$ & 0.0084 & $5.19\times10^{-4}$ & $3.96\times10^{-4}$ \\
6 & 0.0050 & $2.56\times10^{-4}$ & $2.14\times10^{-4}$ & 0.0049 & $2.05\times10^{-4}$ & $1.71\times10^{-4}$ \\
8 & 0.0034 & $1.35\times10^{-4}$ & $1.18\times10^{-4}$ & 0.0034 & $1.07\times10^{-4}$ & $9.38\times10^{-5}$ \\
10 & 0.0025 & $8.25\times10^{-5}$ & $7.41\times10^{-5}$ & 0.0025 & $6.52\times10^{-5}$ & $5.85\times10^{-5}$ \\
14 & 0.0016 & $3.95\times10^{-5}$ & $3.65\times10^{-5}$ & 0.0016 & $3.09\times10^{-5}$ & $2.86\times10^{-5}$ \\
\hline
\multicolumn{2}{|l|}{sum, $1\le k\le15$} & $0.0288$ & & & $0.0247$ & \\
\hline
\end{tabular}
\end{table*}

Table~\ref{tab:exterior} lists the train for $l=1$ and $b=2r_e$. The prompt reflection carries $24\%$ (scalar) and $19\%$ (gauge) of the incident energy. The direct crossing to the far mouth carries $71\%$ and $78\%$. The echoes on the near side carry only $2.9\%$ and $2.5\%$ in total over $k\leq15$, and the first echo has $13\%$ and $14\%$ of the peak amplitude of the prompt reflection. Every echo has tunnelled twice. Its spectrum is filtered by $\kappa^2$ rather than $\kappa$. At late times only $\omega\to0$ survives. By Eq.~\eqref{eq:Em_def} echo $k$ carries
\begin{equation}
    E_k=\frac1\pi\int_0^\infty\omega^2|\tilde\Psi_0|^2|\kappa|^4(1-|\kappa|^2)^{2k-1}d\omega .
\end{equation}
For $k\gg1$ only small $\omega$ contributes, where $|\tilde\Psi_0|^2\simeq2\pi b^2$, $|\kappa|^4\simeq K_l^2\omega^{2\nu}$ and $(1-|\kappa|^2)^{2k-1}\simeq e^{-\varsigma_k\omega^\nu}$ with $\varsigma_k=(2k-1)K_l$. Then
\begin{equation}
    E_k\simeq2b^2K_l^2\int_0^\infty\omega^{2+2\nu}e^{-\varsigma_k\omega^\nu}d\omega .
\end{equation}
The substitution $z=\varsigma_k\omega^\nu$, with $\omega=(z/\varsigma_k)^{1/\nu}$ and $d\omega=\nu^{-1}\varsigma_k^{-1/\nu}z^{1/\nu-1}dz$, gives
\begin{align}
    \int_0^\infty\omega^{2+2\nu}e^{-\varsigma_k\omega^\nu}d\omega&=\frac{\varsigma_k^{-2-3/\nu}}{\nu}\int_0^\infty z^{1+3/\nu}e^{-z}dz\nonumber\\
    &=\frac{\varsigma_k^{-2-3/\nu}}{\nu}\,\Gamma\Big(2+\frac3\nu\Big).
\end{align}
With $K_l^2\varsigma_k^{-2-3/\nu}=K_l^{-3/\nu}(2k-1)^{-2-3/\nu}$,
\begin{equation}
    E_k\simeq\frac{2b^2}{\nu}\,\Gamma\!\left(2+\frac3\nu\right)K_l^{-3/\nu}\,(2k-1)^{-2-3/\nu},\qquad k\gg1.
    \label{eq:latetime_ext}
\end{equation}
This is one power of $k$ steeper than Eq.~\eqref{eq:latetime}. For $l=1$ the asymptotic exponent is $5/2$. Fits over $k=4$--$14$ give $E_k\propto k^{-2.22}$ (scalar) and $k^{-2.25}$ (gauge). The local exponent reaches $2.48$ (scalar) and $2.49$ (gauge) at $k=10^{12}$ (Fig.~\ref{fig:decay}).

\subsection{Cavity resonances and time-scales}
\label{sec:cavity}

A wave that crosses both barriers picks up a factor $\kappa$ at each and the phase $e^{2i\omega D}$ across the cavity. Each additional round trip multiplies it by $\beta^2e^{4i\omega D}$. The geometric series sums to the transfer function
\begin{equation}
    T_{\rm cav}(\omega)=\frac{\kappa^2e^{2i\omega D}}{1-\beta^2e^{4i\omega D}}.\label{eq:Tcav}
\end{equation}
Its poles are the quasinormal modes of the two-barrier cavity. They satisfy $\beta^2e^{4i\omega D}=1$. Since $\beta$ varies only on the scale of the barrier, neighbouring poles differ in $4\omega D$ by $2\pi$. Their spacing is $\Delta\omega=\pi/2D$, with $D$ of Eq.~\eqref{eq:D_def}. It is $1.99\times10^{-35}/r_e$ for set~1 and $1.99\times10^{-27}/r_e$ for set~2. This is $2\pi$ divided by the round-trip time $4D$. The condition splits into $\beta e^{2i\omega D}=+1$, the modes even under $r_*\to-r_*$, and $\beta e^{2i\omega D}=-1$, the odd modes. The symmetric pulse of Eq.~\eqref{eq:initial_data} excites only the even ones. Summing the series \eqref{eq:echo_series} gives the signal $\tfrac12\kappa e^{i\omega D}\tilde\Psi_0/(1-\beta e^{2i\omega D})$, whose poles $\beta e^{2i\omega D}=1$ are spaced by $\pi/D=2\Delta\omega$. This is $2\pi$ divided by $2D$, the spacing of the echoes.

The sharpness of the resonances is measured by the Fabry--P\'erot finesse \cite{Born:1999ory},
\begin{equation}
    \mathcal{F}=\frac{\pi|\beta|}{1-|\beta|^2}=\frac{\pi\sqrt{1-|\kappa|^2}}{|\kappa|^2},
    \label{eq:finesse}
\end{equation}
the ratio of the spacing $\Delta\omega$ to the width of one resonance. For the $l=1$ scalar it is $6.2\times10^4$ at $\omega r_e=0.15$, $3.1$ at the barrier top and $0.2$ at $\omega r_e=0.55$. Above the top $\mathcal F<1$, and the resonances merge into a smooth curve. The finesse does not involve $D$ and is the same for both sets.

The pulse of Eq.~\eqref{eq:initial_data} covers a band of width about $1/b$, which contains $1/(b\,\Delta\omega)$ resonances, $2.5\times10^{34}$ for set~1 and $2.5\times10^{26}$ for set~2. As a sum over modes its evolution would need all their amplitudes. In time it is the short list of echoes of Eq.~\eqref{eq:echo_series}. The two descriptions are related by a Fourier transform, as for the quasi-bound states of exotic compact objects \cite{Cardoso:2016oxy,Mark:2017dnq}.

The amplitudes $\beta$ and $\kappa$ change only on the scale of the barrier, $10^{-2}/r_e$ to $10^{-1}/r_e$. For set~1 a window of $10^3$ resonances spans $2\times10^{-32}/r_e$, across which they change by a relative $2\times10^{-30}$. We therefore fix them at a carrier frequency $\omega_0$ and let only the phase $4(\omega-\omega_0)D$ vary. This is also the only way to evaluate $T_{\rm cav}$ numerically. Double precision keeps no digits of the fractional part of $\omega L_{\rm wh}$, which is about $10^{34}$ for set~1 and $10^{26}$ for set~2.

On resonance the whole wormhole becomes transparent, $|T_{\rm cav}|^2=1$, even at frequencies where a single barrier transmits only $10^{-6}$.

The low-frequency analysis of \cite{Freivogel:2026ujn} gives an independent check, and we use none of its results as input. Its single-mouth absorption probability at $l=0$ is $4(\omega r_e)^2$, the $l=0$ case of Eq.~\eqref{eq:lowfreq}. For the whole cavity it gives, for the $l=0$ scalar,
\begin{equation}
    |T_{\rm cav}|^2=\frac{4y^2}{4y^2+\sin^2(\pi\omega L_{\rm wh})\,(y^2-1)^2},
    \qquad y=\omega^2r_e^2 .
    \label{eq:Tcav_analytic}
\end{equation}
We compare it with Eq.~\eqref{eq:Tcav} built from our $\beta$ and $\kappa$ at $l=0$, frozen at the carrier frequency that fixes $y$, while the phase $\pi\omega L_{\rm wh}$ runs over one period. The phase grid resolves the width $|\kappa|^2/2$ of the resonance. At $\omega r_e=2\times10^{-3}$ the two curves differ by at most $1.3\times10^{-2}$, near half maximum. The difference comes from the single-barrier transmission. Eq.~\eqref{eq:Tcav_analytic} is the Fabry--P\'erot form of Eq.~\eqref{eq:Tcav} with $|\kappa|^2=4y/(1+y)^2$. Our $|\kappa|^2$ is larger by the factor $1.0260$, and the Coulomb factor of Eq.~\eqref{eq:sommerfeld} is $S_0=1.0254$. With $4y/(1+y)^2$ multiplied by $S_0$ the two curves agree to $2.9\times10^{-4}$. At $\omega r_e=4\times10^{-3}$ the difference is $2.6\times10^{-2}$ without $S_0$ and $1.0\times10^{-3}$ with it. At $10^{-2}$ it is $6.7\times10^{-2}$ and $5.2\times10^{-3}$. The remainder comes from terms of order $(\omega r_e)^2$, with logarithms, that neither closed form contains. The comparison does not involve $D$ and holds for both sets.

The resonances also sit where the throat alone would put them. As $\omega\to0$ the throat-side solution $w_-+\beta w_+$ must match the static solution $\Xi\propto|x|^{l+1}$, which has no term in $|x|^{-l}$. With $h^{(1)}_l(z)\simeq-i(2l-1)!!\,z^{-l-1}$ at small $z$ \cite[Ch.~10]{Abramowitz:1964} this gives $\beta\to(-1)^{l+1}$, that is $\phi_\beta\to(l+1)\pi$ for the phase $\phi_\beta$ of $\beta$. Numerically $(\phi_\beta-(l+1)\pi)/\pi$ lies between $-0.007$ and $-0.009$ at $\omega r_e=10^{-3}$, for $l=0,1,2$ and both fields. With $D=\pi L_{\rm wh}/2$ the phase condition of Eq.~\eqref{eq:qnm_cavity} then puts the resonances at integer $\omega L_{\rm wh}$, the normal modes \eqref{eq:ads2_modes} of the well \eqref{eq:throat_PT}. Eq.~\eqref{eq:Tcav_analytic} has its total transmission at the same frequencies.

For set~1 the mouth is small, $r_e=2.8359$~cm, and $L_{\rm wh}=1.4255\times10^{35}$~cm (Table~\ref{tab:bg_params}). The barrier-top period is $2\pi/\sqrt{V_0}\simeq15.9\,r_e=45$~cm$/c=1.5\times10^{-9}$~s, and the individual echoes are nanosecond wave packets. The echoes are spaced by $2D=\pi L_{\text{wh}}=4.5\times10^{35}$~cm. In time this is $1.5\times10^{25}$~s or $4.7\times10^{17}$~yr, more than seven orders of magnitude longer than the age of the Universe, $13.8$~Gyr. A signal that enters and leaves through the same mouth needs $4D=2\pi L_{\rm wh}$, or $9.5\times10^{17}$~yr.

Set~2 has a mouth $10^8$ times smaller, $r_e=2.8359\times10^{-8}$~cm, and a throat $10^{16}$ times shorter, $L_{\rm wh}=1.4255\times10^{19}$~cm. The barrier-top period is $1.5\times10^{-17}$~s. The echoes are spaced by $2D=1.49\times10^{9}$~s, or $47.3$~yr. A signal that returns through the same mouth needs $94.7$~yr. The crossing time of set~2 is thus comparable to a human lifetime, while one echo lasts tens of attoseconds.

The ratio of the echo spacing $2D$ to the barrier-top period is $9.9\times10^{33}$ for set~1 and $9.9\times10^{25}$ for set~2. It is about $0.2L_{\text{wh}}/r_e$, and it grows as $L_{\text{wh}}/r_e=16\pi q/g^2=16\sqrt\pi\,r_e/(g\,\ell_p)$, the size of the mouth in Planck units.

The mouth of set~1 is centimetre-sized, and that of set~2 is of atomic size. A mouth of astrophysical size makes every hierarchy worse. With $r_e=1$~km and $g=1$, Eq.~\eqref{eq:closed_form_bg} gives $q\simeq3.5\times10^{37}$, an echo spacing of about $6\times10^{26}$~yr, a binding energy of $2.5\times10^{-13}$~eV and $d_{\min}\simeq7.5\times10^{3}$~Gpc, larger than the Hubble radius $c/H_0\simeq4.4$~Gpc \cite{Planck:2018vyg}. No choice of $q$ brings the echo spacing close to the duration of one echo, since their ratio is of order $16\pi q/g^2$. For every $q$ and $g$ of order one the echo band lies above $\omega_*$ (Sec.~\ref{sec:energy}). In this class of wormhole only the direct pulse and the ringdown of a single mouth can be seen. The echoes computed here describe its late-time dynamics in linear theory and cannot be detected.

\subsection{Quasinormal modes of the cavity}
\label{sec:qnm_cavity}

Writing $\omega=\omega_R+i\omega_I$ and $\beta=|\beta|e^{i\phi_\beta}$, the pole condition of Eq.~\eqref{eq:Tcav}, $\beta^2e^{4i\omega D}=1$, separates into a modulus and a phase part. The modulus gives $|\beta|^2e^{-4\omega_ID}=1$, and
\begin{equation}
    \omega_I=\frac{\ln|\beta(\omega_R)|}{2D}\ (<0),
    \qquad
    2\phi_\beta(\omega_R)+4\omega_R D=2\pi n .
    \label{eq:qnm_cavity}
\end{equation}
The second relation fixes the phase over one round trip. We evaluate Eq.~\eqref{eq:qnm_cavity} at real $\omega_R$, with $\ln|\beta|=\tfrac12\ln(1-|\kappa|^2)$, which avoids the cancellation in $1-|\beta|$. The damping is at most of order $1/D$, about $10^{-35}/r_e$ for set~1 and $10^{-27}/r_e$ for set~2, while $\beta$ changes only over $0.1/r_e$. Evaluating $\beta$ at $\omega_R$ instead of the complex $\omega$ therefore changes it by a relative $10^{-34}$ for set~1 and $10^{-26}$ for set~2, and Eq.~\eqref{eq:qnm_cavity} is an explicit formula.

Each mode has the quality factor $Q=\omega_R/2|\omega_I|$ and the ringdown time $t_{\rm d}=1/|\omega_I|$. It survives $N_{\rm echo}=t_{\rm d}/2D$ crossings. With $2D|\omega_I|=|\ln|\beta||$ from Eq.~\eqref{eq:qnm_cavity},
\begin{equation}
    Q=\frac{\omega_RD}{|\ln|\beta||},\qquad N_{\rm echo}=\frac{1}{|\ln|\beta||}.
    \label{eq:Q_exact}
\end{equation}
Below the barrier top $\ln|\beta|=\tfrac12\ln(1-|\kappa|^2)\simeq-|\kappa|^2/2$, and
\begin{equation}
    Q\simeq\frac{2\omega_RD}{|\kappa|^2},\qquad N_{\rm echo}\simeq\frac{2}{|\kappa|^2}.
    \label{eq:Q_kappa}
\end{equation}
The very large values of $Q$ come from $D/r_e$, which is known exactly. Their relative precision is that of $|\kappa|^2$, about $10^{-6}$ (Sec.~\ref{sec:numerovB}). For the scalar field at $l=1$, Eq.~\eqref{eq:Q_exact} with $|\kappa|^2=0.619$ at $\omega=\sqrt{V_0}$ (Sec.~\ref{sec:transmission}) gives $Q=6.5\times10^{34}$ for set~1. With $|\kappa|^2=1.39\times10^{-8}$ at $\omega r_e=0.05$ (Table~\ref{tab:transmission}) it gives $Q=5.7\times10^{41}$ and $N_{\rm echo}=1.4\times10^{8}$. Since $Q\propto D$, set~2 has $Q$ smaller by $10^{8}$, $6.5\times10^{26}$ at the barrier top and $5.7\times10^{33}$ at $\omega r_e=0.05$. The ringdown times $1/|\omega_I|=2N_{\rm echo}D$ are $3.1\times10^{25}$~s and $2.1\times10^{33}$~s for set~1, and $3.1\times10^{9}$~s and $2.1\times10^{17}$~s for set~2. The number of crossings $N_{\rm echo}$ is the same for both sets. At low frequency the gauge values are about four times smaller, since the gauge field transmits four times more there (Sec.~\ref{sec:transmission}). Above the top the modes barely resonate. At $\omega r_e=0.55$, $|\kappa|^2=0.996$ gives $N_{\rm echo}=0.36$. The scalar mode leaks out within about a third of one crossing, and the resonances merge into a smooth curve. All of these modes lie above $\omega_*$ (Sec.~\ref{sec:onequantum}). Below $\omega_*$ the cavity modes are the resonances at integer $\omega L_{\rm wh}$ of Sec.~\ref{sec:cavity}, with still larger $Q$.

For $|\beta|\to1$ the width $2|\omega_I|$ equals $\Delta\omega/\mathcal F$. Since $T_{\rm cav}$ depends on $\beta$ only through the combination $\beta^2e^{4i\omega D}$, its poles exist only by virtue of the round trip between the two barriers, and they vanish if $\beta\to0$. They are properties of the cavity as a whole. They are a different object from the photon-sphere QNM of Sec.~\ref{sec:qnm_photonsphere}, which is a pole of $\kappa(\omega)$ for a single barrier in isolation. The long-lived branch below the barrier top has the same qualitative form as for photon-sphere exotic compact objects with a Minkowski core \cite{Zhang:2025ygb}.

\section{Energy budget and the regime of validity}
\label{sec:energy}

So far both fields have been treated as classical waves of arbitrarily small amplitude. Real signals are made of quanta, and a quantum that falls into the throat adds its energy to the wormhole. MMP point out that an object with a Compton wavelength of order $r_e$ carries more than the binding energy of the throat \cite{Maldacena:2018gjk}. The largest energy one can send through is of order the binding energy \cite{Maldacena:2020sxe}. The scattering analysis of \cite{Freivogel:2026ujn} takes the binding energy to be of order $1/r_e$ and restricts its probes to $\omega r_e\ll1$ for this reason. Here we give the exact ratio and place the echo band relative to it. We then turn the bound into a lifetime in thermal radiation. In this section we use $\hbar=c=1$ and $G_N=\ell_p^2$ and keep $g$ explicit. Every relation follows from the closed forms \eqref{eq:closed_form_bg}.

\subsection{Binding energy}
\label{sec:binding}

Each mouth is lighter than an extremal black hole of the same charge by $|\epsilon|$, Eq.~\eqref{eq:epsilon_def}. If a mouth gains this much energy it reaches extremality, and any more gives it a horizon. The total deficit of the two mouths is
\begin{equation}
    E_b=\frac{2|\epsilon|}{G_N}=\frac{g^3}{256\pi^{3/2}\,q\,\ell_p}=\frac{q}{16L_{\rm wh}}.
    \label{eq:Eb}
\end{equation}
The second and third forms follow from $\epsilon$ and $L_{\rm wh}$ of Eq.~\eqref{eq:closed_form_bg} with $G_N=\ell_p^2$,
\begin{align}
    \frac{2|\epsilon|}{\ell_p^2}&=\frac{2}{\ell_p^2}\,\frac{g^3\ell_p}{512\pi^{3/2}q}=\frac{g^3}{256\pi^{3/2}q\ell_p},\nonumber\\
    \frac{q}{16L_{\rm wh}}&=\frac{q\,g^3}{16\cdot16\pi^{3/2}q^2\ell_p}=\frac{g^3}{256\pi^{3/2}q\ell_p}.
\end{align} The last form is the magnitude of the binding energy $E_{\rm bind}=-c_2/(16\ell)$ of \cite{Maldacena:2020sxe}, with central charge $c_2=q$ and $\ell=L_{\rm wh}$. It also equals the magnitude of the minimum energy $E_{\min}=-G_Nq^2/(256r_e^3)$ of MMP, since their throat length $\ell=16r_e^3/(G_Nq)$ is $L_{\rm wh}$ \cite{Maldacena:2018gjk}.

\subsection{One quantum}
\label{sec:onequantum}

With $q\ell_p=g\,r_e/\sqrt\pi$ from Eq.~\eqref{eq:closed_form_bg}, the binding energy is
\begin{equation}
    E_b=\frac{g^3\sqrt\pi}{256\pi^{3/2}\,g\,r_e}=\frac{g^2}{256\pi r_e}.
\end{equation} A quantum of frequency $\omega$ carries the energy $\hbar\omega$. Its ratio to the binding energy is
\begin{equation}
    \frac{\hbar\omega}{E_b}=\frac{256\pi}{g^2}\,\omega r_e .
    \label{eq:quantum_ratio}
\end{equation}
The charge drops out. This holds at fixed $\omega r_e$. At a fixed physical frequency the ratio grows as $r_e$, that is as $q$. A photon of $1$~GHz carries $480\,E_b$ for set~1 and $4.8\times10^{-6}E_b$ for set~2. One quantum carries exactly $E_b$ at the threshold frequency
\begin{equation}
    \omega_*=\frac{g^2}{256\pi\,r_e}\simeq1.24\times10^{-3}\,\frac{g^2}{r_e}.
    \label{eq:omega_star}
\end{equation}
With $\hbar c=1.973\times10^{-5}$~eV\,cm and $r_e=2.836$~cm, set~1 has $E_b=\hbar\omega_*=8.7\times10^{-9}$~eV. This is the energy of a photon of frequency $E_b/2\pi\hbar=2.1$~MHz, or a thermal energy at $E_b/k_B=1.0\times10^{-4}$~K. Set~2 has a mouth $10^8$ times smaller and $E_b=0.87$~eV. This is a near-infrared photon of frequency $2.1\times10^{14}$~Hz and wavelength $1.43~\mu$m, or a thermal energy at $1.0\times10^{4}$~K. Eq.~\eqref{eq:quantum_ratio} gives $\hbar\omega/E_b=804\,\omega r_e/g^2$. At the photon-sphere frequency of the $l=1$ scalar mode, $\omega r_e=0.3776$ (Table~\ref{tab:qnm_ps}), one quantum carries $304/g^2$ times $E_b$. For the gauge field, $\omega r_e=0.3353$ gives $270/g^2$. The $l=1$ scalar echoes have spectral centroids between $0.749\sqrt{V_0}$ and $0.955\sqrt{V_0}$, that is $\omega r_e=0.296$--$0.378$ (Sec.~\ref{sec:echo_results}). This echo band corresponds to $238/g^2$ to $304/g^2$. The $l=0$ scalar echoes sit lower, since its barrier is lower. Their spectral centroids fall from $121/g^2$ times $E_b$ for $m=1$ to $57/g^2$ for $m=14$. For $l=2$ they lie between $433/g^2$ and $497/g^2$. Even the trapped cavity mode at $\omega r_e=0.05$ (Sec.~\ref{sec:qnm_cavity}) needs $804\times0.05/g^2=40/g^2$ times $E_b$ per quantum.

\begin{figure}
    \centering
    \includegraphics[width=\columnwidth]{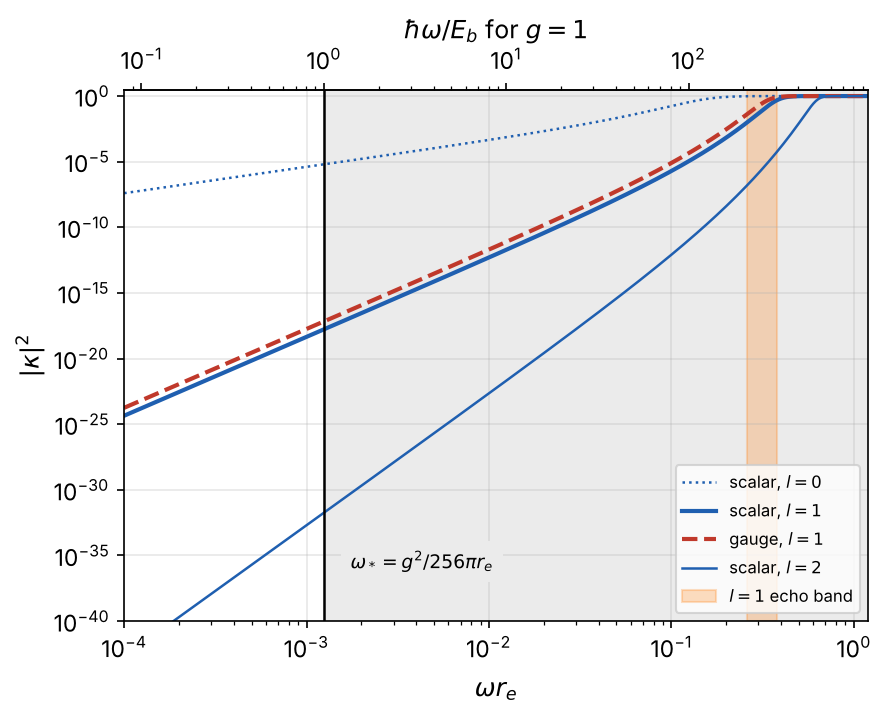}
    \caption{Single-barrier transmission against $\omega r_e$ (bottom axis) and against the energy of one quantum in units of the binding energy, $\hbar\omega/E_b$, for $g=1$ (top axis). The vertical line marks the threshold $\omega_*$ of Eq.~\eqref{eq:omega_star}. Quanta in the shaded region carry more than the binding energy. The orange band is the $l=1$ echo band. Both axes are the same for both parameter sets.}
    \label{fig:binding}
\end{figure}

\subsection{Frequency of the echoes}

The background is static. A wave keeps its frequency through every reflection and transmission, and an echo has the frequency of the wave that went in. The barrier acts as a band-pass filter. A wave well above the barrier top passes through both mouths without being trapped. A wave well below the top is reflected at the first mouth and never enters. Only near the top does a wave both enter the throat and stay trapped long enough to come back. The echoes computed in Sec.~\ref{sec:echoes} sit just below the light-ring frequency for this reason. We call this range the echo band. For the $l=1$ scalar it spans $\omega r_e=0.296$--$0.378$, that is $238$ to $304$ times $\omega_*$ for $g=1$ (Fig.~\ref{fig:binding}).

\subsection{Throat frame}

At the throat centre $\sqrt f=r_e/L_{\rm wh}$, and an infalling quantum is blueshifted by $L_{\rm wh}/r_e$, the boost factor of \cite{Maldacena:2020sxe}. Its proper wavelength there is $\lambda_c=(2\pi/\omega)(r_e/L_{\rm wh})$. With $r_e/L_{\rm wh}=g^2/(16\pi q)$ and $r_e/\ell_p=\sqrt\pi q/g$ from Eq.~\eqref{eq:closed_form_bg},
\begin{equation}
    \frac{\lambda_c}{\ell_p}=\frac{\sqrt\pi}{8}\,\frac{g}{\omega r_e},
    \label{eq:lambda_c}
\end{equation}
again independent of $q$. At $\omega r_e=0.3776$, the $l=1$ scalar photon-sphere frequency, it equals $0.59g$, a local energy of about $11/g$ Planck energies. A quantum in the echo band is outside the semiclassical description for this reason alone. At the threshold $\omega_*$ Eq.~\eqref{eq:lambda_c} gives $\lambda_c/\ell_p=32\pi^{3/2}/g\simeq178/g$. The throat-centre pulse of Sec.~\ref{sec:bounces}, with $b=2r_e$, has the proper width $g\ell_p/8\sqrt\pi\simeq0.07g\,\ell_p$. The boost does not change the energy budget. Both $\hbar\omega$ and $E_b$ are conserved energies measured far away, and a quantum that enters adds $\hbar\omega$ to the mass of the wormhole.

\subsection{Consequences for the echoes}
\label{sec:consequences}

For a classical signal the relevant quantity is the total energy that enters the throat. It must stay below $E_b$, which is $8.7\times10^{-9}$~eV for set~1 and $0.87$~eV for set~2. A classical wave consists of many quanta. It exceeds this energy once more than $E_b/\hbar\omega$ of its quanta have entered. For single quanta the threshold applies quantum by quantum. In the echo band every quantum that tunnels into the throat carries between $57/g^2$ and $497/g^2$ times $E_b$ (Sec.~\ref{sec:onequantum}). The wormhole is then no longer in its ground state, and it is expected to turn into two near-extremal black holes \cite{Maldacena:2018gjk,Maldacena:2020sxe}. We do not compute how fast this happens. The only time-scale of the throat seen from outside is $L_{\rm wh}$, the same order as the crossing time. Whether a first crossing can complete is open. The fixed-background linear theory does not apply in this band, and the echo train cannot form. The echo trains and the trapped cavity modes of Sec.~\ref{sec:echoes} are therefore formal results of linear theory. The prompt reflection and the ringdown of the mouth need no energy to enter the throat and are not affected. The bound \eqref{eq:quantum_ratio} uses only the energy of one quantum and the background. It applies to gravitational waves as well as to the fields studied here.

\subsection{Quanta below the threshold}
\label{sec:below}

Quanta with $\omega<\omega_*$ carry less than the binding energy, and the energy argument does not forbid their echoes. The barrier is almost opaque to them, however. An echo must tunnel into the throat and out again, and its probability carries the factor $|\kappa|^4$. At $\omega_*r_e=1.243\times10^{-3}$ ($g=1$), Eq.~\eqref{eq:sommerfeld} gives
\begin{align}
    |\kappa|^2&=4(\omega_*r_e)^2S_0=6.3\times10^{-6}, & &\text{scalar, }l=0,\nonumber\\
    |\kappa|^2&=\tfrac49(\omega_*r_e)^6S_1=1.7\times10^{-18}, & &\text{scalar, }l=1,\nonumber\\
    |\kappa|^2&=\tfrac{16}{9}(\omega_*r_e)^6S_1=6.7\times10^{-18}, & &\text{gauge, }l=1.
    \label{eq:kappa_star}
\end{align}
The echo probability $|\kappa|^4$ is then $3.9\times10^{-11}$, $2.8\times10^{-36}$ and $4.5\times10^{-35}$. It falls as $\omega^{8l+4}$ at lower frequency.

A signal sent into one mouth first reaches the far mouth after one crossing, with probability $|\kappa|^4$. It first returns to the sender after one round trip, with probability $|\kappa|^4|\beta|^2\simeq|\kappa|^4$. Over all round trips each side receives about $|\kappa|^2/2$ (Sec.~\ref{sec:exterior}). At $\omega_*$ this total is $3.1\times10^{-6}$ for the $l=0$ scalar and $8\times10^{-19}$ for the $l=1$ scalar, and it falls as $\omega^{4l+2}$ at lower frequency. For single quanta these are detection probabilities, and for a classical pulse they are fractions of its energy.

Each crossing of the throat takes the time $2D$ and keeps the fraction $1-|\kappa|^2$ of the trapped energy. A quantum that does get in stays trapped for about $1/|\kappa|^2$ crossings, that is for the time
\begin{equation}
    t_{\rm trap}=\frac{2D}{|\kappa|^2}.
    \label{eq:ttrap}
\end{equation}
With $2D=1.5\times10^{25}$~s for set~1 (Sec.~\ref{sec:cavity}) this is $2.4\times10^{30}$~s for the $l=0$ scalar at $\omega_*$ and $2.2\times10^{42}$~s for the $l=1$ gauge field. With $2D=1.5\times10^{9}$~s for set~2 it is $2.4\times10^{14}$~s, or $7.5\times10^{6}$~yr, and $2.2\times10^{26}$~s. It is longer at lower frequency, where $|\kappa|^2$ is smaller. These times hold for an undisturbed wormhole. For set~2 they exceed the mean time between X-ray photons, $t_X=8.7\times10^{2}$~yr of Eq.~\eqref{eq:tX}. Within $t_X$ the far mouth emits nine times, at the times $2D+4kD$ with $k=0,\dots,8$, and passes on $3.6\times10^{-10}$ of the $l=0$ signal at $\omega_*$. At these frequencies the wavelength, $2\pi/\omega_*\simeq5000\,r_e$, is far larger than the mouth. It is $1.4\times10^{4}$~cm for set~1 and $1.4~\mu$m for set~2. This is the low-frequency regime of \cite{Freivogel:2026ujn}.

\subsection{Lifetime in a thermal bath}
\label{sec:lifetime}

The energy bound becomes a lifetime once the wormhole sits in radiation. Infalling quanta are blueshifted by $L_{\rm wh}/r_e$ at the centre, and quanta that lose energy inside accumulate and add positive energy \cite{Maldacena:2020sxe}. Consider a bath of massless quanta at temperature $T$ that illuminates both mouths. Each partial wave is a one-dimensional channel, and each $l$ contributes $2l+1$ of them. A thermal channel carries $\bar n(\omega)\,d\omega/2\pi$ incoming quanta per unit time in each frequency interval, and a fraction $|\kappa_l|^2$ of them crosses the barrier. Per mouth and per polarisation, quanta pass one barrier into the throat at the rate
\begin{align}
    \mathcal R&=\int_0^\infty\frac{d\omega}{2\pi}\,\bar n(\omega)\,\Sigma(\omega),\nonumber\\
    \Sigma(\omega)&=\sum_l(2l+1)\,|\kappa_l(\omega)|^2,
    \label{eq:bath_rate}
\end{align}
with $\bar n=1/(e^{\hbar\omega/k_BT}-1)$. Here $\Sigma$ is the absorption cross-section in units of $\pi/\omega^2$. At high frequency a wave behaves like a ray. Every ray with impact parameter below the critical value $b_c=r_{ph}/\sqrt{f(r_{ph})}=4r_e$ of the photon sphere is captured \cite{Chandrasekhar:1985kt}, and every other ray is scattered. The cross-section then approaches the area $\pi b_c^2$ of a disk of radius $b_c$, which we call its geometric value. In units of $\pi/\omega^2$ it is $\Sigma\simeq16\,\omega^2r_e^2$. Our sums over $l$ give $\Sigma/16\omega^2r_e^2=0.990$ (scalar) and $0.934$ (gauge) at $\omega r_e=1$, and $1.000$ and $0.998$ at $\omega r_e=6$.

The wormhole can collapse in two ways. A single quantum above $\omega_*$ enters the throat, or quanta below $\omega_*$ bring in a total energy $E_b$. We therefore split the rate \eqref{eq:bath_rate} at $\omega_*$ and define
\begin{align}
    \mathcal R_>&=\int_{\omega_*}^\infty\frac{d\omega}{2\pi}\,\bar n\,\Sigma,\nonumber\\
    P_<&=\int_0^{\omega_*}\frac{d\omega}{2\pi}\,\hbar\omega\,\bar n\,\Sigma .
    \label{eq:bath_split}
\end{align}
Per mouth and per polarisation, $\mathcal R_>$ is the number of quanta above $\omega_*$ and $P_<$ the energy below $\omega_*$ that enter per unit time. With two mouths and $N_{\rm pol}$ polarisations, quanta above $\omega_*$ enter at the rate $\Upsilon_>=2N_{\rm pol}\mathcal R_>$. Here $N_{\rm pol}=1$ for the scalar and $2$ for the gauge field. A quantum below $\omega_*$ that enters does not stay forever. Each crossing of the throat takes the time $2D$, and at each barrier the quantum leaves with probability $|\kappa_l|^2$. It escapes after about $2D/|\kappa_l(\omega)|^2$, Eq.~\eqref{eq:ttrap}. The energy trapped in the throat after a time $t$ is then
\begin{align}
    E_<(t)&=2N_{\rm pol}\int_0^{\omega_*}\frac{d\omega}{2\pi}\,\hbar\omega\,\bar n\nonumber\\
    &\quad\times\sum_l(2l+1)\,2D\left[1-e^{-t|\kappa_l|^2/2D}\right].
    \label{eq:E_trap}
\end{align}
For $t\ll2D/|\kappa_l|^2$ the bracket is $t|\kappa_l|^2/2D$, and $E_<=2N_{\rm pol}P_<\,t$. For $t\gg2D/|\kappa_l|^2$ each channel holds the occupation $\bar n$ of the bath. Its energy is then that of a one-dimensional thermal gas in a box of length $2D$. We define $\Upsilon_<$ as the inverse of the time at which $E_<$ reaches $E_b$, and $\Upsilon_<=0$ if it never does. Without escape this time is $E_b/(2N_{\rm pol}P_<)$ with $P_<$ of Eq.~\eqref{eq:bath_split}, and $\Upsilon_<=2N_{\rm pol}P_</E_b$. We take the lifetime as the inverse of the total rate,
\begin{equation}
    t_{\rm life}=\frac{1}{\Upsilon_>+\Upsilon_<}.
    \label{eq:lifetime}
\end{equation}
This is exact when one of the two rates dominates. When both act, the mean time to collapse is $(1-e^{-\Upsilon_>/\Upsilon_<})/\Upsilon_>$, and Eq.~\eqref{eq:lifetime} differs from it by at most a factor $1.3$.

Both limits of Eq.~\eqref{eq:lifetime} follow in closed form when escape is neglected. We set $\hbar=c=k_B=1$. At high temperature almost every absorbed quantum lies above $\omega_*$ and far above the barrier top. The cross-section then takes its geometric value, $\Sigma\simeq16\omega^2r_e^2$, the capture area $\pi b_c^2$ of the photon sphere. The thermal integrals are
\begin{equation}
    \int_0^\infty\omega^2\bar n\,d\omega=2\zeta(3)T^3,\qquad \int_0^\infty\omega^3\bar n\,d\omega=\frac{\pi^4T^4}{15}.
    \label{eq:thermal_int}
\end{equation}
The first gives $t_{\rm life}\simeq1/(2N_{\rm pol}\mathcal R_>)$. Restoring $\hbar$, $c$ and $k_B$, this is
\begin{equation}
    t_{\rm life}\simeq\frac{\pi}{32\zeta(3)N_{\rm pol}}\,\frac{r_e}{c}\left(\frac{\hbar c}{k_BTr_e}\right)^3
    \label{eq:life_high}
\end{equation}
for $k_BT\gg\hbar c/r_e$. This holds for both fields, with $N_{\rm pol}=1$ for the scalar and $2$ for the gauge field, and includes every $l$. At low temperature in a scalar bath only the $l=0$ channel matters, with $\Sigma\simeq4\omega^2r_e^2$. The second integral of Eq.~\eqref{eq:thermal_int} and $E_b$ of Eq.~\eqref{eq:Eb} give $t_{\rm life}\simeq E_b/(2P_<)$, that is
\begin{equation}
    t_{\rm life}\simeq\frac{15g^2}{1024\pi^4}\,\frac{r_e}{c}\left(\frac{\hbar c}{k_BTr_e}\right)^4
    \label{eq:life_low}
\end{equation}
for $k_BT\ll\hbar\omega_*$. A vector field has no radiating monopole. Its axial mode is built on $\partial_\theta Y_l$, which vanishes for $l=0$, and its polar $l=0$ mode is a static Coulomb field (\ref{app:gp}). The lowest gauge channel is therefore $l=1$, with $\Sigma\simeq3K_1(\omega r_e)^6=\tfrac{16}{3}(\omega r_e)^6$ from Eq.~\eqref{eq:lowfreq}. With $\int_0^\infty\omega^7\bar n\,d\omega=8\pi^8T^8/15$ and $N_{\rm pol}=2$ the same steps give
\begin{equation}
    t_{\rm life}\simeq\frac{45g^2}{65536\pi^8}\,\frac{r_e}{c}\left(\frac{\hbar c}{k_BTr_e}\right)^8
    \label{eq:life_low_gauge}
\end{equation}
for $k_BT\ll\hbar\omega_*$. In both baths each higher channel is weaker by a factor of order $(\omega r_e)^4$.

\begin{figure}
    \centering
    \includegraphics[width=\columnwidth]{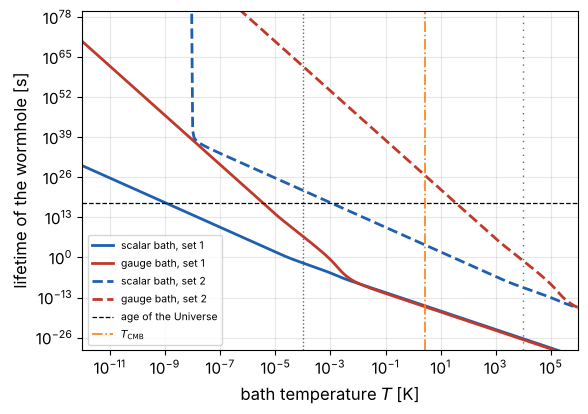}
    \caption{Lifetime of the wormhole in a thermal bath of massless scalar or gauge quanta, Eq.~\eqref{eq:lifetime} with the escape of Eq.~\eqref{eq:E_trap}, against the bath temperature. Solid curves are set~1 and dashed curves set~2. The dotted vertical lines mark the threshold temperatures $E_b/k_B$ of the two sets, $10^{-4}$~K and $10^{4}$~K. The vertical rise of the dashed blue curve is at $T_{\rm eq}=9.9\times10^{-9}$~K, Eq.~\eqref{eq:Teq}. The dash-dotted line is the temperature of the cosmic microwave background, and the dashed line the age of the Universe.}
    \label{fig:lifetime}
\end{figure}

Figure~\ref{fig:lifetime} shows the lifetime for both sets. For set~1, in a bath at the temperature of the cosmic microwave background, $2.7$~K, the wormhole lasts $2\times10^{-16}$~s in a scalar bath and $1\times10^{-16}$~s in a gauge bath. At the threshold temperature $E_b/k_B\simeq10^{-4}$~K it lasts $1.6\times10^{-2}$~s and $7.5\times10^{6}$~s. It survives for the age of the Universe only below $1.1\times10^{-9}$~K in a scalar bath and $3.6\times10^{-6}$~K in a gauge bath.

The lowest gauge channel, $l=1$, tunnels as $|\kappa|^2\propto\omega^6$, Eq.~\eqref{eq:lowfreq}. Above $E_b/k_B$ the gauge lifetime therefore falls as $T^{-7}$, set by quanta near $\omega\simeq6k_BT/\hbar$. Near $5$~mK these quanta reach the barrier top $0.354/r_e$, with $\hbar c/(k_Br_e)=0.081$~K. The cross-section approaches its geometric value $\pi b_c^2$, and the lifetime bends to $T^{-3}$. Just below the top $|\kappa|^2$ rises faster than $\omega^6$ (Table~\ref{tab:powerlaw}), and the local slope $d\ln t_{\rm life}/d\ln T$ briefly reaches $-10$ at $1.8\times10^{-3}$~K. We call this point the kink of the gauge curve. The scalar $l=0$ channel has $\Sigma\propto\omega^2$ at both low and high frequency. Between the two $\Sigma$ rises from $4\omega^2r_e^2$ to $16\omega^2r_e^2$, and the scalar slope steepens only to $-3.8$ near $1.6\times10^{-3}$~K.

Both curves also bend near $E_b/k_B$, from $T^{-4}$ to $T^{-3}$ for the scalar and from $T^{-8}$ to $T^{-7}$ for the gauge field. There the main cause of collapse passes from the accumulated energy of quanta below $\omega_*$ to single quanta above it. For set~1 escape lengthens the time to accumulate $E_b$ by less than $2\times10^{-14}$ in relative terms between $10^{-12}$ and $10^{6}$~K.

Without escape, $t_{\rm life}c/r_e$ depends on the temperature only through $k_BTr_e/\hbar c$. The curves of set~2 are those of set~1 moved to $10^{8}$ times higher temperature and $10^{8}$ times shorter time. They show the same features. With $\hbar c/(k_Br_e)=8.1\times10^{6}$~K, the bend lies near $E_b/k_B=10^{4}$~K. The gauge quanta reach the barrier top near $5\times10^{5}$~K, and the gauge kink, with slope $-10$, lies at $1.8\times10^{5}$~K. At the threshold temperature $E_b/k_B\simeq10^{4}$~K set~2 lasts $1.6\times10^{-10}$~s in a scalar bath and $7.5\times10^{-2}$~s in a gauge bath. The cosmic microwave background lies far below this threshold, and only the accumulated energy of quanta below $\omega_*$ acts. Set~2 then lasts $1.1\times10^{4}$~s in a scalar bath and $4.1\times10^{26}$~s, or $1.3\times10^{19}$~yr, in a gauge bath. Equations~\eqref{eq:life_low} and~\eqref{eq:life_low_gauge} reproduce both values. It survives for the age of the Universe below $1.1\times10^{-3}$~K in a scalar bath and $36$~K in a gauge bath.

Escape matters for set~2 only at the lowest temperatures. It lengthens the lifetime by more than a factor $2$ below $1.8\times10^{-8}$~K (scalar) and $3.5\times10^{-8}$~K (gauge). There the lifetime already exceeds the age of the Universe by many orders of magnitude. Let $l_0$ be the lowest channel, $l_0=0$ for the scalar and $l_0=1$ for the gauge field. At late times this channel of Eq.~\eqref{eq:E_trap} holds $N_{\rm pol}(2l_0+1)\pi D(k_BT)^2/3\hbar c$, the energy of a thermal gas of length $2D$. It stays below $E_b$ for $T<T_{\rm eq}$, with
\begin{equation}
    k_BT_{\rm eq}=\sqrt{\frac{3\hbar c\,E_b}{N_{\rm pol}(2l_0+1)\pi D}}.
    \label{eq:Teq}
\end{equation}
Eq.~\eqref{eq:Teq} gives $T_{\rm eq}=9.9\times10^{-9}$~K (scalar) and $4.0\times10^{-9}$~K (gauge) for set~2, and $9.9\times10^{-21}$~K and $4.0\times10^{-21}$~K for set~1. As $T$ falls towards $T_{\rm eq}$ the time to reach $E_b$ grows without bound, since the lowest frequencies of the channel $l_0$ fill last. Below $T_{\rm eq}$ this channel saturates below $E_b$, and the next one fills only on a time-scale longer by the factor $(\omega r_e)^{-4}$. The lifetime then jumps by many orders of magnitude. For set~2 in a scalar bath it is $7.5\times10^{36}$~s at $2T_{\rm eq}$, $4.4\times10^{40}$~s at $1.02\,T_{\rm eq}$ and $1.0\times10^{94}$~s at $0.98\,T_{\rm eq}$. This is the vertical rise of the dashed blue curve near $10^{-8}$~K in Fig.~\ref{fig:lifetime}. In a gauge bath the lifetime of set~2 goes from $4.0\times10^{107}$~s at $1.02\,T_{\rm eq}$ to $1.9\times10^{157}$~s at $0.98\,T_{\rm eq}$, above the range of the figure. For set~1 the jump lies below the range of the figure. There the scalar lifetime rises from $4.4\times10^{64}$~s at $1.02\,T_{\rm eq}$ to $1.0\times10^{134}$~s at $0.98\,T_{\rm eq}$, and the gauge lifetime from $4.0\times10^{147}$~s to $1.9\times10^{213}$~s. The solid curves therefore show no jump in Fig.~\ref{fig:lifetime}.

Their features are the bend near $E_b/k_B=10^{-4}$~K, the gauge kink at $1.8\times10^{-3}$~K and the mild steepening of the scalar curve near the same temperature. At every jump the lifetime already exceeds the age of the Universe by more than twenty orders of magnitude. The jump changes no conclusion about survival. It marks the temperature below which the lowest channel alone can never store $E_b$. Since $D\simeq\pi L_{\rm wh}/2$, $T_{\rm eq}$ is of the order of the geometric mean of the threshold temperature $E_b/k_B$ and the gap $\hbar c/(k_BL_{\rm wh})$ of the throat.

The throat must also not fill with thermal excitations. The orbital motion of the mouths gives them an Unruh-like temperature proportional to the orbital frequency $\Omega$. MMP require it to stay far below the energy gap of the throat, $\hbar c/(k_BL_{\rm wh})$, which is the condition \eqref{eq:slow_orbit}. The gap is $1.6\times10^{-36}$~K for set~1 and $1.6\times10^{-20}$~K for set~2. A hotter throat fills with excitations, which likely ends with the formation of a horizon \cite{Maldacena:2018gjk}. The survival temperatures above exceed this gap. Our lifetimes therefore assume that the bath does not heat the throat fermions.

The bath that describes real photons depends on how the wormhole is realised. In \cite{Maldacena:2020sxe} the throat is supported by a dark U(1) that has its own charged fermions. Ordinary photons then have no background field and no charged matter in the throat. They behave as our test gauge field, and the gauge-bath lifetime above is the lifetime in the cosmic microwave background for either set. If the wormhole's U(1) were electromagnetism itself, its photon would couple to the throat fermions and its barrier would change (Sec.~\ref{sec:em_pert}). At frequencies far above the barrier the absorption of any massless field approaches the same geometric limit, however. For set~1 the thermal energy $k_BT$ of the cosmic microwave background is $2.7\times10^{4}$ times the binding energy, and the short lifetime does not depend on the barrier. For set~2 it is $2.7\times10^{-4}$ times $E_b$, and the lifetime would depend on the barrier of the coupled photon.

The lifetimes of this section count massless quanta only. Slow massive particles are not included. A hydrogen atom at rest, with $m_Hc^2=938.8$~MeV, carries $1.1\times10^{17}$ times $E_b$ for set~1 and $1.1\times10^{9}$ times for set~2. Its capture needs the geodesics and the wave equation of a massive particle, which we do not study. Massive particles can only add to the rates of this section, and the lifetimes given here are upper bounds.

Our lifetimes in the cosmic microwave background refer to the dark realisation. The background gauge field and the charged fermions of MMP then belong to a dark sector. They couple to ordinary matter and radiation, and to the test scalar, only through gravity. We assume that the throat fermions do not thermalise and stay in their ground state. Gravity could still heat them, and a bath of dark-sector quanta would heat them directly. We do not compute either effect.

The cosmic microwave background is not the only photon bath. For set~2 it lies below the threshold, but a single photon above $E_b=0.87$~eV that enters the throat is expected to destroy the wormhole (Sec.~\ref{sec:consequences}). An isotropic photon intensity $N(E)$ sends
\begin{equation}
    \mathcal R_X=4\pi\int N(E)\,\sigma(E)\,dE,\qquad \sigma=\frac{\pi\Sigma}{\omega^2},
    \label{eq:xray_rate}
\end{equation}
photons per unit time into one mouth. Here $\sigma$ is the capture cross-section, with $\Sigma$ of Eq.~\eqref{eq:bath_rate}. Between $0.3$ and $7$~keV the extragalactic X-ray background is the power law \cite{Cappelluti:2017miu}
\begin{equation}
    N_1(E)=10.91\left(\frac{E}{\rm keV}\right)^{-1.45}
    \label{eq:N_chandra}
\end{equation}
in photons cm$^{-2}$\,s$^{-1}$\,sr$^{-1}$\,keV$^{-1}$. Between $1.5$ and $200$~keV it is fitted in the same units by \cite{Moretti:2008hs}
\begin{equation}
    N_2(E)=\frac{0.109}{(E/E_B)^{1.40}+(E/E_B)^{2.88}}
    \label{eq:N_swift}
\end{equation}
with $E_B=29.0$~keV. We use Eq.~\eqref{eq:N_chandra} up to $7$~keV and Eq.~\eqref{eq:N_swift} above it.

For set~2 a photon of $0.3$~keV has $\omega r_e=0.43$, above the $l=1$ barrier top, and carries $347\,E_b$. From $0.3$ to $200$~keV the cross-section lies between $0.73$ and $1.05$ times its geometric value
\begin{equation}
    \pi b_c^2=16\pi r_e^2=4.0\times10^{-14}~{\rm cm}^2 .
\end{equation}
With two mouths the mean time between photons is $t_X=1/(2\mathcal R_X)$, with $\mathcal R_X$ of Eq.~\eqref{eq:xray_rate}. The two bands give
\begin{align}
    t_X&=3.3\times10^{10}~{\rm s}=1.0\times10^{3}~{\rm yr}, & &0.3\text{--}7~{\rm keV},\nonumber\\
    t_X&=2.8\times10^{10}~{\rm s}=8.7\times10^{2}~{\rm yr}, & &0.3\text{--}200~{\rm keV}.
    \label{eq:tX}
\end{align}
The photons between $7$ and $200$~keV raise the rate by $18\%$. Between $2$ and $7$~keV the two fits overlap, and their rates differ by $13\%$. This measures the uncertainty of the input. The lifetime of $1.3\times10^{19}$~yr therefore holds only for a wormhole shielded from photons above $E_b$. Without such a shield the life of set~2 is at most $t_X$. This is an upper bound from photons alone.

Photons outside this band change the result little. Extending Eq.~\eqref{eq:N_swift} from $200$~keV to $10$~MeV adds $0.12\%$ to the rate. Below $0.3$~keV neither fit applies. A mouth captures efficiently only above the $l=1$ barrier top, which is $0.25$~keV for set~2, and extending Eq.~\eqref{eq:N_chandra} down to $0.1$~keV would add $20\%$. Starlight lies far below the barrier. Below the top $\sigma/E$ rises with $E$. The rate is therefore largest if the whole optical background of $24$~nW\,m$^{-2}$\,sr$^{-1}$ \cite{Driver:2016krv} sits at the Lyman limit, $13.6$~eV, and even then it adds less than $0.2\%$. Photons below $E_b$ only add energy. The optical and infrared backgrounds together, $50$~nW\,m$^{-2}$\,sr$^{-1}$ \cite{Driver:2016krv}, need more than $10^{9}$~yr to bring in $E_b$. We neglect all of these. For set~1 the cosmic microwave background already destroys the wormhole within $10^{-16}$~s.

\section{Discussion and conclusion}
\label{sec:conclusion}

We have studied a massless scalar and a test U(1) field on the MMP wormhole. Each mouth acts on a wave as the barrier of an extremal Reissner--Nordstr\"om black hole, and the throat acts as a cavity of half-length $D\simeq\pi L_{\rm wh}/2$, Eq.~\eqref{eq:D_closed}. By Eq.~\eqref{eq:quantum_ratio} a quantum of frequency $\omega$ carries $\hbar\omega/E_b=256\pi\,\omega r_e/g^2$ times the binding energy. In units of $1/r_e$ this ratio does not depend on the charge, and a quantum carries less than $E_b$ only below the threshold frequency $\omega_*$, Eq.~\eqref{eq:omega_star}. The echoes sit just below the light-ring frequency of the mouth. For the echoes $m=1$ to $14$ their spectral centroids give $57/g^2$ to $121/g^2$ times $E_b$ for $l=0$, $238/g^2$ to $304/g^2$ for the $l=1$ scalar and $433/g^2$ to $497/g^2$ for $l=2$ (Sec.~\ref{sec:onequantum}). A single quantum in this band is expected to turn the wormhole into two black holes \cite{Maldacena:2018gjk,Maldacena:2020sxe}. The fixed background then does not survive, and the echo trains and the trapped cavity modes are formal results of linear theory.

Below $\omega_*$ the wormhole survives the signal. Suppose an observer sends a low-frequency signal into one mouth from outside, and a second observer waits at the other mouth. Each barrier transmits with probability $|\kappa|^2$, given at low frequency by Eq.~\eqref{eq:sommerfeld}, and reflects with probability $|\beta|^2=1-|\kappa|^2$. The signal first reaches the second observer after one crossing of the throat, with probability $|\kappa|^4$. Every later round trip inside multiplies this by $|\beta|^4$. Over all round trips the far mouth receives the fraction of Eq.~\eqref{eq:far_fraction}, about $|\kappa|^2/2$ (Sec.~\ref{sec:exterior}). The sender first receives a part of the signal that has been through the throat after one round trip, with probability $|\kappa|^4|\beta|^2\simeq|\kappa|^4$. Over all echoes the sender also receives about $|\kappa|^2/2$.

For the $l=0$ scalar at $\omega_*$ the first passage has probability $3.9\times10^{-11}$ and the total is $3.1\times10^{-6}$. For the $l=1$ scalar they are $2.8\times10^{-36}$ and $8\times10^{-19}$ (Sec.~\ref{sec:below}). Both fall at lower frequency, the first as $\omega^{8l+4}$ and the total as $\omega^{4l+2}$. For single quanta these are detection probabilities, and for a classical pulse they are fractions of its energy. Collecting the total takes about $1/|\kappa|^2$ crossings.

For set~1 one crossing takes $2D=\pi L_{\rm wh}\simeq4.7\times10^{17}$~yr and one round trip $9.5\times10^{17}$~yr (Sec.~\ref{sec:cavity}). The total for the $l=0$ scalar at $\omega_*$ takes $2.4\times10^{30}$~s. For set~2 the first passage arrives after $47$~yr and the first return to the sender after $95$~yr. Collecting the total would take $7.5\times10^{6}$~yr. Photons of the X-ray background limit the life of set~2 to about $8.7\times10^{2}$~yr, Eq.~\eqref{eq:tX}. Within this bound the far mouth emits at most nine times and passes on at most $3.6\times10^{-10}$ of the $l=0$ signal. The threshold frequency of set~2 is $2.1\times10^{14}$~Hz, a wavelength of $1.4~\mu$m, and every radio quantum lies below this frequency. Behind a horizon nothing would reach the second observer, and nothing would return to the sender. Only these weak signals let a test field tell a mouth from an extremal black hole.

The low-frequency transmission through one barrier follows in closed form, $|\kappa|^2\simeq K_l(\omega r_e)^{4l+2}S_l$, Eq.~\eqref{eq:sommerfeld}. The prefactor $K_l$ agrees with earlier results \cite{Maldacena:1997ih,Gubser:1997cm,Crispino:2000jx}. The Coulomb factor $S_l$ is new as far as we know. The logarithm in the tortoise coordinate acts on a wave as an attractive Coulomb potential. On one side of the barrier this multiplies $|\kappa|^2$ by the factor $[C_l(\eta)/C_l(0)]^2$, with $C_l$ of Eq.~\eqref{eq:Cl}. The inversion symmetry puts the same Coulomb problem on the other side, and $S_l$ is the square of this factor. The numerics reproduce the exponent, the prefactor and $S_l$, and the lineshape of the cavity resonances confirms $S_0$ (Sec.~\ref{sec:cavity}).

This law fixes the late-time decay of the echoes in linear theory. For a pulse released in the throat the echo energies fall as $E_m\propto m^{-1-3/(4l+2)}$ with echo number $m$, Eq.~\eqref{eq:latetime}. A pulse sent in from outside tunnels twice, and its echoes fall one power faster, $E_k\propto(2k-1)^{-2-3/(4l+2)}$, Eq.~\eqref{eq:latetime_ext}. Earlier studies obtained power laws for echo trains from fits \cite{Wang:2018gin,Abedi:2026gxm}. Our derivation uses only the low-frequency form of $|\kappa|^2$ and a cavity without absorption, and it applies to any such cavity. For $l=1$ the asymptotic law is reached only for $m\gtrsim10^{12}$. Below the barrier top the quality factor of the cavity modes grows as $2\omega_RD/|\kappa|^2$, Eq.~\eqref{eq:Q_kappa}.

The same energy budget limits the life of the wormhole in thermal radiation. Quanta above $\omega_*$ destroy it one at a time, and quanta below $\omega_*$ once their energy adds up to $E_b$. The lifetime has closed-form limits at high and low temperature, Eqs.~\eqref{eq:life_high}, \eqref{eq:life_low} and~\eqref{eq:life_low_gauge}. At low temperature only the lowest channel acts, $l=0$ for the scalar and $l=1$ for the gauge field. Quanta below $\omega_*$ also escape after $2D/|\kappa_l|^2$, and below the temperature $T_{\rm eq}$ of Eq.~\eqref{eq:Teq} the lowest channel cannot store $E_b$. If the throat is supported by a dark sector (Sec.~\ref{sec:lifetime}), set~1 lasts about $10^{-16}$~s in the cosmic microwave background. It survives for the age of the Universe only below $1.1\times10^{-9}$~K in a scalar bath and $3.6\times10^{-6}$~K in a gauge bath. Set~2 lasts $4.1\times10^{26}$~s in the cosmic microwave background. In a gauge bath it survives for the age of the Universe below $36$~K. Photons of the X-ray background still destroy set~2 within about $8.7\times10^{2}$~yr. All these lifetimes count massless quanta only. Slow massive particles can only shorten them, and the lifetimes are upper bounds.

The two parameter sets differ only in the charge, $q=10^{33}$ and $q=10^{25}$. Every single-barrier quantity, the energy ratio \eqref{eq:quantum_ratio} and the probabilities above are the same for both. Lengths scale as $r_e\propto q$, and the time-scales of the throat as $L_{\rm wh}\propto q^2$. Only set~2 fits inside a universe with our cosmological constant. Its echo band still lies above $\omega_*$, and the echo trains cannot form for either set.

The ringdown of a mouth does not probe the throat. The single-barrier problem is the extremal black-hole problem, and the photon-sphere modes reproduce the spectrum of \cite{Onozawa:1995vu} to within $3\times10^{-6}$ (Table~\ref{tab:qnm_ps}). At $l=1$ the quality factor is $Q\simeq2$, and the mouth rings down within about one cycle (Sec.~\ref{sec:qnm_photonsphere}). The prompt ringdown of a horizonless compact object is in general that of its photon sphere as well \cite{Cardoso:2016rao}. Neither the echoes nor the ringdown of this class of solution supplies an echo template for gravitational-wave searches \cite{LIGOScientific:2020tif,Abedi:2016hgu}.

These results rest on the amplitudes of one barrier. A direct time-domain evolution across the cavity, of length $2D\simeq1.6\times10^{35}r_e$ for set~1 and $1.6\times10^{27}r_e$ for set~2, is out of reach. We computed the amplitudes by Numerov integration with exact Riccati--Hankel boundary conditions and summed the reflections in closed form. Changing the step, the domain or $N_\omega$ alters $|\kappa|^2$ by less than $7\times10^{-7}$ (Table~\ref{tab:convergence}), and a 20-digit recomputation agrees to $2.5\times10^{-7}$ down to $|\kappa|^2\sim10^{-30}$. An independent leapfrog evolution reproduces their phases (Fig.~\ref{fig:fdtd}). The leading-order parabolic WKB formula \eqref{eq:wkb_par} fails below the barrier top. Used in the echo calculation, it overestimates the echo amplitudes by a factor $3.1$ at $m=14$ (Sec.~\ref{sec:echo_results}).

\subsection{Future directions}

The most direct extension is to the coupled gauge--gravity perturbations, which give the response of the mouth without assuming a test field. Their master equations, including the response of the fermions, are derived in \cite{Sadhukhan:2026czl,Sadhukhan:2026mcu}. The echoes and quasinormal modes of that system remain to be computed. The background magnetic field is odd under parity. Axial gravitational perturbations therefore couple to polar electromagnetic ones, and polar gravitational perturbations to axial electromagnetic ones \cite{DeFelice:2023rra}. This reverses the pairing familiar from the electrically charged black hole \cite{Zerilli:1974ai}. In the coupled system the two parity sectors are no longer isospectral. Their echo trains would differ, and the difference would carry the imprint of the fermions. The contour integration of \ref{app:contour} can also be extended to the overtones $n\ge1$. A massive scalar field is a further extension. It would give the capture of slow massive particles, which the lifetimes of Sec.~\ref{sec:lifetime} leave out. A quantum treatment of a single quantum entering the throat would make the collapse argument of Sec.~\ref{sec:energy} quantitative.

\paragraph{Data Availability Statement} This manuscript has no associated data. [Authors' comment: No external data were used. Every number, table and data figure in this paper is generated by the notebook cited in the Code Availability Statement.]

\paragraph{Code Availability Statement} This manuscript has associated code/software in a data repository. [Authors' comment: The notebook that reproduces all numerical results is available on Zenodo \cite{Mondal:2026code}, \url{https://doi.org/10.5281/zenodo.24578801}.]

\appendix

\section{Scalar and vector master equations}
\label{app:scalar}
\label{app:gp}

We use the metric of Eq.~\eqref{eq:static_form} with $dr_*=dr/f$ and a prime for $d/dr$.

\paragraph{Scalar field.} With $\Phi=R(r)Y_{lm}e^{-i\omega t}$ and $\sqrt{-g}=g_{\theta\theta}\sin\theta$, the Klein--Gordon equation gives
\begin{equation}
    \frac{d}{dr}\!\left(g_{\theta\theta}f\frac{dR}{dr}\right) + \left[\frac{\omega^2 g_{\theta\theta}}{f} - l(l+1)\right]R = 0.
    \label{eq:radial_scalar_general}
\end{equation}
With $Z=g_{\theta\theta}^{1/2}$ and $\chi=ZR$ the first-derivative terms cancel, and
\begin{equation}
    \frac{d^2\chi}{dr_*^2} + \Big[\omega^2 - V_s\Big]\chi = 0,\qquad V_s=f\frac{l(l+1)}{g_{\theta\theta}}+\frac{Z_{,r_*r_*}}{Z}.
    \label{eq:master_general_scalar}
\end{equation}
Since $Z_{,r_*r_*}/Z=f(fZ')'/Z$ with $Z'=g_{\theta\theta}'/2Z$,
\begin{align}
    V_s = \;&f\frac{l(l+1)}{g_{\theta\theta}} + \frac{f f'\,g_{\theta\theta}'}{2\,g_{\theta\theta}} \nonumber\\
    &+ \frac{f^2\,g_{\theta\theta}''}{2\,g_{\theta\theta}} - \frac{f^2\,(g_{\theta\theta}')^2}{4\,g_{\theta\theta}^2}.
    \label{eq:Vs_general}
\end{align}
With $g_{\theta\theta}=r^2$ and $g_{\theta\theta}=h_{\rm th}$ Eq.~\eqref{eq:Vs_general} gives Eqs.~\eqref{eq:mouth_potential_scalar} and~\eqref{eq:throat_potential_scalar}. The stress tensor of $\Phi$ is quadratic in $\Phi$, and the metric perturbation it sources does not enter at this order.

\paragraph{Vector field.} The field has no background value, and $\mathcal A_\mu$ and $F_{\mu\nu}=\partial_\mu\mathcal A_\nu-\partial_\nu\mathcal A_\mu$ are of first order. We project the vacuum Maxwell equations
\begin{equation}
    \partial_\mu \left( \sqrt{-g} \, g^{\mu\alpha} g^{\nu\beta} F_{\alpha\beta} \right) = 0
    \label{eq:maxwell_appendix}
\end{equation}
onto the vector harmonics of \cite{Zerilli:1974ai}. Spherical symmetry makes the radial equations independent of $m$, and we set $m=0$ with $Y_l\equiv Y_{l0}(\theta)$. Then $\partial_\theta(\sin\theta\,\partial_\theta Y_l)=-l(l+1)\sin\theta\,Y_l$.

The axial mode is
\begin{equation}
    \mathcal A_\mu=e^{-i\omega t}\Psi(r)\,(0,0,0,-\sin\theta\,\partial_\theta Y_l).
\end{equation}
Its only field components are $F_{t\phi}$, $F_{r\phi}$ and $F_{\theta\phi}=l(l+1)\Psi\,e^{-i\omega t}\sin\theta\,Y_l$. The $\nu=\phi$ component of Eq.~\eqref{eq:maxwell_appendix} carries the common factor $e^{-i\omega t}\partial_\theta Y_l$ and gives $(\omega^2/f)\Psi+(f\Psi')'-l(l+1)\Psi/g_{\theta\theta}=0$, that is
\begin{equation}
    \frac{d^2\Psi}{dr_*^2} + \left[ \omega^2 - V_g \right] \Psi = 0, \qquad V_g = f \frac{l(l+1)}{g_{\theta\theta}}.
    \label{eq:axial_master}
\end{equation}
The polar mode is $\mathcal A_\mu=e^{-i\omega t}(\mathcal U Y_l,\mathcal W Y_l,0,0)$. The $\nu=\theta$ and $\nu=r$ components of Eq.~\eqref{eq:maxwell_appendix} give
\begin{align}
    &f(f\mathcal W)'=-i\omega\,\mathcal U,\nonumber\\
    &-i\omega\,g_{\theta\theta}\,\mathcal U'+\big[\omega^2g_{\theta\theta}-l(l+1)f\big]\mathcal W=0,
\end{align}
and the $\nu=t$ component then holds identically, since $\partial_\nu\partial_\mu(\sqrt{-g}F^{\mu\nu})\equiv0$. Eliminating $\mathcal U$ gives
\begin{equation}
    \big[f(f\mathcal W)'\big]'+\left[\omega^2-\frac{l(l+1)f}{g_{\theta\theta}}\right]\mathcal W=0 .
\end{equation}
With $P=f\mathcal W$ and $f\,d/dr=d/dr_*$ this is Eq.~\eqref{eq:axial_master} for $P$. The two parities share the same potential and spectrum, the polar--axial isospectrality of a test Maxwell field on a spherically symmetric background \cite{Zerilli:1974ai}. With $g_{\theta\theta}=r^2$ and $g_{\theta\theta}=r_e^2(1+\phi)$ the potential of Eq.~\eqref{eq:axial_master} gives Eqs.~\eqref{eq:gauge_mpot} and~\eqref{eq:gauge_thpot}. For $l=0$ the axial mode vanishes, since $\partial_\theta Y_0=0$. The polar mode then has the single component $F_{tr}\propto i\omega\mathcal W+\mathcal U'$, and the $\nu=r$ equation sets it to zero for $\omega\neq0$. Only the static Coulomb field remains, and the test vector field has no propagating $l=0$ mode.

\section{Complex-contour computation of the photon-sphere modes}
\label{app:contour}

A quasinormal mode is a solution of Eq.~\eqref{eq:master_x} that is purely outgoing, $\psi\propto e^{i\omega|x|}$, on both sides of the barrier. It exists only at complex $\omega$ with ${\rm Im}\,\omega<0$, and the outgoing wave then grows along the real $x$ axis. A shooting method started at large real $x$ is swamped by the incoming solution. We follow the complex-coordinate approach \cite{Andersson:1992prs,Glampedakis:2003dn} and continue $x$ onto the ray $x=x_c+s\,e^{i\vartheta}$, with $x_c=20r_e$, $s\ge0$ and $\vartheta>-\arg\omega$. On the ray the outgoing wave decays, and we seed it at the far end with $\psi=1$, $\psi'=i\omega$. In this appendix a prime denotes $d/dx$. The potential is continued through the inverse $u(x)$ of Eq.~\eqref{eq:mouth_tortoise_closed_form}, found by Newton iteration at each complex $x$. Its nearest branch points, at $x=\pm2\pi i\,r_e$, stay well away from the contour. The angle $\vartheta$ lies between $36^\circ$ and $60^\circ$, and the ray is long enough that the incoming solution is suppressed by $e^{-120}$.

The symmetry of Sec.~\ref{sec:global_potentials} replaces the outgoing condition on the throat side by a condition at the peak. A solution outgoing on the right must match the mirror image of a solution outgoing on the left. This requires $\psi'(0)=0$ for even modes, which include the fundamental, and $\psi(0)=0$ for odd modes. The method thus never reaches the degenerate horizon of the extremal geometry. We integrate with a fourth-order Runge--Kutta scheme with step $0.01r_e$, down the ray and then along the real axis to $x=0$, and find the root of $\psi'(0)/\psi(0)$ by a secant iteration. It starts from the leading-order Schutz--Will estimate \cite{Schutz:1985km}
\begin{equation}
    \omega_{\rm WKB}^2 = V_0 - i\left(n+\tfrac12\right)\sqrt{-2V_0''},
    \label{eq:qnm_wkb}
\end{equation}
with $n$ the overtone number. Halving the step, moving $x_c$ to $35r_e$ or raising $\vartheta$ by $0.2$ changes $\omega r_e$ by at most $3\times10^{-11}$ for the scalar modes at $l=0,1,2$ and the gauge mode at $l=1$. We compute the fundamental modes $n=0$ only.

As an independent check we fire a narrow Gaussian pulse ($b=r_e$) at a single barrier, evolve it with the leapfrog scheme \eqref{eq:leapfrog}, and fit the transmitted ringdown to a damped cosine in four time windows between $18r_e$ and $70r_e$. This shares no code with the contour integration. The fit agrees with the contour values to within $0.54\%$ in $\omega_R$ and $4.8\%$ in $|\omega_I|$ (Table~\ref{tab:qnm_ps}). Its accuracy is limited by the direct pulse in early windows and by the late-time tail \cite{Price:1972pw} in late ones.

\section{Difference schemes, validation and convergence}
\label{app:numerics}

Equation \eqref{eq:master_x} has no first-derivative term. Writing $W_j=\omega^2-V(x_j)$ on a uniform grid $x_j=-X+jh$, the three-point recursion
\begin{align}
    \left(1+\tfrac{h^2}{12}W_{j-1}\right)\chi_{j-1}
    =\;&2\left(1-\tfrac{5h^2}{12}W_j\right)\chi_j \nonumber\\
    &-\left(1+\tfrac{h^2}{12}W_{j+1}\right)\chi_{j+1}
    \label{eq:numerov}
\end{align}
has a local truncation error of order $h^6$. This is two orders better than the plain three-point difference scheme, whose local error is of order $h^4$, because the leading error term is absorbed into the coefficients using the differential equation itself.

For the time-domain checks we evolve $\partial_t^2\Psi=\partial_x^2\Psi-V\Psi$. Writing $\Psi_j^n\equiv\Psi(x_j,t_n)$ on the uniform grid $x_j=x_0+j\Delta x$, $t_n=n\Delta t$, the standard second-order central difference in both $t$ and $x$ gives the explicit update rule
\begin{align}
    \Psi_j^{n+1}=\;&2\Psi_j^n-\Psi_j^{n-1}+\left(\frac{\Delta t}{\Delta x}\right)^2\left(\Psi_{j+1}^n-2\Psi_j^n+\Psi_{j-1}^n\right) \nonumber\\
    &-\Delta t^2\,V(x_j)\,\Psi_j^n.
    \label{eq:leapfrog}
\end{align}
The scheme is started from the initial data and its time derivative, encoded through the fictitious value $\Psi_j^{-1}=\Psi(x_j+\Delta t,0)$ for the right-moving pulse. Stability requires the Courant number $\Delta t/\Delta x\le1$, and we use $\Delta t/\Delta x=0.5$.

\subsection{Time-domain check}
\label{sec:fdtd}

The Numerov coefficients are complex, and their phases carry the dispersion of the reflection and transmission processes. We test modulus and phase together with a method that shares no code with the frequency-domain solver. We launch a right-moving Gaussian pulse $\Psi(x,0)=\exp[-(x-x_{\rm p})^2/2b^2]$ with $b=2r_e$ at $x_{\rm p}=-350r_e$, on the throat side of the barrier. We evolve it with the leapfrog scheme \eqref{eq:leapfrog}, with $\Delta x=0.025r_e$ and $\Delta t=0.0125r_e$, on a domain large enough that no boundary reflection reaches the detectors during the run. Two detectors record the field. The transmitted signal is recorded at $x=x_{\rm det}$ and the reflected signal at $x=-x_{\rm det}$, with $x_{\rm det}=350r_e$. The second detector sits at the launch point. The incident pulse has passed it long before the reflection returns.

We compare the two recorded signals with a prediction built entirely from the Numerov amplitudes $\beta(\omega),\kappa(\omega)$, with no time evolution at all. Each frequency component of the launched pulse scatters according to $\beta$ and $\kappa$, and the scattered components are summed back by a Fourier transform. With $\tilde\Psi_0(\omega)=b\sqrt{2\pi}\,e^{-b^2\omega^2/2}$ the predicted waveforms are
\begin{align}
    \Psi_{T}(\tau)&=\int\frac{d\omega}{2\pi}\,\tilde\Psi_0(\omega)\,\kappa(\omega)\,\frac{w_+(x_{\rm det})}{w_-(x_{\rm p})}\,e^{-i\omega(\tau+x_{\rm det}-x_{\rm p})},\nonumber\\
    \Psi_{R}(\tau)&=\int\frac{d\omega}{2\pi}\,\tilde\Psi_0(\omega)\,\beta(\omega)\,\frac{w_+(-x_{\rm det})}{w_-(x_{\rm p})}\,e^{-i\omega(\tau+x_{\rm det}-x_{\rm p})}.
    \label{eq:fdtd_pred}
\end{align}
The local time $\tau$ is measured from the arrival time of a freely propagating pulse at each detector. The ratio of Riccati--Hankel functions carries each component exactly through the centrifugal tails between the barrier and the launch and detector points. Far from the barrier it reduces to $e^{i\omega(x_{\rm det}-x_{\rm p})}$.

\begin{figure*}
    \centering
    \includegraphics[width=0.24\textwidth]{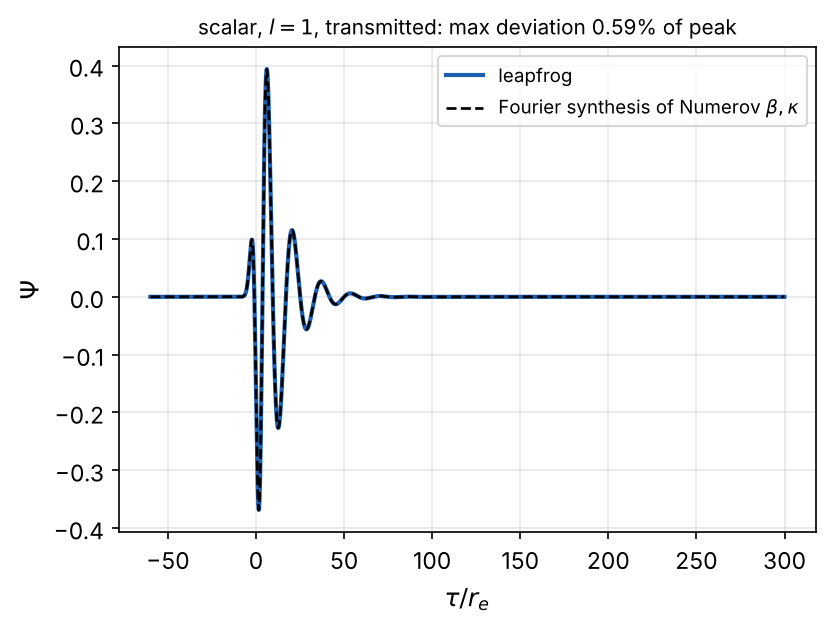}\hfill
    \includegraphics[width=0.24\textwidth]{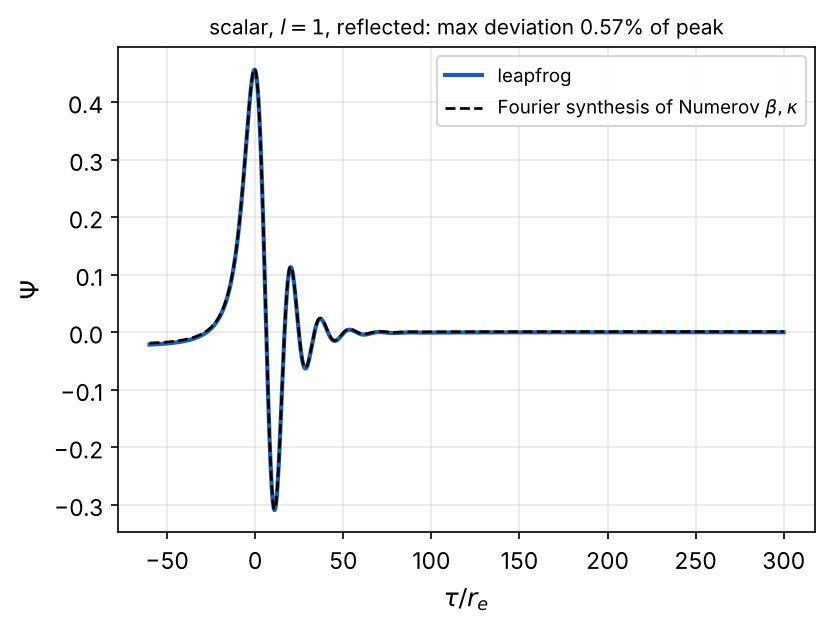}\hfill
    \includegraphics[width=0.24\textwidth]{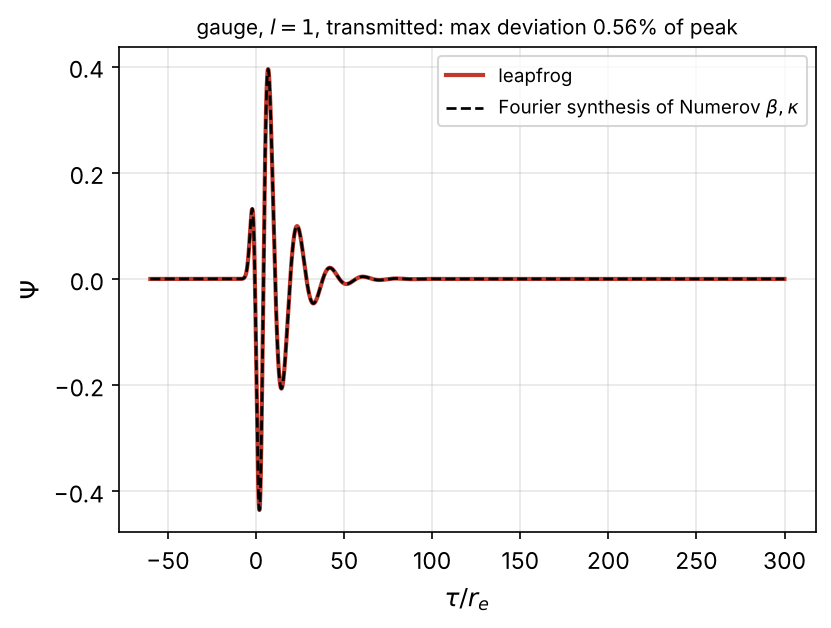}\hfill
    \includegraphics[width=0.24\textwidth]{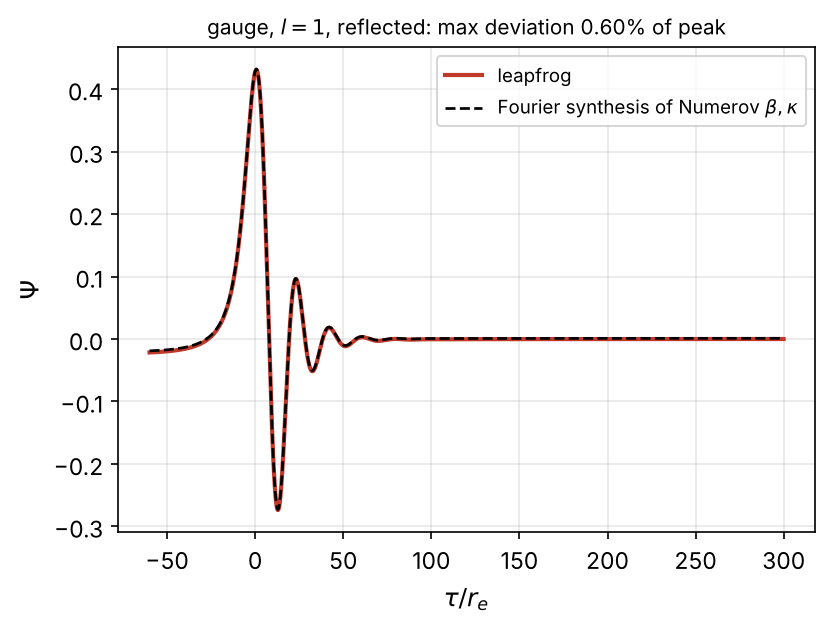}
    \caption{Independent time-domain check of the single-barrier scattering at $l=1$. From left to right, the scalar transmitted and reflected pulses and the gauge transmitted and reflected pulses. A Gaussian pulse is evolved through the barrier with a leapfrog scheme (coloured) and compared with the waveform predicted from the Numerov $\beta(\omega),\kappa(\omega)$ by Fourier synthesis (dashed). The local time $\tau$, in units of $r_e$, is measured from the free-propagation arrival time. In units of $r_e$ it is the same for both parameter sets.}
    \label{fig:fdtd}
\end{figure*}

The evolution and the prediction of Eq.~\eqref{eq:fdtd_pred} agree to $0.6\%$ of the peak for both the transmitted and the reflected pulse, for both fields (Fig.~\ref{fig:fdtd}). This confirms that the dispersive phases of $\beta$ and $\kappa$, which the real-phase WKB treatment of Sec.~\ref{sec:wkb} discards, are physical and correctly captured by Numerov.

\subsection{Exactly solvable test}

We test the solver on the P\"oschl--Teller barrier, $V=V_0\,{\rm sech}^2(ax)$ with $V_0=0.02/r_e^2$ and $a=0.08/r_e$. Its transmission coefficient is known exactly \cite[$\S$25, Problem~4]{Landau:1991wop},
\begin{equation}
    |\kappa(\omega)|^2_{\rm exact}=\frac{\sinh^2(\pi\omega/a)}{\sinh^2(\pi\omega/a)+\cosh^2\!\left(\dfrac{\pi}{2}\sqrt{4V_0/a^2-1}\right)}.
    \label{eq:PT_exact}
\end{equation}
The scheme reproduces Eq.~\eqref{eq:PT_exact} to $6\times10^{-12}$ and unitarity to $2\times10^{-11}$ over $0.01\le\omega r_e\le0.6$. The test uses the grid step $h=0.05\,r_e$ of the MMP runs. The boundary conditions are imposed at $|x|=600\,r_e$, where $V<10^{-42}/r_e^2$ and the free solutions at $l=0$ are plane waves.

\subsection{Convergence}

Table~\ref{tab:convergence} shows how the Numerov amplitudes $\beta$ and $\kappa$ at $l=1$ change when the step $h$, $X_{\min}$ or $N_\omega$ is halved or doubled, with the geometry fixed. The relative change of $|\kappa|^2$ stays below $7\times10^{-7}$. The phase of $\kappa$ changes by less than $4\times10^{-4}$~rad, except for $h=0.1\,r_e$, where it reaches $6\times10^{-3}$~rad at the highest frequencies.

As an independent check we integrate Eq.~\eqref{eq:master_x} in the variable $\ln u$, where the potential needs no inversion of Eq.~\eqref{eq:mouth_tortoise_closed_form}. We use the adaptive eighth-order Dormand--Prince method \cite{Hairer:1993}, in its DOP853 implementation in SciPy, at a relative tolerance of $10^{-12}$, and fit $A\,w_-+B\,w_+$ to $\chi$ and $d\chi/dx$ at the single point $x=-X$. At 14 frequencies between $\omega r_e=0.002$ and $1.5$ this agrees with Numerov to $2\times10^{-6}$ in $|\kappa|^2$. The phases agree to $6\times10^{-5}$~rad for $\omega r_e\le1$ and to $4\times10^{-4}$~rad at $\omega r_e=1.5$. This holds down to $|\kappa|^2\sim10^{-17}$. Below it rounding in double precision could matter. We therefore recompute $|\kappa|^2$ in 20-digit arithmetic with the Taylor-series integrator \texttt{odefun} of the mpmath library, for both fields at $l=1$ and $l=2$ and $\omega r_e=0.002$, $0.01$, $0.05$ and $0.2$. The two results agree to $2.5\times10^{-7}$ down to $|\kappa|^2=2.1\times10^{-30}$. Rounding limits neither $|\kappa|^2$ nor the quality factors of Sec.~\ref{sec:qnm_cavity}.

\begin{table*}
\centering
\renewcommand{\arraystretch}{1.3}
\setlength{\arrayrulewidth}{0.6pt}
\caption{Convergence of the Numerov amplitudes $\beta$ and $\kappa$ of one barrier at $l=1$ with respect to the numerical settings of the solver. The geometry and the parameters $q$, $\ell_p$, $g$ and $\delta$ are fixed. Each row changes one setting from the values used for all results in the paper ($h=0.05$, $X_{\min}=2000$, $N_\omega=400$). $h$ is the grid step, and the boundary conditions are imposed at $|x|=X(\omega)=\max(X_{\min},N_\omega/\omega)$. The entries are the maximum over $0.002\le\omega r_e\le1.5$ of the relative change of $|\kappa|^2$ and of the change of $\arg\kappa$ (in rad). The last column is the maximum unitarity defect $\big||\beta|^2+|\kappa|^2-1\big|$ of the run with the changed setting. $h$ and $X_{\min}$ are in units of $r_e$, and $N_\omega$ is dimensionless. The entries are the same for both parameter sets.}
\label{tab:convergence}
\begin{tabular}{|c|l|c|c|c|}
\hline
field & \hcell{changed\\parameter} & \hcell{max relative\\change of $|\kappa|^2$} & \hcell{max change of\\$\arg\kappa$ (rad)} & \hcell{max unitarity\\defect} \\
\hline
\multirow{6}{*}{scalar} & $h=0.1$ & $2.7\times10^{-7}$ & $5.9\times10^{-3}$ & $2\times10^{-10}$ \\
 & $h=0.025$ & $2.4\times10^{-7}$ & $3.7\times10^{-4}$ & $5\times10^{-10}$ \\
 & $X_{\min}=1000$ & $1.8\times10^{-7}$ & $1.8\times10^{-4}$ & $5\times10^{-10}$ \\
 & $X_{\min}=4000$ & $1.2\times10^{-7}$ & $3.9\times10^{-4}$ & $5\times10^{-10}$ \\
 & $N_\omega=200$ & $6.9\times10^{-7}$ & $5.4\times10^{-5}$ & $3\times10^{-10}$ \\
 & $N_\omega=800$ & $1.7\times10^{-7}$ & $2.5\times10^{-5}$ & $5\times10^{-10}$ \\
\hline
\multirow{6}{*}{gauge} & $h=0.1$ & $2.6\times10^{-7}$ & $5.9\times10^{-3}$ & $9\times10^{-11}$ \\
 & $h=0.025$ & $2.4\times10^{-7}$ & $3.7\times10^{-4}$ & $5\times10^{-10}$ \\
 & $X_{\min}=1000$ & $1.9\times10^{-7}$ & $1.9\times10^{-4}$ & $3\times10^{-10}$ \\
 & $X_{\min}=4000$ & $1.3\times10^{-7}$ & $3.9\times10^{-4}$ & $3\times10^{-10}$ \\
 & $N_\omega=200$ & $6.3\times10^{-7}$ & $5.2\times10^{-5}$ & $2\times10^{-10}$ \\
 & $N_\omega=800$ & $1.7\times10^{-7}$ & $2.4\times10^{-5}$ & $3\times10^{-10}$ \\
\hline
\end{tabular}
\end{table*}

\subsection{Absorption sum}

The sum $\Sigma$ of Eq.~\eqref{eq:bath_rate} runs over $l\le44$. For $l\le2$ we use the Numerov amplitudes at every frequency. For $l>2$ we use them for $\omega r_e\ge0.1$ and Eq.~\eqref{eq:lowfreq} below. We tabulate $\Sigma$ on $140$ log-spaced frequencies between $\omega r_e=0.002$ and $6$ and interpolate. Below this range we use Eq.~\eqref{eq:sommerfeld} for $l\le1$, and above it the geometric value $16\omega^2r_e^2$. For set~2 the X-ray rate above $4.2$~keV therefore uses the geometric value. The trapped energy of Eq.~\eqref{eq:E_trap} uses Eq.~\eqref{eq:sommerfeld} for $l\le7$.

\section{Low-frequency matching}
\label{app:lowfreq}

We set $r_e=1$ and $u=r-1$. Then $r=1+u$, $f=u^2/(1+u)^2$ and $d/dx=f\,d/du$. A prime denotes $d/du$.

\subsection{Static solutions}

At $\omega=0$ the master equation \eqref{eq:master_x} reads $d^2\chi/dx^2=V\chi$. It has two independent solutions, which we call the static solutions. We define $\Xi_1$ as the static solution that vanishes for $u\to0$, on the throat side, and $\Xi_2$ as the one that decays for $u\to\infty$. A superscript marks the field, $(s)$ for the scalar and $(g)$ for the gauge field.

For the scalar the master function is $\chi=rR=(1+u)R$. With $g_{\theta\theta}=r^2$ we have $g_{\theta\theta}f=u^2$, and Eq.~\eqref{eq:radial_scalar_general} at $\omega=0$ becomes $(u^2R')'=l(l+1)R$. Its solutions are $R=u^l$ and $R=u^{-l-1}$. Multiplying them by $1+u$ gives
\begin{equation}
    \Xi_1^{(s)}=(1+u)\,u^{l},\qquad \Xi_2^{(s)}=(1+u)\,u^{-l-1}.
    \label{eq:static_s}
\end{equation}
For the gauge field the master function is $\Psi$ itself. Eq.~\eqref{eq:axial_master} at $\omega=0$ reads $\big[u^2(1+u)^{-2}\Psi'\big]'=l(l+1)(1+u)^{-2}\Psi$. Its solutions follow from the scalar ones as $\Psi=(1+u)^2\,[uR/(1+u)]'$, which one checks by substitution. With $R=u^l$ and $R=u^{-l-1}$ this gives
\begin{equation}
    \Xi_1^{(g)}=u^{l}(lu+l+1),\qquad \Xi_2^{(g)}=u^{-l-1}\big(l+(l+1)u\big),
    \label{eq:static}
\end{equation}
where we drop an overall sign of $\Xi_2^{(g)}$. We omit the superscript when a statement holds for both fields. For both, $\Xi_1$ vanishes as $u^l$ at small $u$ and grows as $u^{l+1}$ at large $u$, while $\Xi_2$ grows as $u^{-l-1}$ at small $u$ and decays as $u^{-l}$ at large $u$. The Couch--Torrence map $u\to1/u$ exchanges $\Xi_1$ and $\Xi_2$ up to normalisation.

Let $p_1$ and $p_2$ be the coefficients of $u^{l+1}$ in $\Xi_1$ and of $u^{-l}$ in $\Xi_2$ at large $u$. Let $\bar p_1$ and $\bar p_2$ be the coefficients of $u^{l}$ in $\Xi_1$ and of $u^{-l-1}$ in $\Xi_2$ at small $u$. For the scalar all four are $1$. For the gauge field $p_1=l$, $p_2=l+1$, $\bar p_1=l+1$ and $\bar p_2=l$. The solution $\Xi$ of Sec.~\ref{sec:transmission} is $\Xi_2/p_2$.

\subsection{Far regions}

In the variable $u$ the master equation \eqref{eq:master_x} reads $f(f\chi')'+(\omega^2-V)\chi=0$. We write $\chi=f^{-1/2}\hat\chi$. Then
\begin{equation}
    f(f\chi')'=f^{3/2}\Big[\hat\chi''+\Big(\frac{f'^2}{4f^2}-\frac{f''}{2f}\Big)\hat\chi\Big],
\end{equation}
and the first derivative of $\hat\chi$ drops out. From $\ln f=2\ln u-2\ln(1+u)$,
\begin{align}
    \frac{f'}{f}&=\frac{2}{u(1+u)},\nonumber\\
    \frac{f''}{f}&=\Big(\frac{f'}{f}\Big)'+\frac{f'^2}{f^2}=\frac{2-4u}{u^2(1+u)^2},\nonumber\\
    \frac{f'^2}{4f^2}-\frac{f''}{2f}&=\frac{1-(1-2u)}{u^2(1+u)^2}=\frac{2}{u(1+u)^2}.
\end{align}
Dividing by $f^{3/2}$ gives $\hat\chi''+[(\omega^2-V)/f^2+2/u(1+u)^2]\hat\chi=0$, with $\omega^2/f^2=\omega^2(1+u)^4/u^4$. For the scalar Eq.~\eqref{eq:V_of_u} gives $V/f^2=l(l+1)/u^2+2/[u(1+u)^2]$, and the two terms $2/[u(1+u)^2]$ cancel. For the gauge field $V/f^2=l(l+1)/u^2$. Hence
\begin{equation}
    \hat\chi''+\left[\omega^2\frac{(1+u)^4}{u^4}-\frac{l(l+1)}{u^2}+U(u)\right]\hat\chi=0,
    \label{eq:phi_eq}
\end{equation}
with $U=0$ for the scalar and $U=2/[u(1+u)^2]$ for the gauge field. For $u\gg1$ we have $f\to1$ and $\hat\chi\simeq\chi$. Sec.~\ref{sec:coulomb} shows that in the variable $\xi=\omega u$ Eq.~\eqref{eq:phi_eq} becomes the Coulomb equation \eqref{eq:coulomb} with $\eta=-2\omega$, up to terms of relative order $\omega^2$. Its solutions are the Coulomb functions $F_l(\eta,\xi)$ and $G_l(\eta,\xi)$, and $H^\pm_l=G_l\pm iF_l$ are the outgoing and incoming waves \cite[Ch.~14]{Abramowitz:1964}. For $\xi\ll1$,
\begin{equation}
    F_l\simeq C_l(\eta)\,\xi^{l+1},\qquad G_l\simeq\frac{\xi^{-l}}{(2l+1)C_l(\eta)},
\end{equation}
with $C_l(\eta)$ of Eq.~\eqref{eq:Cl}.
The master equation \eqref{eq:master_x} is invariant under $x\to-x$. For $u\ll1$ the function $\chi$ therefore obeys the Coulomb equation \eqref{eq:coulomb} in the variable $\bar\xi=\omega/u$.

\subsection{Matching}

On the right the solution is purely outgoing, $\chi=A_{\rm out}H^+_l(\eta,\omega u)$. On the left $x\simeq-1/u$, and the incident wave $e^{i\omega x}$ is $H^-_l(\eta,\omega/u)$. There $\chi=A_{\rm in}H^-_l(\eta,\omega/u)+A_{\rm ref}H^+_l(\eta,\omega/u)$. In the overlap regions $1\ll u\ll1/\omega$ and $\omega\ll u\ll1$ both far solutions take their small-$\xi$ forms. Both must equal one combination $\mathcal M_2\Xi_2+\mathcal M_1\Xi_1$ of the static solutions of Eqs.~\eqref{eq:static_s} and~\eqref{eq:static}, with constants $\mathcal M_1$ and $\mathcal M_2$. At large $u$ the powers $u^{-l}$ and $u^{l+1}$ give
\begin{align}
    \mathcal M_2\,p_2&=\frac{A_{\rm out}\,\omega^{-l}}{(2l+1)C_l},\nonumber\\
    \mathcal M_1\,p_1&=iA_{\rm out}C_l\,\omega^{l+1}.
\end{align}
At small $u$ we write $A_{\rm in}H^-_l+A_{\rm ref}H^+_l=(A_{\rm in}+A_{\rm ref})G_l+i(A_{\rm ref}-A_{\rm in})F_l$. The powers $u^{-l-1}$ and $u^{l}$ give
\begin{align}
    \mathcal M_2\,\bar p_2&=iC_l\,\omega^{l+1}(A_{\rm ref}-A_{\rm in}),\nonumber\\
    \mathcal M_1\,\bar p_1&=\frac{(A_{\rm in}+A_{\rm ref})\,\omega^{-l}}{(2l+1)C_l}.
\end{align}
Eliminating $\mathcal M_1$ and $\mathcal M_2$ between the two pairs gives
\begin{align}
    A_{\rm ref}-A_{\rm in}&=\frac{\mathcal M_2\,\bar p_2}{iC_l\,\omega^{l+1}}\nonumber\\
    &=-\frac{i\,\bar p_2}{(2l+1)p_2C_l^2}\,\omega^{-2l-1}A_{\rm out},\nonumber\\
    A_{\rm ref}+A_{\rm in}&=(2l+1)C_l\,\omega^{l}\,\mathcal M_1\,\bar p_1\nonumber\\
    &=\frac{i(2l+1)C_l^2\bar p_1}{p_1}\,\omega^{2l+1}A_{\rm out}.
\end{align}
Half the second minus half the first is
\begin{align}
    \frac{A_{\rm in}}{A_{\rm out}}=\frac i2\bigg[&\frac{(2l+1)C_l^2\bar p_1}{p_1}\,\omega^{2l+1}\nonumber\\
    &+\frac{\bar p_2}{(2l+1)p_2C_l^2}\,\omega^{-2l-1}\bigg].
\end{align}
As $\xi\to\infty$ the Coulomb functions approach \cite[Ch.~14]{Abramowitz:1964}
\begin{align}
    F_l&\simeq\sin\varphi_l,\qquad G_l\simeq\cos\varphi_l,\nonumber\\
    \varphi_l&=\xi-\eta\ln2\xi-\frac{l\pi}{2}+\sigma_l,
    \label{eq:coulomb_phase}
\end{align}
with the constant $\sigma_l=\arg\Gamma(l+1+i\eta)$. Then $H^\pm_l\simeq e^{\pm i\varphi_l}$, and $|H^\pm_l|\to1$ only in this limit. At finite $\xi$ the modulus differs from one, and it diverges as $\xi\to0$. Far from the barrier on both sides each $H^\pm_l$ is a wave of unit amplitude, and $|\kappa|=|A_{\rm out}/A_{\rm in}|$. The phase $\xi-\eta\ln2\xi$ differs from $\omega|x|$ only by a constant, since the term $-\eta\ln2\xi=2\omega\ln2\xi$ reproduces the logarithm $2\omega\ln u$ of the tortoise coordinate. The constant changes the phase of $\kappa$ but not $|\kappa|$. The second term in the bracket dominates as $\omega\to0$, and
\begin{align}
    |\kappa|^2=\Big|\frac{A_{\rm out}}{A_{\rm in}}\Big|^2&\simeq\Big[\frac{2(2l+1)p_2C_l^2}{\bar p_2}\,\omega^{2l+1}\Big]^2\nonumber\\
    &=4(2l+1)^2\left(\frac{p_2}{\bar p_2}\right)^2C_l(\eta)^4\,\omega^{4l+2}.
\end{align}
At $\eta=0$ we have $C_l(0)=2^l\,l!/(2l+1)!=1/(2l+1)!!$, since $(2l+1)!=2^l\,l!\,(2l+1)!!$. With $\mu_l=\bar p_2/p_2$ and $(2l+1)!!=(2l+1)(2l-1)!!$,
\begin{align}
    4(2l+1)^2\,\frac{C_l(0)^4}{\mu_l^2}&=\frac{4(2l+1)^2}{\mu_l^2[(2l+1)!!]^4}\nonumber\\
    &=\frac{4}{\big[(2l-1)!!\,(2l+1)!!\,\mu_l\big]^2},
\end{align}
which is $K_l$ of Eq.~\eqref{eq:lowfreq}. The solution $\Xi=\Xi_2/p_2$ decays as $u^{-l}\simeq x^{-l}$ on the right and grows as $\mu_lu^{-l-1}\simeq\mu_l|x|^{l+1}$ on the left. Keeping $\eta=-2\omega$ multiplies $|\kappa|^2$ by $[C_l(\eta)/C_l(0)]^4=[e^{-\pi\eta/2}|\Gamma(l+1+i\eta)|/l!]^4$, from Eq.~\eqref{eq:Cl}. This is the Coulomb factor $S_l$ of Eq.~\eqref{eq:sommerfeld}. The matching neglects terms of relative order $\omega^2$, with logarithms. For the gauge field the static solutions contain the short-range term $U$ in the overlap region. In the far region it is $U/\omega^2\simeq2\omega/\xi^3$, formally of first order in $\omega$ like the Coulomb term. The numerics of Sec.~\ref{sec:coulomb} show that it leaves no correction linear in $\omega$, and that the remainder of both fields grows as $\omega^2\ln\omega$.

\section{Tortoise length of the throat}
\label{app:D}

We set $r_e=1$ and evaluate
\begin{equation}
    J=L_{\rm wh}\int_0^{\tilde\rho}\frac{d\rho}{1+\rho^2+\gamma},
    \label{eq:J_int}
\end{equation}
with $\gamma\propto\alpha$ of Eq.~\eqref{eq:gam}. Up to the junction $|\gamma|\ll1+\rho^2$, since
\begin{equation}
    \frac{|\gamma|}{\rho^2}\simeq\frac{\pi}{2}\,\alpha\rho\le\frac{\pi}{2}\,\alpha\tilde\rho\sim10^{-6}
\end{equation}
(Table~\ref{tab:bg_params}). To first order in $\alpha$ the integrand is $1/(1+\rho^2)-\gamma/(1+\rho^2)^2$, and
\begin{align}
    J&=L_{\rm wh}\arctan\tilde\rho+L_{\rm wh}\alpha\,\mathcal I(\tilde\rho),\nonumber\\
    \mathcal I&=\int_0^{\tilde\rho}\frac{\rho^2+\rho(3+\rho^2)\arctan\rho-\ln(1+\rho^2)}{(1+\rho^2)^2}\,d\rho .
\end{align}
We substitute $\rho=\tan\theta$ and integrate the term $\theta\tan\theta$, which diverges at $\theta=\pi/2$, by parts. With $\Theta=\arctan\tilde\rho$,
\begin{align}
    \mathcal I=-\Theta\ln\cos\Theta+\int_0^{\Theta}\Big[&\sin^2\theta+\theta\sin2\theta\nonumber\\
    &+(2+\cos2\theta)\ln\cos\theta\Big]d\theta .
    \label{eq:I_theta}
\end{align}
For large $\tilde\rho$ we have $\cos\Theta=(1+\tilde\rho^2)^{-1/2}$ and $\Theta\to\pi/2$. The boundary term is then $\tfrac{\Theta}{2}\ln(1+\tilde\rho^2)\to\tfrac{\pi}{2}\ln\tilde\rho$. The integrand has only a logarithmic singularity at $\theta=\pi/2$, and we may set $\Theta=\pi/2$ in the integral of Eq.~\eqref{eq:I_theta}. With $\int_0^{\pi/2}\ln\cos\theta\,d\theta=-\tfrac{\pi}{2}\ln2$ the integral is $\tfrac{3\pi}{4}-\pi\ln2$, and
\begin{equation}
    \mathcal I=\frac{\pi}{2}\ln\tilde\rho+\pi\left(\frac34-\ln2\right)
\end{equation}
up to terms that vanish as $\tilde\rho\to\infty$. With $L_{\rm wh}\alpha=4/\pi$ from Eq.~\eqref{eq:bg_identities}, the correction to $J$ is $\tfrac{4}{\pi}\mathcal I=2\ln\tilde\rho+3-4\ln2$. With $\arctan\tilde\rho=\pi/2-1/\tilde\rho+O(\tilde\rho^{-3})$ and $\tilde\rho=L_{\rm wh}\delta$, Eq.~\eqref{eq:coord_rel},
\begin{equation}
    J=\frac{\pi}{2}L_{\rm wh}-\frac1\delta+2\ln\tilde\rho+3-4\ln2 .
\end{equation}
Adding $z_{\rm peak}=1/\delta-2\ln\delta-\delta$ gives
\begin{align}
    D&=\frac{\pi}{2}L_{\rm wh}+2\ln\frac{\tilde\rho}{\delta}+3-4\ln2-\delta\nonumber\\
     &=\frac{\pi}{2}L_{\rm wh}+2\ln\frac{L_{\rm wh}}{4}+3-\delta,
\end{align}
which is Eq.~\eqref{eq:D_closed} with the term $-\delta$ kept. The next term of the expansion of the integrand is $+\gamma^2/(1+\rho^2)^3$. It is of second order in $\alpha$, but the range of integration is long. At large $\rho$, Eq.~\eqref{eq:largeg} gives $\gamma\simeq-\tfrac{\pi}{2}\alpha\rho^3$, and the term tends to the constant $\pi^2\alpha^2/4$. Its contribution to $J$ is
\begin{equation}
    J_2=L_{\rm wh}\int_0^{\tilde\rho}\frac{\gamma^2\,d\rho}{(1+\rho^2)^3}\simeq\frac{\pi^2}{4}\,L_{\rm wh}\alpha\cdot\alpha\tilde\rho=4\delta,
\end{equation}
since $L_{\rm wh}\alpha=4/\pi$ and $\alpha\tilde\rho=\alpha L_{\rm wh}\delta=4\delta/\pi$. The small factor $\alpha^2$ is compensated by $L_{\rm wh}\tilde\rho=L_{\rm wh}^2\delta$. The terms of third and higher order in $\gamma$ are of order $\delta^2$. Together with the term $-\delta$ of $z_{\rm peak}$ in Sec.~\ref{sec:global_potentials}, $J_2$ shifts $D$ by $+3\delta\,r_e=3\times10^{-6}r_e$ from Eq.~\eqref{eq:D_closed}. This is the difference between the quadrature and Eq.~\eqref{eq:D_closed}.

\bibliographystyle{spphys-arxiv}
\bibliography{biblio}

\end{document}